\pdfoutput=1
\newcommand*{\ATLASLATEXPATH}{}
\documentclass[PAPER, cernpreprint=true, texlive=2016, UKenglish]{\ATLASLATEXPATH atlasdoc}
 
\usepackage{\ATLASLATEXPATH atlaspackage}
\usepackage{\ATLASLATEXPATH atlasbiblatex}
 
\usepackage{\ATLASLATEXPATH atlasphysics}
 
\graphicspath{{logos/}{figures/}}
 
\usepackage{ANA-TOPQ-2016-05-PAPER-defs}
\usepackage{float}

\AtlasTitle{Measurement of $K_S^0$ and $\Lambda^0$ production in $t \bar{t}$ dileptonic events in $pp$ collisions at $\sqrt{s} =$ 7 TeV with the ATLAS detector}

\AtlasAbstract{
Measurements of $K_S^0$ and $\Lambda^0$ production in $t\bar{t}$ final states have been performed. They are based on a data sample with integrated  luminosity of 4.6 $\mathrm{fb}^{-1}$ from proton--proton collisions at a centre-of-mass energy of 7 TeV, collected in 2011 with the ATLAS detector at the Large Hadron Collider. Neutral strange particles are separated into three classes, depending on whether they are contained in a jet, with or without a $b$-tag, or not associated with a selected jet. The aim is to look for differences in their main kinematic distributions. A comparison of data with several Monte Carlo simulations using different hadronisation  and fragmentation schemes, colour reconnection models and different tunes for the underlying event has been made. The production of neutral strange particles in $t\bar{t}$ dileptonic events is found to be well described by current Monte Carlo models for $K_S^0$ and $\Lambda^0$ production within jets, but not for those produced outside jets.}
 
\author{The ATLAS Collaboration}
 
\AtlasRefCode{TOPQ-2016-05}

\PreprintIdNumber{CERN-EP-2019-112}
 
\AtlasJournal{Eur. Phys. J. C}
\AtlasJournalRef{Eur. Phys. J. C 79 (2019) 1017}
\AtlasDOI{10.1140/epjc/s10052-019-7512-y}

\AtlasCoverSupportingNote{ATL-COM-PHYS-2015-282}{https://cds.cern.ch/record/2010273}

\AtlasCoverAnalysisTeam{Sergio Calvente, Javier Llorente, Fernando Barreiro}

\AtlasCoverEdBoardMember{Alfred Goshaw}
\AtlasCoverEdBoardMember{Thorsten Kuhl (chair)}
\AtlasCoverEdBoardMember{Dominique Pallin}
\AtlasCoverEdBoardMember{Daniel Scheirich}

\AtlasCoverEgroupEditors{atlas-TOPQ-2016-05-editors@cern.ch}
 
\AtlasCoverEgroupEdBoard{atlas-TOPQ-2016-05-editorial-board@cern.ch}

\hypersetup{pdftitle={ATLAS document},pdfauthor={The ATLAS Collaboration}}
 
\begin{document}
 
\maketitle
 
\tableofcontents

\clearpage
\section{Introduction and motivation}
\label{sec:intro}
 
Neutral strange particle production has been studied in collider experiments using $\ee$ \cite{PLUTO,JADE,TPC,MARK,HRS,CELLO,TASSO,ALEPH,DELPHI,OPAL,L3}, $pp$ \cite{minimum_bias,lambdas,CMS,CMS2,ALICE, LHCb, LHCb2, STAR1},
$\ppbar$ \cite{UA5,UA1,Tevatron}, $ep$ \cite{ZEUS,H1} and heavy-ion collisions \cite{STAR2,STAR3}, as well as in fixed-target experiments \cite{fixed1,fixed2,fixed3,fixed4,fixed5,fixed6,fixed7,fixed8,fixed9,fixed10,fixed11,fixed12,fixed13,fixed14}. These measurements provide interesting tests of theoretical jet fragmentation functions \cite{kramer} and can be used to validate and tune the values of empirical parameters used in the parton shower and fragmentation parts of the Monte Carlo (MC) models. Since the mass of the strange quark is comparable to the QCD scale parameter $\Lambda_{\mathrm{QCD}}$, perturbative calculations cannot be performed. These models must be highly accurate to constrain the underlying event (UE) effects in the high transverse momentum ($\pt$) production investigated at the Large Hadron Collider (LHC). In particular the ratio $\gamma_{\mathrm{s}}=s/u$, giving the suppression factor of strange to non-strange meson production in the hadronic final states, is measured to be larger in $pp$ collisions than in $\ee$ annihilation. A review is given in Ref.~\cite{rope}.
 
It was suggested \cite{Nielsen} that the suppression factor $\gamma_{\mathrm{s}}$ would be significantly larger, or even tend to unity, in nucleus--nucleus collisions because the many strings produced within the Lund fragmentation scheme in a limited phase space may interact, giving rise to the formation of `colour ropes'. Recent data from RHIC \cite{STAR2,STAR4} tend to support these ideas and show that neutral strange particle production is enhanced. In $pp$ collisions at LHC energies, many overlapping strings due to multi-parton interactions are also expected to come into play, so that higher rates of strange meson and baryon production are expected \cite{rope}. This effect was confirmed recently by the ALICE \cite{ALICE} and CMS \cite{CMS2} collaborations.
 
The measurements presented in this paper are useful contributions for future determinations of \Vts. The prospects for directly measuring the CKM matrix element \Vts are discussed in Refs.~\cite{ABL,ichep}. The idea is to measure the fraction of $pp\to \ttbar\to \Wboson^+ b \Wboson^- \bar{s} (\Wboson^+ s \Wboson^- \bar{b})$ in $\ttbar$ decays. Since this is small compared with the dominant background $pp\to \ttbar\to \Wboson^+ b \Wboson^- \bar{b}$, a good understanding of neutral strange particle production inside $b$-jets in $\ttbar$ final states is needed for a future direct measurement of this matrix element.
 
Studies of neutral strange particle production at the LHC have been carried out using minimum-bias events at low luminosities \cite{minimum_bias,lambdas,CMS,CMS2,ALICE, LHCb, LHCb2}. The aim of this paper is to extend the studies to $\ttbar$ production, which is known to be a copious source of high-\pt jets, especially $b$-jets. In doing so, three cases are considered depending on whether the neutral strange particles are
embedded in jets, with or without a $b$-tag, or not associated with any selected jet.
 
In current MC generators the production of neutral strange particles within jets in top quark decays exhibits little sensitivity to initial-state radiation effects, different choices of parton distribution functions (PDF) or UE effects. In contrast, neutral strange particle production outside jets is more sensitive to details of the parton shower's initial- and final-state radiation, the fragmentation scheme and multi-parton interactions (MPI). They are also very sensitive to the ratio $\gamma_{\mathrm{s}}$ of strange to up quarks.

This analysis was performed using a $\ttbar$ event sample collected with the ATLAS detector in the 2011 running period with $pp$ collisions at $\sqrt{s} = $ 7~\TeV. These data are less affected by multiple $pp$ interactions within the same (in time) or nearby (out of time) bunch crossings, or pile-up, than data collected later.
 
This paper is organised as follows. Section~\ref{ATLAS} gives a brief description of the ATLAS detector. Section~\ref{MC} is devoted to the MC samples used. Section~\ref{Data} explains the data sample and the event selection criteria.
Section~\ref{selection} is dedicated to the reconstruction and selection of neutral strange particles, as well as the background subtraction procedure. Section~\ref{reco} shows the results at the detector level compared with MC generator simulations. Neutral strange particle production is studied in terms of distributions of transverse momentum, pseudorapidity, energy and multiplicity for the three cases stated above. Section~\ref{eff} discusses the efficiency correction calculations and the statistical error propagation. Section~\ref{syst} gives details of the main systematic uncertainties. Section~\ref{unfold} shows the results corrected to the particle level compared with the predictions of different MC models, thus checking the model-dependence of neutral strange particle production in these events. Finally, Section~\ref{sum} presents a summary and conclusions.

\section{The ATLAS detector}
\label{ATLAS}
 
The ATLAS detector is described in detail in Ref.~\cite{detector}. All of its subsystems are relevant for this analysis, including the inner detector (ID), the electromagnetic and hadronic calorimeters and the muon spectrometer.
 
The inner detector, located within a 2 $\si{\tesla}$ axial magnetic field, is used to measure the momentum of charged particles. Its $\eta$--$\phi$ coverage includes the full azimuthal range $-\pi \leq \phi \leq \pi$ and the pseudorapidity range $\left|\eta\right| < 2.5$.\footnote{ATLAS uses a right-handed coordinate system with its origin at the nominal interaction point (IP) in the centre of the detector and the $z$-axis along the beam pipe. The $x$-axis points from the IP to the centre of the LHC ring, and the $y$-axis points upward. Cylindrical coordinates $(r,\phi)$ are used in the transverse plane, $\phi$ being the azimuthal angle around the beam pipe. The pseudorapidity is defined in terms of the polar angle $\theta$ as $\eta=-\ln\tan(\theta/2)$.} The inner detector includes a silicon pixel detector (Pixel), a silicon microstrip tracker (SCT) and a transition radiation tracker (TRT). The calorimeter system covers the pseudorapidity range $\left|\eta\right| < 4.9$. The electromagnetic section, covering the region $\left|\eta\right| < 3.2$, uses liquid argon as the active material in barrel and endcap calorimeters with accordion-shaped electrodes and lead absorbers. The hadronic calorimeter system consists of a steel/scintillator-tile barrel calorimeter ($\left|\eta\right| < 1.7$) and a copper liquid-argon endcap ($1.7 < \left|\eta\right| < 3.2$). In addition, a forward calorimeter consisting of liquid argon with copper and tungsten for the absorbers extends the pseudorapidity coverage to $\left|\eta\right| = 4.9$. The muon spectrometer, located inside a toroidal magnetic field, provides triggering and muon tracking capabilities in the ranges $\left|\eta\right| < 2.4$ and $\left|\eta\right| < 2.7$ respectively. This allows identification of muons with momenta above 3~$\GeV$ and precision determination of the muon transverse momentum up to 1~\TeV. In this analysis muons reconstructed in the muon spectrometer are matched with well-measured tracks from the inner detector.
 
The trigger system \cite{atlasTrigger} uses three consecutive levels: level 1 (L1), level 2 (L2) and the event filter (EF). The L1 triggers are hardware-based and use coarse detector information to identify regions of interest, whereas the L2 triggers are based on fast online data reconstruction algorithms. Finally, the EF triggers use offline data reconstruction algorithms. For this analysis, events are required to pass a single-electron or single-muon trigger.

\section{Monte Carlo event simulation}
\label{MC}
 
The MC generators used to describe particle production in $pp$ collisions differ in the approximations used to calculate the underlying short-distance QCD process, in the manner parton showers are used to take into account higher-order effects and in the fragmentation scheme responsible for long-distance effects. The generated events were passed through a detailed \textsc{Geant}~4 simulation \cite{geant} of the ATLAS detector  \cite{atlasInfra}.
 
The baseline $t\bar{t}$ MC sample was produced with the next-to-leading-order (NLO) generator \textsc{PowhegBox} (referred to hereafter as \textsc{Powheg}) \cite{powheg,powheg2, powheg3} for the matrix element calculation with the \textsc{CTEQ66} NLO PDF. The parton shower and hadronisation processes were implemented using \textsc{Pythia6} \cite{pythia} with the \textsc{CTEQ6L} PDF \cite{pdf2}. \textsc{Pythia6} orders the parton shower by $\pT$ and uses the Lund string fragmentation scheme \cite{Lund}. The parton shower and UE effects were modelled using a set of tuned parameters called the \textsc{Perugia2011c} tune \cite{perugia}. Pile-up contributions were accounted  for by generating events with \textsc{Pythia6}, using the \textsc{AMBT2B} minimum bias (MB) tune. These were then overlaid onto the signal events at detector level. The strangeness suppression factor $\gamma_{\mathrm{s}}$ was taken at its default value $\gamma_{\mathrm{s}} = 0.3$ in the \textsc{AMBT2B} tune, while $\gamma_{\mathrm{s}} = 0.2$ was used in the \textsc{Perugia2011c} tune.  The latter was tuned to LEP data.
 
Additional MC samples are used to estimate the hadronisation model dependence of $\Kshort$ and $\Lambda$ production. They are based on \textsc{MC@NLO}+\textsc{Herwig} \cite{mcnlo,Herwig}, which orders the parton showers by angular separation and uses the cluster hadronisation model \cite{cluster} and \textsc{CT10} NLO \cite{pdf1} PDFs. Multi-parton interactions were simulated using \textsc{Jimmy} \cite{Jimmy} with the \textsc{AUET2} tune, while pile-up effects were taken into account as in \textsc{Powheg+Pythia6}. The parameter governing strangeness suppression  in \textsc{Herwig} is not $\gamma_{\mathrm{s}}$, but the probability of producing an $s\bar{s}$-pair when the clusters are fragmented. This parameter was set at its default value, which is equal to that for the other light quarks. The suppression is then given by the $s$-quark mass in the non-perturbative gluon splitting $g \rightarrow s\bar{s}$.
 
The data, corrected for detector effects, are also compared with events from other MC generators at particle level, without detector MC simulation:
\begin{itemize}
\item  \textsc{Sherpa} 2.1.1 \cite{Sherpa}, which uses a different approach than previous generators for the matrix element  calculation up to NLO accuracy with the \textsc{CT10} PDFs, as well as for the parton shower implementation, with cluster hadronisation. \textsc{Sherpa} uses $\gamma_{\mathrm{s}} = 0.4$.
\item \textsc{Powheg} with the \textsc{NNPDF3.0} NLO PDF set \cite{NNPDF3}, interfaced to {Pythia8} \cite{pythia8} with the \textsc{NNPDF2.3} LO PDF set and the A14 tune \cite{A14} for the parton shower, hadronisation and UE modelling.
\item \textsc{Powheg} interfaced to \textsc{Herwig7} (v7.1)\cite{Herwig7} with the \textsc{NNPDF3.0} NLO PDF set and \textsc{H7UE} tune, as default, for the parton shower, hadronisation and UE modelling.
\item \textsc{MadGraph5} a\textsc{MC@NLO} generator (referred to hereafter as a\textsc{MC@NLO}) \cite{aMC@NLO} interfaced to \textsc{Herwig7} as before.
\item The leading-order (LO) \textsc{Acermc} generator \cite{acer} interfaced to \textsc{Pythia6}, with different tunes such as \textsc{Perugia2011c} or \textsc{TuneAPro} \cite{tuneApro} for parton showering and hadronisation, as well as with different colour reconnection (CR) schemes.
\end{itemize}
 
Table~\ref{tab:MCtable} presents a summary of the different signal MC sample tunes used in this analysis.
 
\begin{table}[H]
\caption{Summary of basic generator settings used to simulate the $t\bar{t}$ events.
}
\label{tab:MCtable}
\begin{center}
\begin{tabular}{llll} \hline
MC generator & ME order & PDF & UE tune \\ \hline
\textsc{Powheg}+\textsc{Pythia6} & NLO & \textsc{CTEQ66} NLO & \textsc{Perugia2011c}   \\
\textsc{MC@NLO}+\textsc{Herwig} & NLO &\textsc{CT10} NLO & \textsc{Jimmy}-\textsc{AUET2} \\
\textsc{Sherpa} 2.1.1 & NLO & \textsc{CT10} NLO & \textsc{Sherpa} \\
\textsc{Powheg}+\textsc{Pythia8} & NLO & \textsc{NNPDF3.0} NLO & \textsc{A14}   \\
\textsc{Powheg}+\textsc{Herwig7} & NLO & \textsc{NNPDF3.0} NLO & \textsc{H7UE}   \\
a\textsc{MC@NLO}+\textsc{Herwig7} & NLO & \textsc{NNPDF3.0} NLO & \textsc{H7UE}   \\
\textsc{Acermc}+\textsc{Pythia6} & LO & \textsc{CTEQ6L} & \textsc{Perugia}/\textsc{TuneAPro} \\
& & & (with and w/o CR)  \\ \hline
\end{tabular}
\end{center}
\end{table}
 
Background samples were generated for the production of $\Zboson$ boson in association with jets, including heavy flavours, using the \textsc{Alpgen} \cite{alpgen} generator with the \textsc{CTEQ6L} PDFs \cite{pdf2}, and interfaced with \textsc{Herwig} and \textsc{Jimmy}. The same generator was used for the diboson backgrounds, $\Wboson\Wboson$, $\Wboson\Zboson$ and  $\Zboson\Zboson$, while \textsc{MC@NLO} was used for the simulation of the single-top-quark background in the $\Wboson t$ final state.
 
The MC simulated samples are normalised to their corresponding cross-sections, as described in the following. The $\ttbar$ signal is normalised to the cross-section calculated at approximate next-to-next-to-leading order (NNLO) using the \textsc{Hathor} package \cite{HATHOR}, while for the single-top-quark production cross-section, the calculations in Ref.~\cite{kidonakis3} were used. The $\Zboson$ plus jets cross-sections are taken from \textsc{Alpgen} \cite{alpgen} with additional NNLO $K$-factors as given in Ref.~\cite{fewz}.
 
The simulated events are weighted such that the distribution of the number of interactions per bunch crossing in the simulated samples matches that of the data. The size of the MC samples considered in this analysis exceeds that of the data sample by more than an order of magnitude.

\section{Data sample and event selection}
\label{Data}
The data sample used in this analysis corresponds to an integrated luminosity of 4.6~$\ifb$, collected in 2011. The uncertainty in the integrated luminosity is 1.8\% \cite{intLumi}. The sample consists of data taken while all relevant subdetector systems were operating under stable beam conditions.
 
In order to reduce the jet activity from hadronic $\Wboson^{\pm}$ decay channels, the dileptonic $t \bar{t}$ decay mode is used in this analysis. Events in this decay mode were selected as described in Refs.~\cite{dileptonic,papershapes}, using a trigger based upon a high-$\pT$ electron with a threshold of either $20$ or $22~\GeV$, or a muon with $\pT(\mu)> 18~\GeV$. Events are required to have at least one primary vertex, with five or more tracks with $\pt^{\mathrm{track}}\geq 400~\MeV$. If there is more than one primary vertex, the one with the largest $\sum \pt^2$ is chosen, where the sum is over the transverse momenta of tracks from the vertex.
 
Electron candidates are reconstructed from energy deposits in the calorimeter that are associated with tracks reconstructed  in  the  ID.  The  candidates  must  pass  a  tight  selection \cite{electron}, which uses calorimeter and tracking variables as well as TRT information for $|\eta|<2.0$, and are required to have transverse momentum $\pT (e) > 25~\GeV$ and $|\eta|<2.47$. Electrons in the transition region between the barrel and endcap calorimeters are not considered. Muon  candidates  are  reconstructed  by  searching  for  track segments in different layers of the muon spectrometer. These segments are combined and matched with tracks found in the ID. The candidates are re-fitted using the complete track information from both detector systems and are required to have a good fit for muons with $\pT (\mu) > 20~\GeV$ and $|\eta|<2.5$.
 
The selected events are required to have exactly two isolated charged leptons ($e$ or $\mu$). At least one of them must match with the corresponding trigger object. For electron candidates, the isolation criterion requires that the transverse energy deposited around the electron in the calorimeter in a cone of size\footnote{Angular distance in the $\eta$--$\phi$ plane is defined as $\Delta R = \sqrt{(\Delta \eta)^2 + (\Delta \phi)^2}$.} $\Delta R = 0.2$ is below 3.5~$\GeV$, excluding the electron energy cluster itself. For muon candidates, both the transverse energy in the calorimeter and the transverse momentum in the tracking detector around the muon in a cone of size $\Delta R = 0.3$ must be below 4~$\GeV$. The track isolation is calculated from the scalar sum of the transverse momenta of tracks with $\pT > 1~\GeV$, excluding the muon. Cosmic-ray muons are rejected by a veto on muon candidate pairs back-to-back in the transverse plane and with transverse impact parameter $|\dzero|>0.5$~$\si{\milli\metre}$ relative to the beam axis \cite{cosmics}. The two isolated leptons are required to have opposite charges. For the $ee$ and $\mu\mu$ channels, the invariant mass of the two leptons must be greater than 15~$\GeV$, to reject background from low-mass resonances decaying into lepton pairs, and at least 10~$\GeV$  away from the \Zboson boson mass.
 
Jets are reconstructed with the anti-$k_t$ algorithm \cite{jets} with radius parameter $R = 0.4$. The input objects to the jet algorithm, for both data and detector-level simulation, are topological clusters of energy in the calorimeter \cite{topoclusters}. These clusters are seeded by calorimeter cells with $|E_{\mathrm{cell}}| > 4 \sigma$, with $\sigma$ the RMS of the noise. Neighbouring cells are added and clusters are formed following an iterative procedure.
 
The baseline calibration of these clusters corrects their energy to the electromagnetic energy scale, which is established using test beam measurements for electrons, pions and muons in the electromagnetic and hadronic calorimeters \cite{lampl,aleksa,aharouche}. Effects due to non-compensating calorimeter response, energy losses in dead material, shower leakage, and inefficiencies in energy clustering and jet reconstruction are taken into account. This is done by matching calorimeter jets with MC particle jets in bins of $|\eta|$ and $\pt$. The result is called the jet energy scale (JES), thoroughly discussed in Ref.~\cite{JES}. It is different for $b$-jets and light-flavour jets since they have different particle compositions. More details and a discussion of JES uncertainties are given in Ref.~\cite{atlasJets}. The jet energy resolution (JER) and its uncertainties are discussed in Ref. \cite{JER}.
 
The selected events are required to have at least two jets with $\abseta < 2.5$ and $\pt > 25~\GeV$. In addition, jets are required to have a jet vertex fraction \cite{JVF}, defined as the scalar transverse momentum sum of the tracks that are associated with the jet and originate from the hard-scatter vertex divided by the scalar sum of all associated tracks, greater than $0.75$ in order to minimise pile-up effects. At least one of the jets must be identified as a $b$-tagged jet, using the multivariate MV1 algorithm \cite{MV1} based on the reconstruction of secondary vertices and three-dimensional impact parameter information. The MV1 working point corresponds to a $b$-tagging efficiency of 70\%, calculated using $t\bar{t}$ MC events with an average light-flavour mistag rate of 2\%. Jets overlapping with an accepted electron are removed if the separation is $\Delta R < 0.2$. Electrons are removed if $0.2 < \Delta R < 0.4$. Muons are removed if their separation from a jet is $\Delta R < 0.4$.
 
The reconstruction of the direction and magnitude of the missing transverse momentum ($\met $) is described in Ref.~\cite{met} and begins with the vector sum of the transverse momenta of all jets with $\pT > 20~\GeV$ and $|\eta|<4.5$. The transverse momenta of electron candidates are added. The contributions from all muon candidates and from all calorimeter clusters not belonging to a reconstructed object are also included. The missing transverse momentum is required to be $\met> 60~\GeV$ for the $ee$ and $\mu\mu$ channels, and for the $e\mu$ channel the requirement is $\HT>130~\GeV$, where $\HT$ is the scalar sum of the transverse momenta of the two leptons and the selected jets.
 
\begin{table}[htbp]
\centering
\sisetup{round-mode=places, round-precision=0, retain-explicit-plus=true, group-digits = integer, group-minimum-digits=0}
{
\begin{tabular}{l c c c}
\hline
Selection & \hspace{1.3cm}$ee$\hspace{1.3cm} & \hspace{1.25cm}$\mu \mu$\hspace{1.25cm} & $e \mu$ \\
\hline
Leptons & \multicolumn{3}{c}{Exactly 2 leptons, opposite-sign charge, isolated} \\
Electrons & \multicolumn{3}{c}{ $\ET > 25$ \GeV, $|\eta| < 2.47$, excluding $1.37 < |\eta| < 1.52$} \\
Muons & \multicolumn{3}{c}{ $\pT > 20$ \GeV, $|\eta| < 2.5$}  \\
Jets    & \multicolumn{3}{c}{$\geq$ 2 jets, $\pt > 25$ \GeV, $|\eta| < 2.5$} \\
$b$-tagging  & \multicolumn{3}{c}{$\geq$ 1 $b$-tagged jet at $\epsilon_b$ = 70\% with MV1} \\
$m_{ll}$          & \multicolumn{2}{c}{$|m_{ll}-91~\GeV| > 10$ \GeV, $m_{ll} > 15$ \GeV} & None  \\
$\MET$ or $\HT$ & \multicolumn{2}{c}{$\MET>60$ \GeV} & $\HT > 130$ \GeV  \\
\hline
\end{tabular}
}
\caption{Summary of the event selection criteria for the analysis.}
\label{tab:eventselection}
\end{table}

After applying these selection criteria, which are summarised in Table \ref{tab:eventselection}, a sample of 6926 $t \bar{t}$ candidate events is selected. MC studies indicate that the background contamination in the sample after event selection is $\sim$ 6\%, dominated by single-top-quark events. The background contribution from \Zboson boson production with the \Zboson boson decaying leptonically, in association with jets (including heavy flavours $b\bar{b}$), is at the level of 1\%. An additional source of background where one or more of the reconstructed lepton candidates are non-prompt or misidentified is found to be at the 1\% level with a very large (50\%) statistical uncertainty  \cite{dileptonic, differential}, and is not considered in this analysis. The expected composition of the sample is summarised in Table~\ref{tab:Composition}, where `Diboson' includes the $\Wboson\Wboson$, $\Wboson\Zboson$ and $\Zboson\Zboson$ contributions. The percentages for signal and background processes quoted in Table~\ref{tab:Composition} are in agreement with those quoted in Ref.~\cite{differential}.
 
\begin{table}[H]
\caption{Expected composition of the selected sample in terms of number of events ($N_{\mathrm{evt}}$) and fractions (\%) of different processes. Uncertainties are statistical only.}\label{tab:Composition}
\begin{center}
\begin{tabular}{lrr} \hline
Process & \multicolumn{1}{c}{$N_{\mathrm{evt}}$} & Percentage [\%]\\ \hline
$t \bar{t}$ dileptonic & 6860 $\pm$ 80  & 93.9 \\
Single top            & 300 $\pm$ 20 & 4.1 \\
$\Zboson$ + jets & 77 $\pm$ \ \ 9 & 1.1 \\
Diboson & 61 $\pm$ \ \ 8 & 0.9 \\ \hline
Predicted             & 7300 $\pm$ 90 &  \\
Observed              & \multicolumn{1}{l}{6926} &   \\ \hline
\end{tabular}
\end{center}
\end{table}
 
Figure~\ref{fig:jets} shows the distribution of jet multiplicity and the $\pt$ spectra of all jets, $b$-tagged jets and non-$b$-tagged jets. The jet activity is indeed limited, as 94\% of the selected events contain at most four jets.
The shapes of the normalised distributions in data are in good agreement with the prediction given by the $\ttbar$ \textsc{Powheg}+\textsc{Pythia6} simulation only. The small contributions from processes other than $\ttbar$ are neglected in the following analysis.
 
MC studies show that 99\% of the selected $b$-tagged jets correspond to particle level $b$-jets, while 28\% of jets in the non-$b$-tagged sample are $b$-jets which are not tagged by the MV1 algorithm. These fractions are calculated by matching detector-level jets, $b$-tagged or not, to their corresponding particle-level jets, which are defined in Section~\ref{eff}. These fractions are found to be largely independent of whether non-$\ttbar$ backgrounds are considered or not. Furthermore, the purity of $b$-tagged jets is rather independent of jet $\pT$ as shown in Figure~\ref{fig:b_ptflavour}.

\begin{figure}[H]
\begin{center}
\vspace*{-0.65cm}
\subfloat[]{
\label{fig:jet_Multi}
\includegraphics[height=6.5cm]{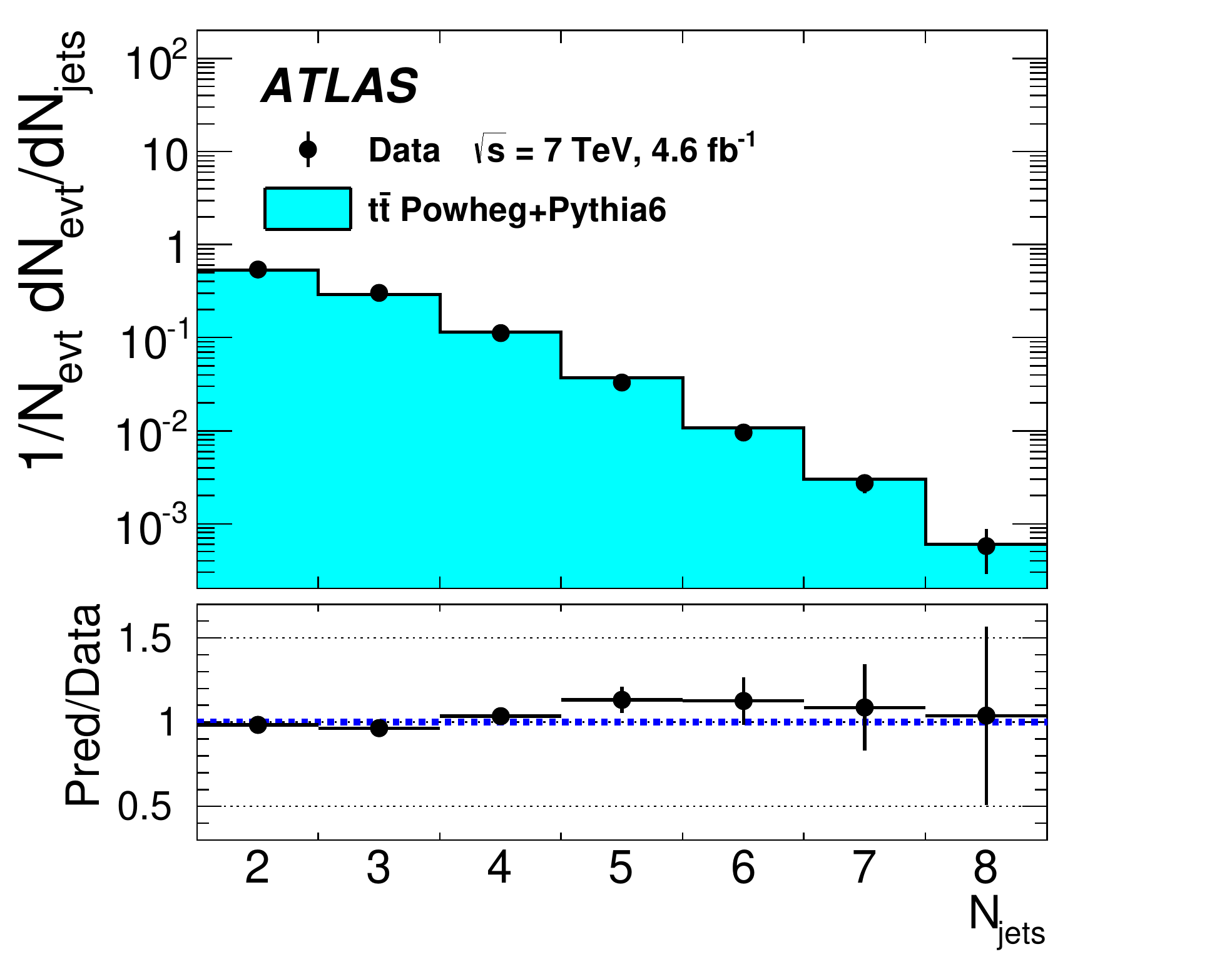}
}
\subfloat[]{
\label{fig:jet_ptflavour}
\includegraphics[height=6.5cm]{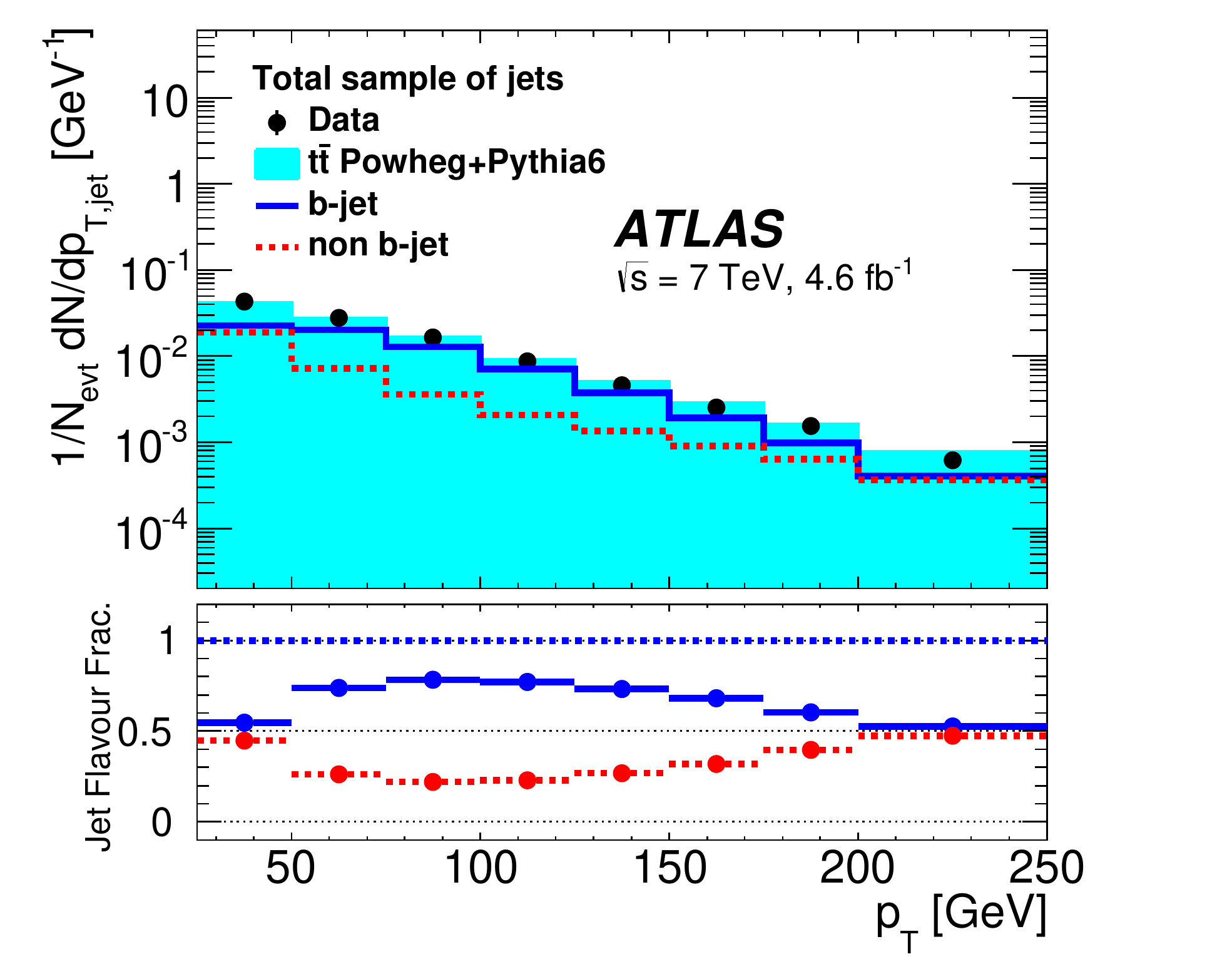}
}
\\
\vspace*{-0.35cm}
\subfloat[]{
\label{fig:b_ptflavour}
\includegraphics[height=6.5cm]{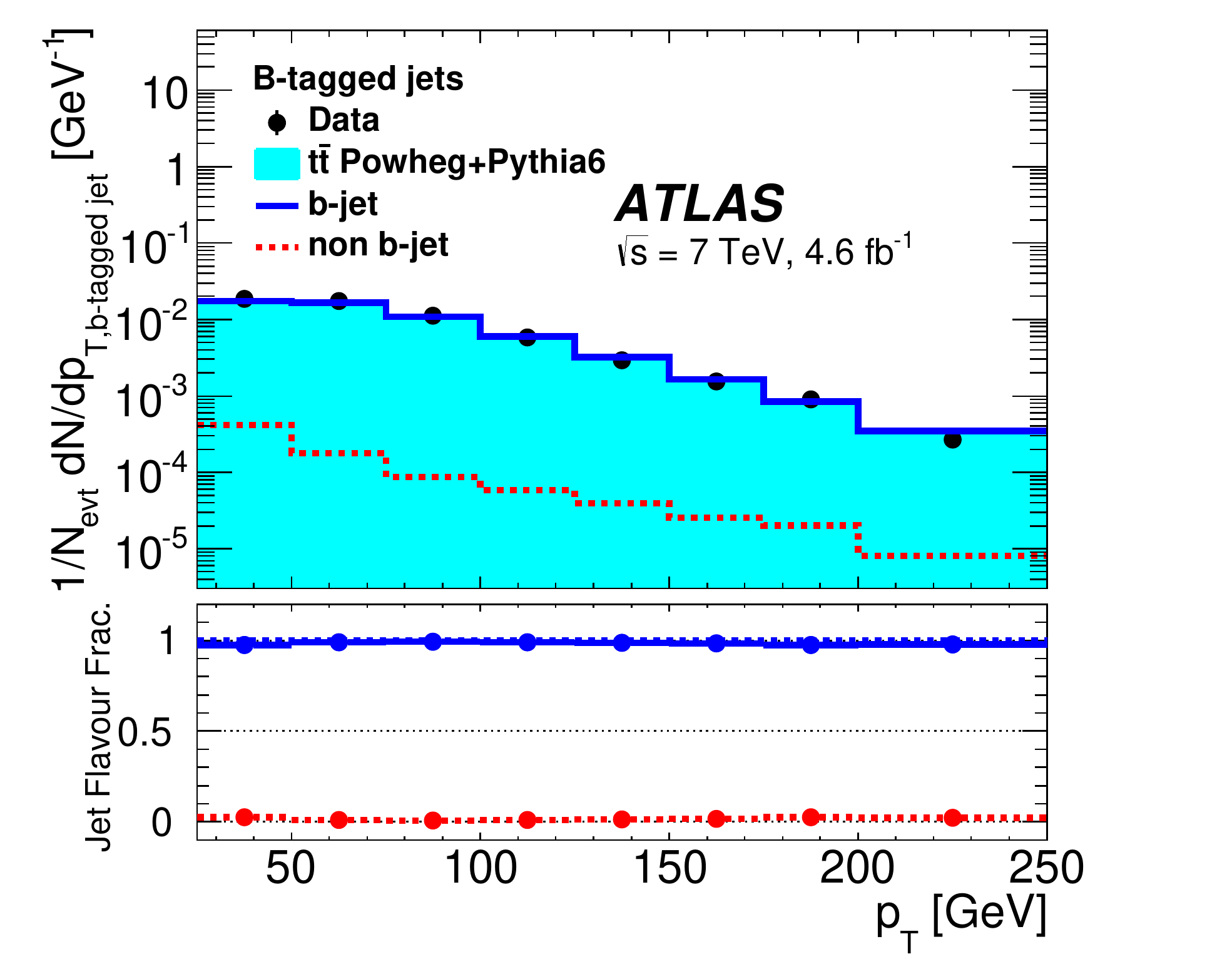}
}
\subfloat[]{
\label{fig:other_ptfraction}
\includegraphics[height=6.5cm]{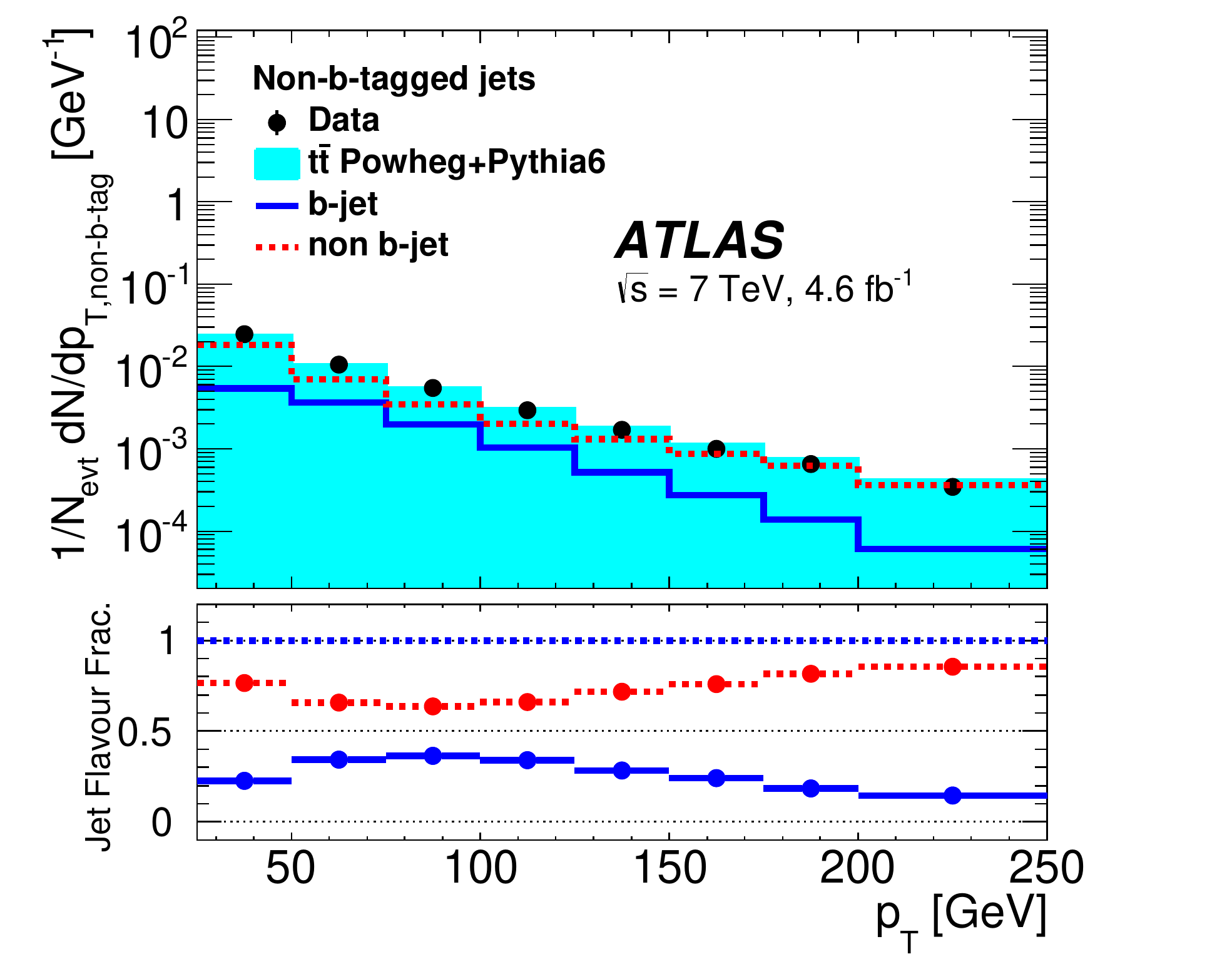}
}
 
\end{center}
\vspace*{-0.35cm}
\caption{
Spectra of (a) jet multiplicity $N_{\mathrm{jets}}$ and (b) jet $\pt$ in data compared with $t\bar{t}$ \textsc{Powheg}+\textsc{Pythia6} predictions at detector level. The expected $b$-jet flavour fractions for (c) $b$-tagged and (d) non-$b$-tagged jets as a function of jet $\pt$ are also compared with data. These distributions are normalised to the total number of selected events in data or MC predictions.
}
\label{fig:jets}
\end{figure}

\section{$\Kshort$ and $\Lambda$ reconstruction}
\label{selection}
 
\subsection{Neutral strange particle reconstruction}
\paragraph{}
Neutral strange hadrons are reconstructed in the $\Kshort \ra \piplus \piminus$
($\Lambda \ra p \piminus$, $\bar{\Lambda} \ra \bar{p} \piplus$) decay mode by identifying two tracks originating from a displaced vertex, thus profiting from the long lifetimes of neutral $K$ mesons ($\Lambda$ baryons) with $c\tau_0 \approx$ 2.7 $\si{\centi\metre}$ ($c\tau_0 \approx$ 7.9 $\si{\centi\metre}$).
 
Tracks are reconstructed within the $\abseta< 2.5 $ acceptance of the ID, as described in Refs.~\cite{ID_tracks1,ID_tracks2}. The $\Kshort$ ($\Lambda$) candidates are oppositely charged track pairs with the transverse momentum of the two-track system $\pT>$ 100 (500)~$\MeV$. The tracks must have at least two hits in the Pixel or SCT detectors, and are fitted to a common vertex. For  $\Kshort$ reconstruction, the pion mass is assumed for both tracks, while the proton and pion masses are assumed for the $\Lambda$ case \footnote{For $\Lambda$ and $\bar{\Lambda}$ decays, the track with the higher $\pt$ is assigned the proton mass and the other track is assigned the pion mass. In the following and due to the sample size, $\Lambda$ refers to the sum of $\Lambda$ and $\bar{\Lambda}$ particles.}. Further requirements on these candidates are given below, and Ref.~\cite{minimum_bias} provides more details:
\begin{itemize}
\item The $\chi^2$ of the two-track vertex fit is required to be less than 15
(with 1 degree of freedom).
\item The transverse flight distance ($R_{xy}$) is defined to be the distance between
the $\Kshort$ ($\Lambda$) decay point and either the secondary $b$-tagged vertex, when the
$\Kshort$ ($\Lambda$) is contained in a jet with a $b$-tag, or the reconstructed primary
vertex when the $\Kshort$ ($\Lambda$) is
contained in a jet without a $b$-tag or is not associated with any selected jet. A requirement
4  $\si{\milli\metre}$ $< R_{xy} <$ 450 $\si{\milli\metre}$
(17 $\si{\milli\metre}$ $< R_{xy} <$ 450 $\si{\milli\metre}$)
ensures that the tracks are reconstructed inside the Pixel+SCT
part of the ID tracker.
\item The angle between the $\Kshort$ ($\Lambda$) momentum vector and the $\Kshort$
($\Lambda$) flight direction (obtained from the line connecting the decay vertex to the primary vertex, or to the secondary
vertex if the $\Kshort$ ($\Lambda$) is inside a $b$-jet) has
to satisfy $\cos \theta_K >$ 0.999 ($\cos \theta_{\Lambda} >$ 0.9998).
\end{itemize}
 
The $\Kshort$ ($\Lambda$) candidates that fulfil these conditions are then separated into three classes: candidates inside a $b$-tagged jet, inside a non-$b$-tagged  jet and outside any jet. To this end the separation $\Delta R$ between the $\Kshort$ ($\Lambda$) line of flight and the jet axis in the $\eta$--$\phi$ plane is calculated. If $\Delta R <$ 0.4, a $\Kshort$ ($\Lambda$) is associated with a jet. Otherwise it is classified as being outside any jet. There are no cases of a single $\Kshort$ ($\Lambda$) being inside two different jets.
 
The mass distributions for three classes of $\Kshort$ and the total sample of $\Lambda$ candidates in data are shown  in Figure~\ref{fig:k0Mass} compared with the \textsc{Powheg}+\textsc{Pythia6} predictions scaled
to the total number of events in the data sample. The three $\Kshort$ mass distributions exhibit a resonance structure, centred around the nominal $\Kshort$ mass, with constant tails extending on both sides, indicating the presence of fake candidates, i.e. track pairs which not being $\Kshort$ or $\Lambda$ decay products have a mass in the signal mass ranges considered. The $\Kshort$ mass distributions are fairly well described by the nominal $t\bar{t}$ MC simulation except for the $\Kshort$ candidates not associated with jets, in which case the MC prediction underestimates the data by roughly $30\%$. Similar features are exhibited by the mass distribution of $\Lambda$ candidates.
 
\begin{figure}[H]
\vspace*{-1.0cm}
\begin{center}
\subfloat[]{
\label{fig:b_k0Mass}
\includegraphics[height=6.8cm]{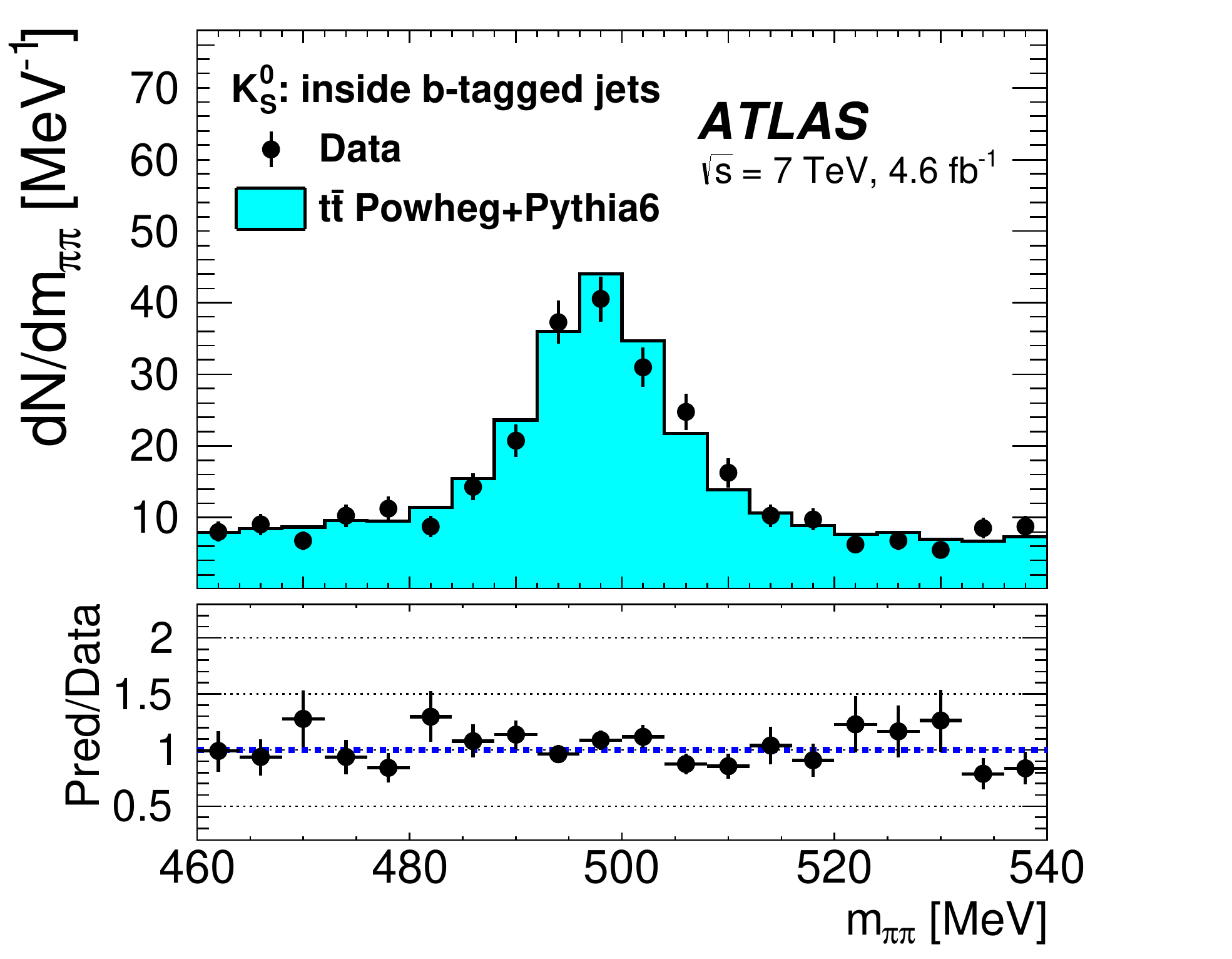}
}
\hspace*{-0.5cm}
\subfloat[]{
\label{fig:j_k0Mass}
\includegraphics[height=6.8cm]{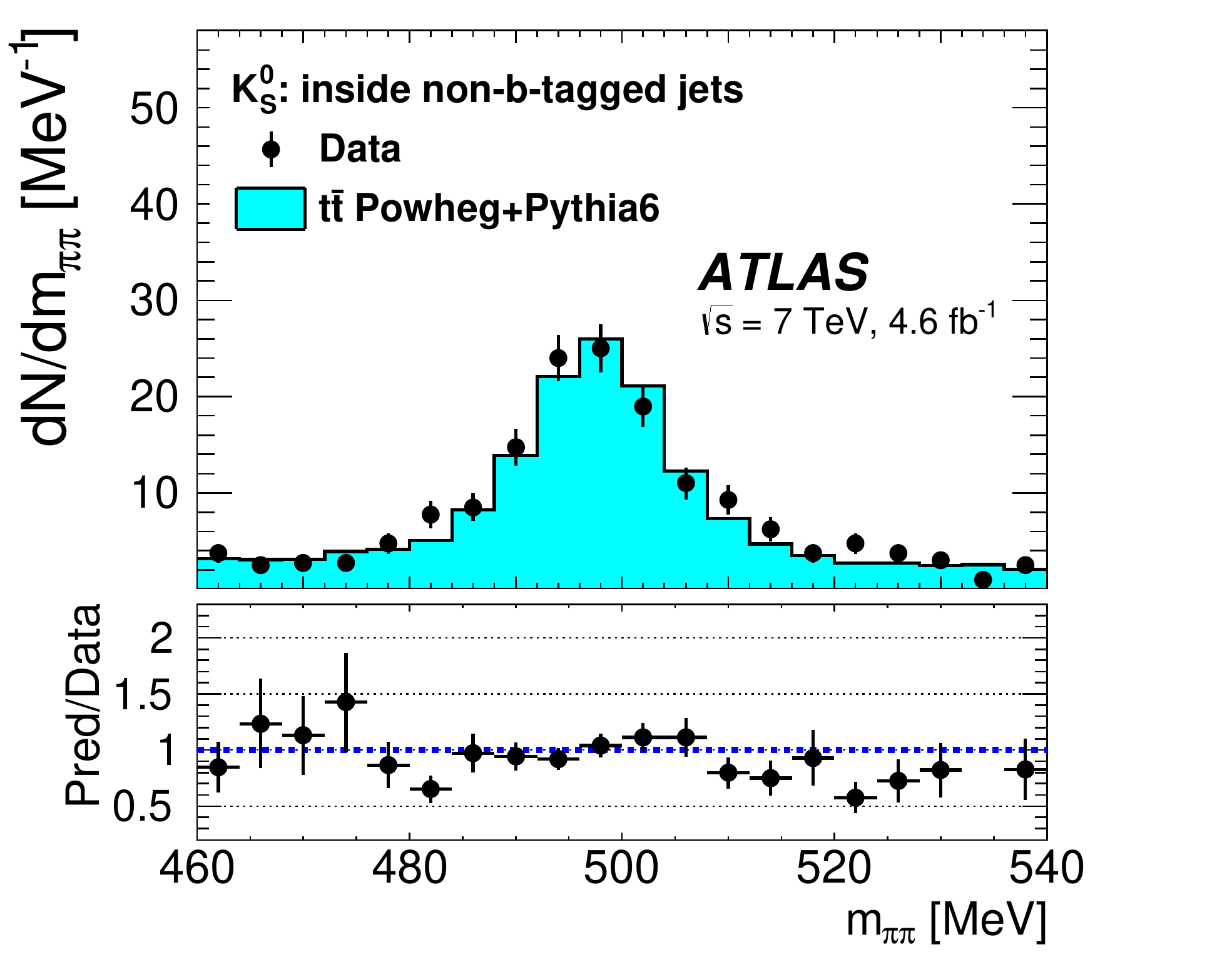}
}
\\
\subfloat[]{
\label{fig:out_k0Mass}
\includegraphics[height=6.8cm]{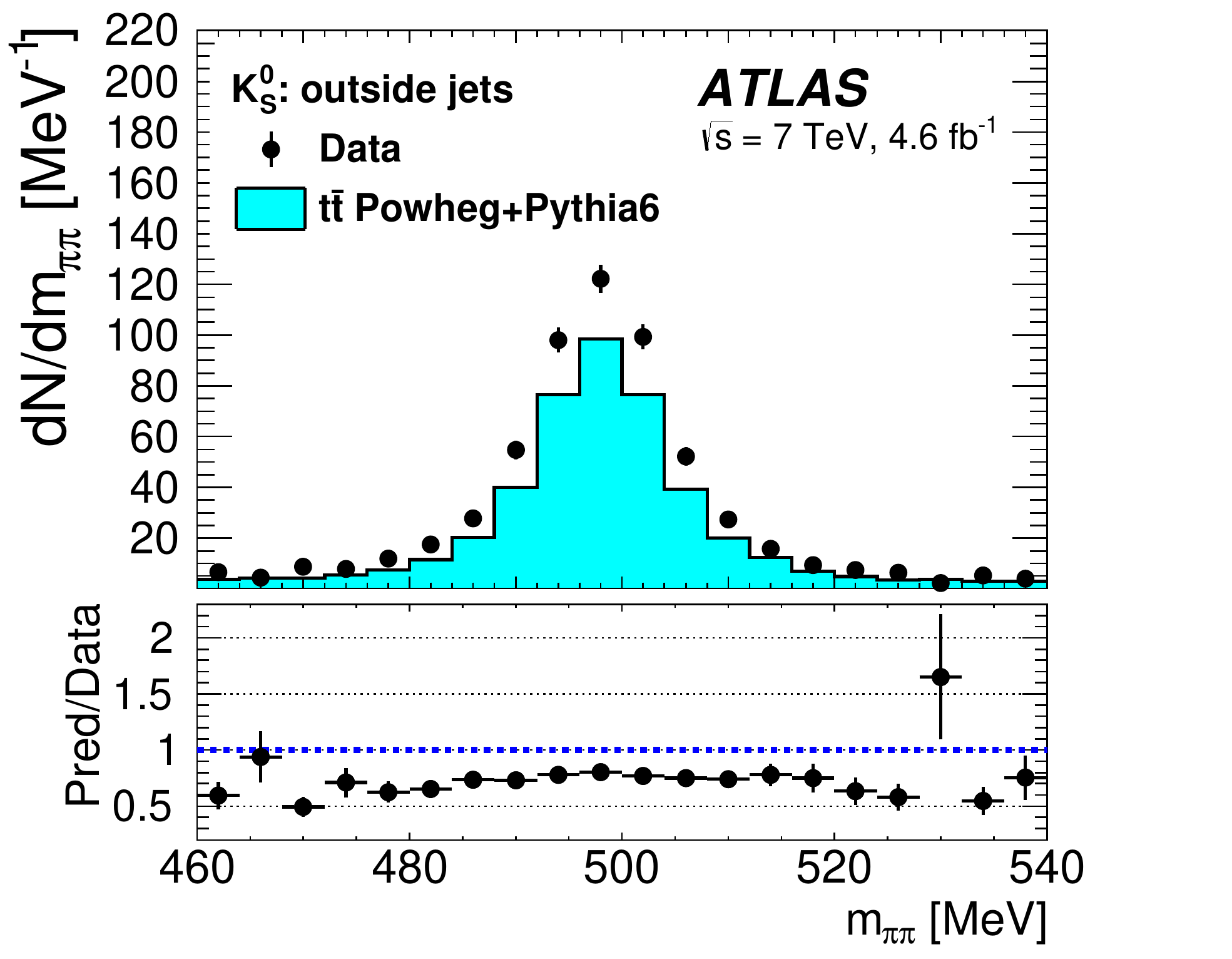}
}
\hspace*{-0.5cm}
\subfloat[]{
\label{fig:all_lambdaMass}
\includegraphics[height=6.8cm]{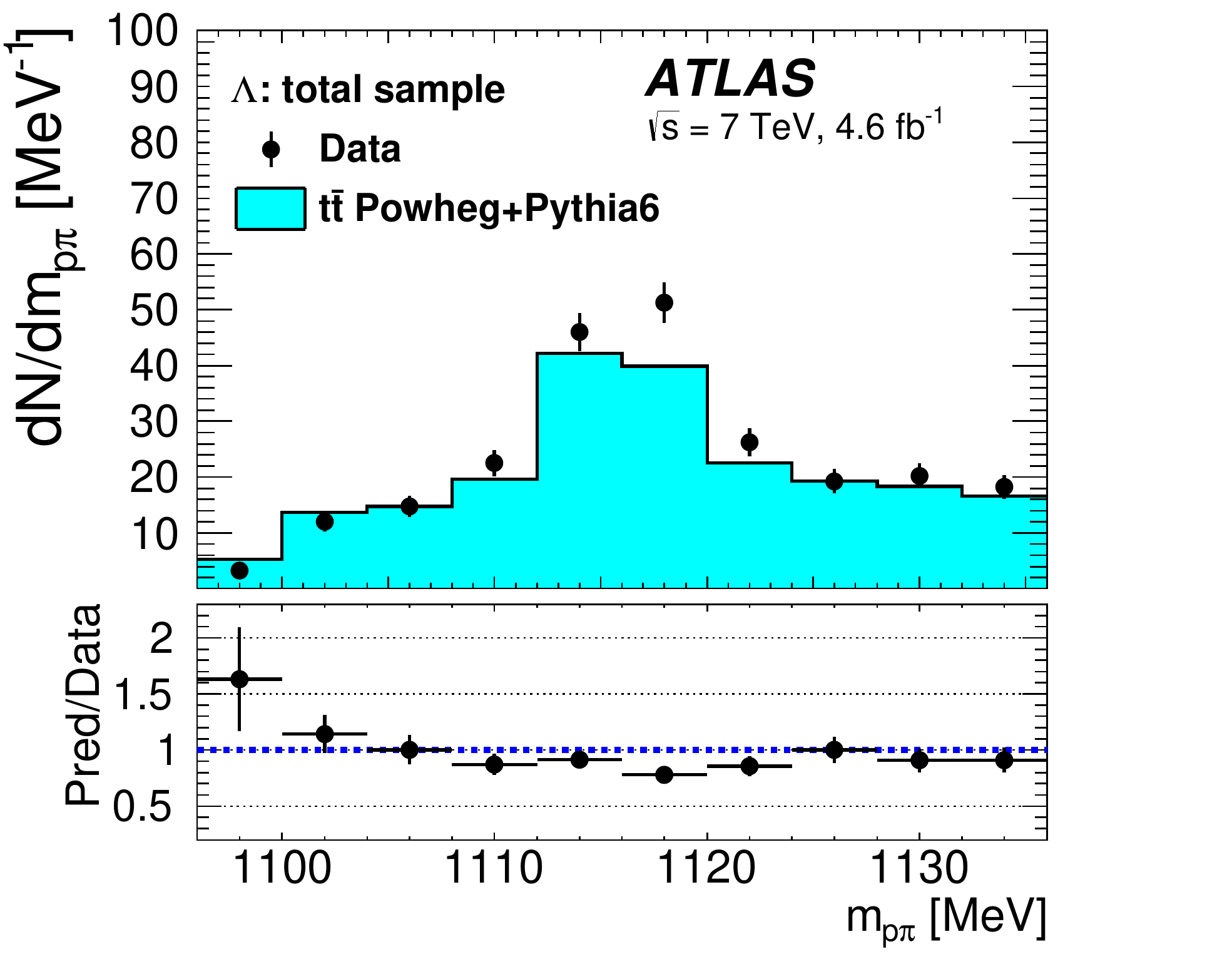}
}
\\
\end{center}
\caption{
$\Kshort$ and $\Lambda$ candidate mass distributions in data compared with \textsc{Powheg}+\textsc{Pythia6} simulation. Three classes are presented for $\Kshort$: (a) inside $b$-tagged jets, (b) inside non-$b$-tagged jets, and (c) outside any jet. The total sample is shown for (d) $\Lambda$ candidates.
}
\label{fig:k0Mass}
\end{figure}
 
\subsection{Background subtraction}\label{back}
\paragraph{}
In order to take into account the background due to fake candidates in the $\Kshort$ ($\Lambda$) mass distributions, a simple sideband subtraction in the reconstructed mass distribution is used. The signal range [480--520]~$\MeV$ ([1106--1126]~$\MeV$) is considered for $\Kshort$  ($\Lambda$) production. The background sidebands are taken to be [460--480] and [520--540]~$\MeV$ ([1096--1106] and [1126--1136]~$\MeV$)  for  $\Kshort$ ($\Lambda$) production. Candidates in the signal (sideband) region are given positive (negative) weights when filling histograms for neutral strange particle spectra such as the transverse momentum, pseudorapidity or energy. The sideband subtraction is applied to both data and detector-level MC samples. It relies on the assumption that the kinematic distributions for fake candidates in the signal region are similar to those in the sidebands.
The validity of this assumption was checked with MC studies. The number of reconstructed events is shown in Table~\ref{tab:MultiReco}. It was checked that after unfolding for detector effects, the results of the analysis are independent of sensible variations of the signal and sideband region widths.
 
The results of this simple sideband subtraction procedure, used as a baseline, is cross-checked by fitting the mass distributions to a Gaussian function centred at the nominal mass plus a constant (linear) shape for the $\Kshort$ ($\Lambda$) background. Choosing a different background shape (constant/linear) changes the results by less than 10\% of the statistical uncertainties. The estimated numbers of signal and background events obtained using the two background methods agree within statistical uncertainties. The limited sample size precludes extending this fitting procedure for signal extraction as a function of the neutral strange particle kinematic variables under study.
 
The resulting $\Kshort$ and $\Lambda$ masses from fits to the total $\pi \pi$ and $p \pi$ mass distributions are $497.8 \pm 0.2~\MeV$ and $1115.8 \pm 0.3~\MeV$, in agreement with the PDG values \cite{PDG}. The $\Kshort$ ($\Lambda$) widths from the fits are $6.83 \pm 0.03~\MeV$ ($4.16 \pm 0.04~\MeV$). The signal mass range includes 99\% (95\%) of the $\Kshort$  ($\Lambda$) signal, which ensures that the sidebands for the background subtraction are not contaminated by signal.
 
As previously noted in Refs.~\cite{minimum_bias,lambdas,CMS,ALICE, LHCb}, fitting to a double-Gaussian function with a common mean value improves the quality of the fits. This was also tried in this analysis. The resulting Gaussian mean values coincide with those obtained from a single-Gaussian fit and the numbers of signal and background events are stable within statistical uncertainties.
 
\begin{table}[H]
\caption{The numbers of $\Kshort$ and $\Lambda$ particles ($N_{K}$ and $N_{\Lambda}$) reconstructed in the data after sideband background subtraction for each class and for the total sample with their statistical uncertainties.  }
\label{tab:MultiReco}
\begin{center}
\begin{tabular}{l r r} \hline
Class & \multicolumn{1}{ c }{$N_{K}$} &  \multicolumn{1}{ c }{$N_{\Lambda}$} \\ \hline
Inside $b$-tagged jets & 530 $\pm$ 34 & 115 $\pm$ 19 \\
Inside non-$b$-tagged jets & 391 $\pm$ 25 & 65 $\pm$ 14 \\
Outside any jet & 1837 $\pm$ 49  & 183 $\pm$ 18  \\
Total sample & 2758 $\pm$ 69 & 363 $\pm$ 31 \\ \hline
\end{tabular}
\end{center}
\end{table}

\section{Results at detector level}
\label{reco}
\paragraph{}
Neutral strange particle production is studied as a function of the transverse momentum $\pt$, the energy $E$, the pseudorapidity $\eta$, the transverse flight distance $R_{xy}$ and the multiplicity $N$. For  $\Kshort$ and $\Lambda$ production inside jets, the energy fraction, $x_{K,\Lambda} = E_{K,\Lambda}/E_{\mathrm{jet}}$, is also considered.
 
For this purpose, the reconstructed $\Kshort$ and $\Lambda$ mass distributions are obtained in different bins of the variables under study, and the numbers of signal events after proper sideband subtraction are determined, as  discussed in Section~\ref{back}. They are normalised to the total number of events in data or MC generator fulfilling the dileptonic $t \bar{t}$ selection criteria presented in Table~\ref{tab:Composition}. No attempt is made to subtract the non-$t \bar{t}$ background because the normalised $\Kshort$ spectra for the $t\bar{t}$ signal are found in MC predictions to be compatible with those for single-top-quark events, which form the main background.
 
\subsection{$\Kshort$ production at detector level}
\paragraph{}
The kinematic distributions for $\Kshort$ production are displayed in Figures~\ref{fig:b_results_pythia} to \ref{fig:out_results}. They are separated into the three different classes defined in Section~\ref{selection} and compared with two different MC models, namely \textsc{Powheg+Pythia6+Perugia2011c} and \textsc{MC@NLO}+\textsc{Herwig}+\textsc{Jimmy}. The data show both the statistical as well as the total systematic uncertainties. The total uncertainties are obtained as the sum in quadrature of the statistical and detector level systematic uncertainties, namely those due to tracking, JES and JER. The systematic uncertainties are discussed in detail in Section~\ref{syst}.
 
As shown in Table~\ref{tab:MultiReco}, approximately two-thirds of the total $\Kshort$ sample are not associated with jets, the remaining one-third being roughly equally distributed between $b$-tagged and non-$b$-tagged jets. For those inside jets, the $\Kshort$ spectra do not show a strong dependence on whether the jets are $b$-tagged or not.  On the other hand, $\Kshort$ candidates not associated with jets are softer in $\pt$ or energy than those embedded in jets and their pseudorapidity distribution is constant over a wider central plateau. The $\Kshort$ multiplicity inside $b$-tagged jets, Figure~\ref{fig:b_k0Multi}, is similar to that inside non-$b$-tagged jets, Figure~\ref{fig:j_k0Multi}, while that outside any jet falls off less steeply, Figure~\ref{fig:out_k0Multi}.
 
The gross features of the data are described fairly well by both MC simulations, except for $\Kshort$ production not associated with any jet. Here the MC simulations predict roughly 30\% fewer $\Kshort$ than observed in data, while the shapes of the distributions exhibit fair agreement but for the multiplicity distribution.
 
\begin{figure}[H]
\begin{center}
\vspace*{-0.65cm}
\subfloat[]{
\label{fig:b_k0Pt}
\includegraphics[height=6.5cm]{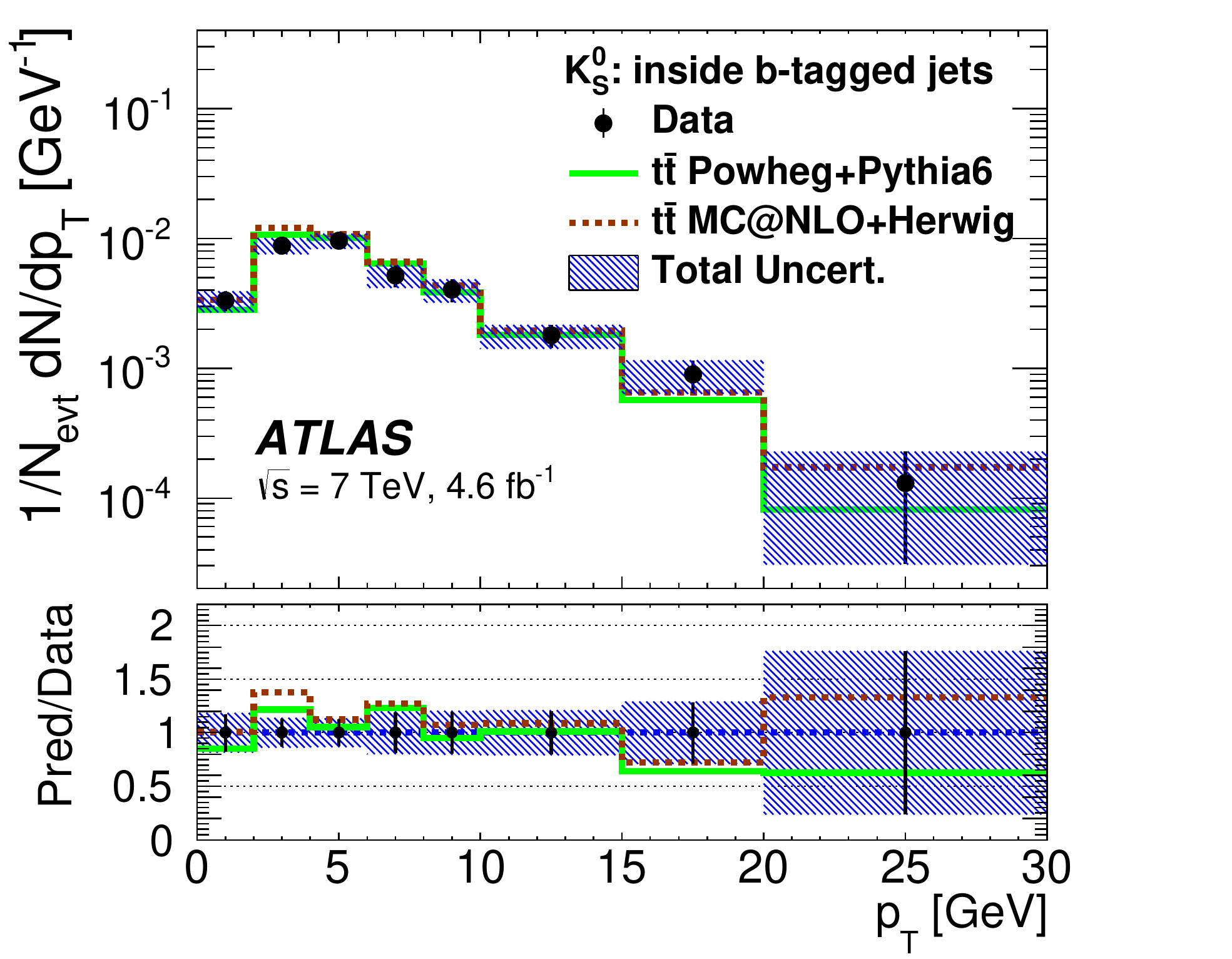}
}
\subfloat[]{
\label{fig:b_k0x}
\includegraphics[height=6.5cm]{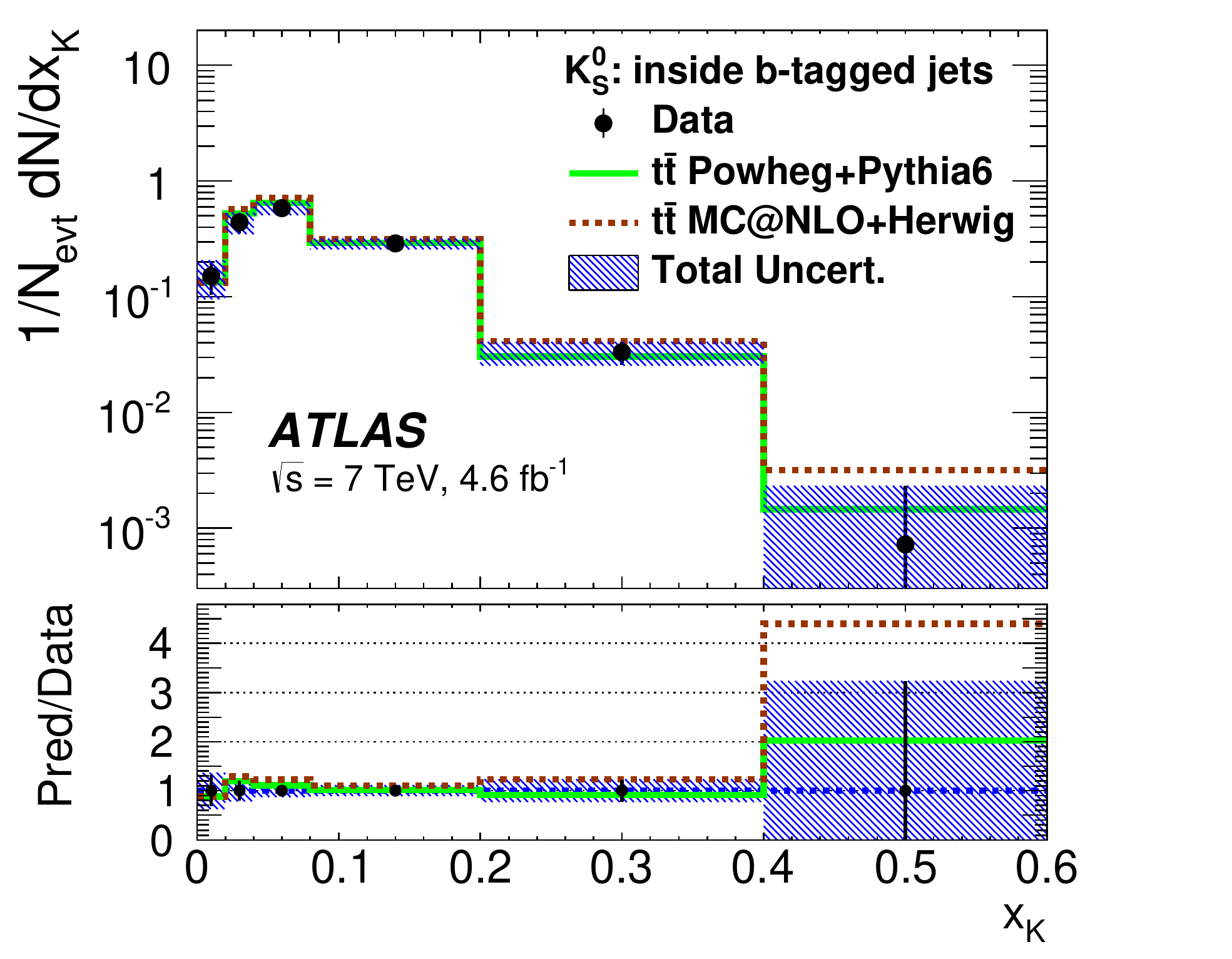}
}
\\
\subfloat[]{
\label{fig:b_k0E}
\includegraphics[height=6.5cm]{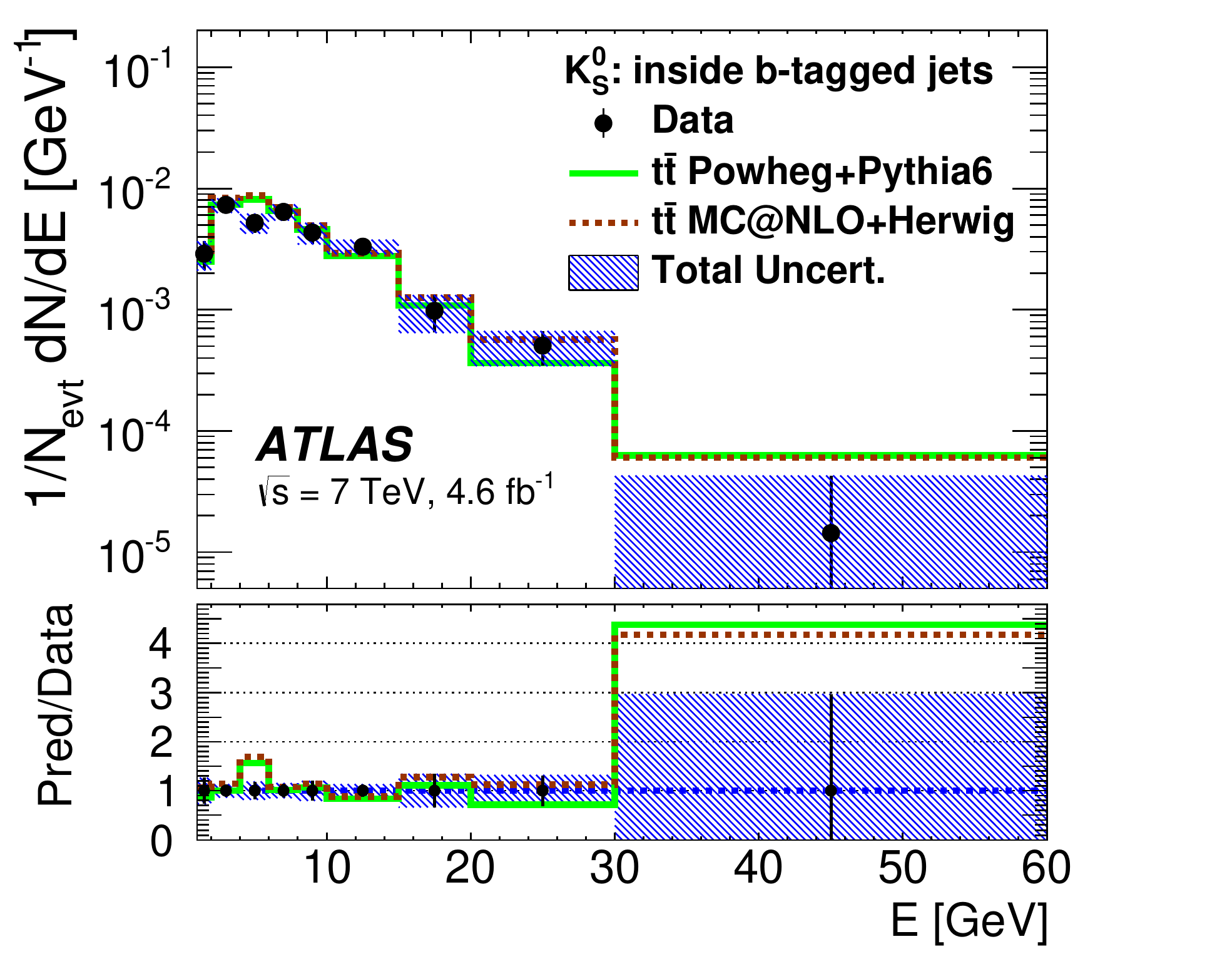}
}
\subfloat[]{
\label{fig:b_k0Eta}
\includegraphics[height=6.5cm]{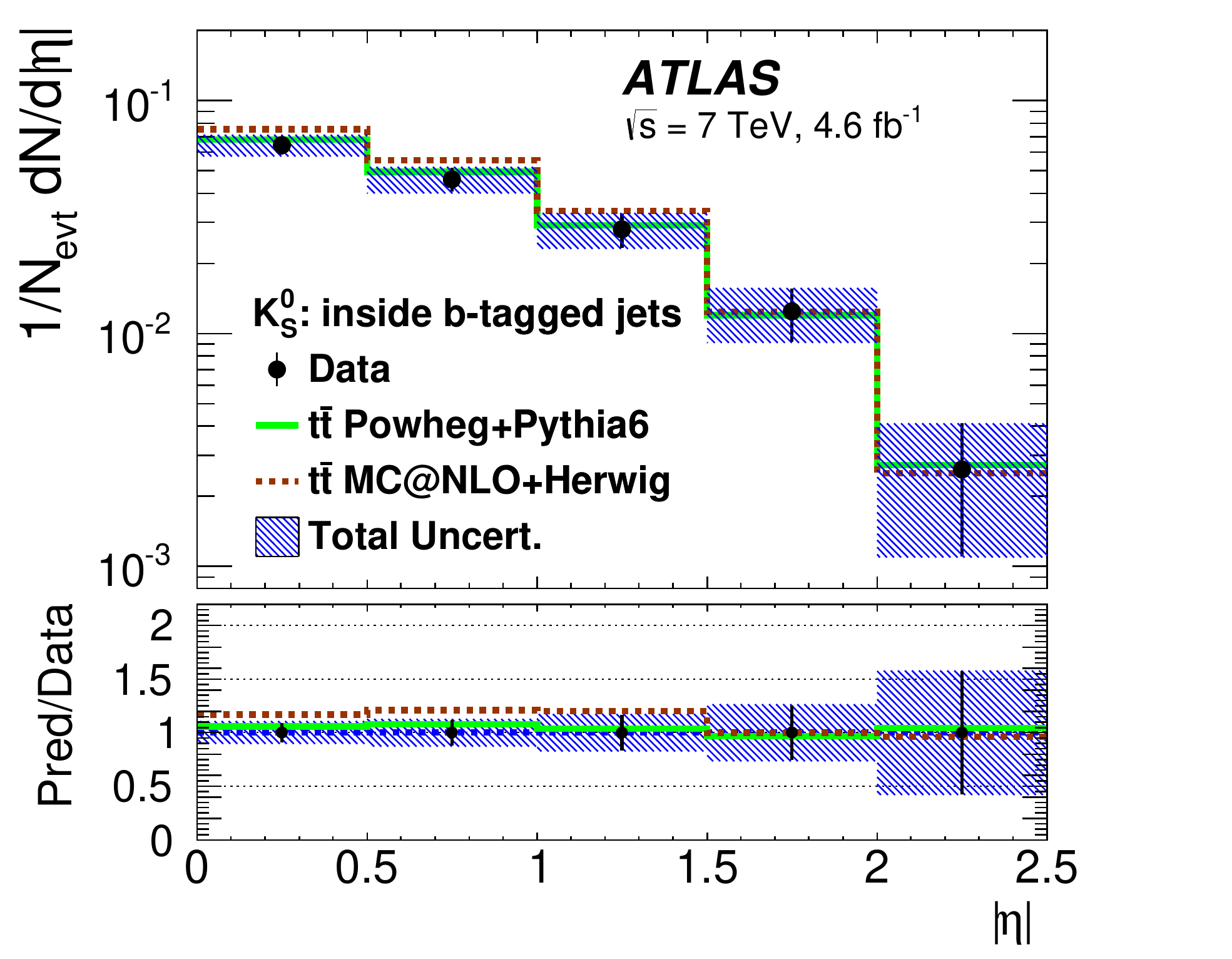}
}
\\
\subfloat[]{
\label{fig:b_k0R}
\includegraphics[height=6.5cm]{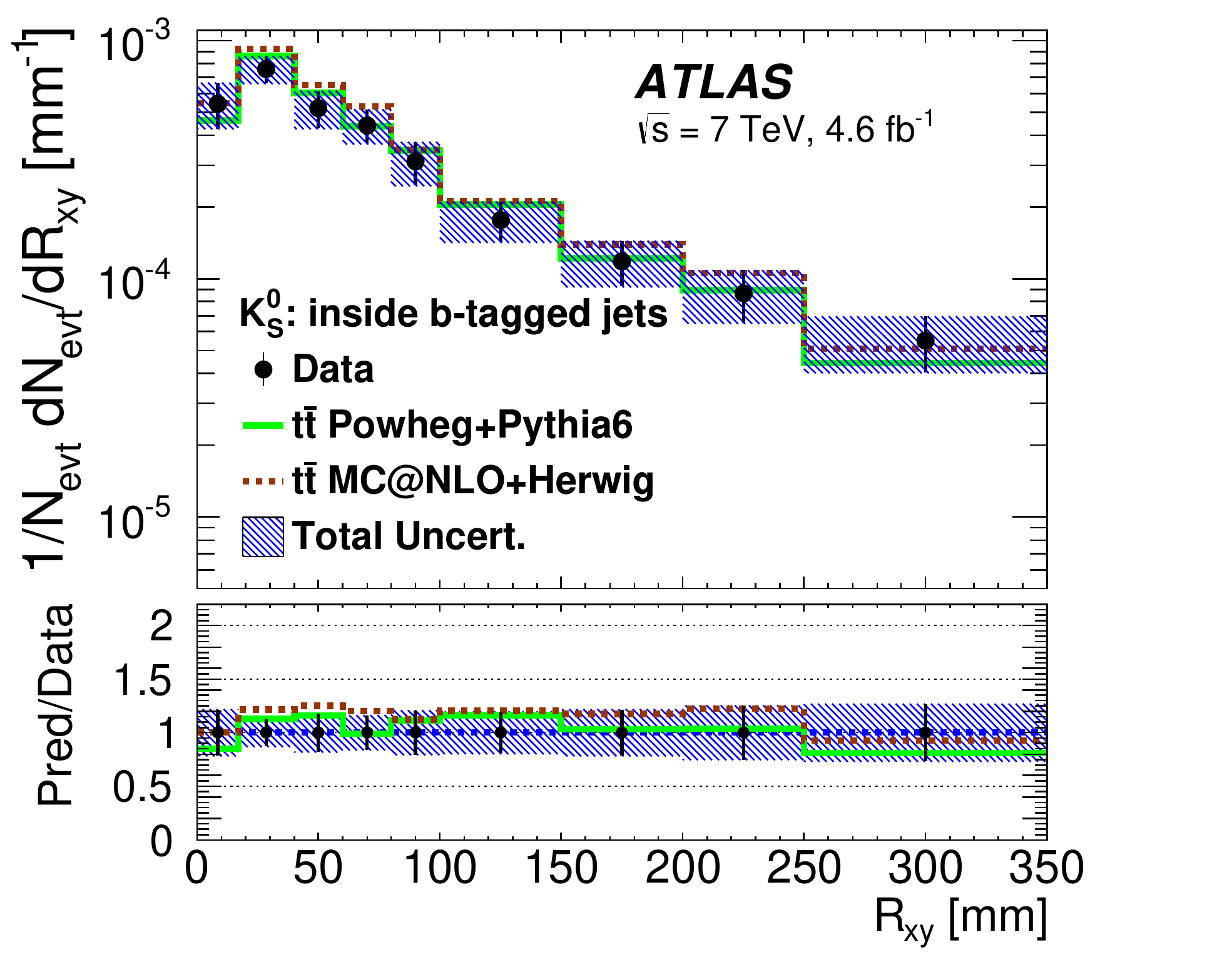}
}
\subfloat[]{
\label{fig:b_k0Multi}
\includegraphics[height=6.5cm]{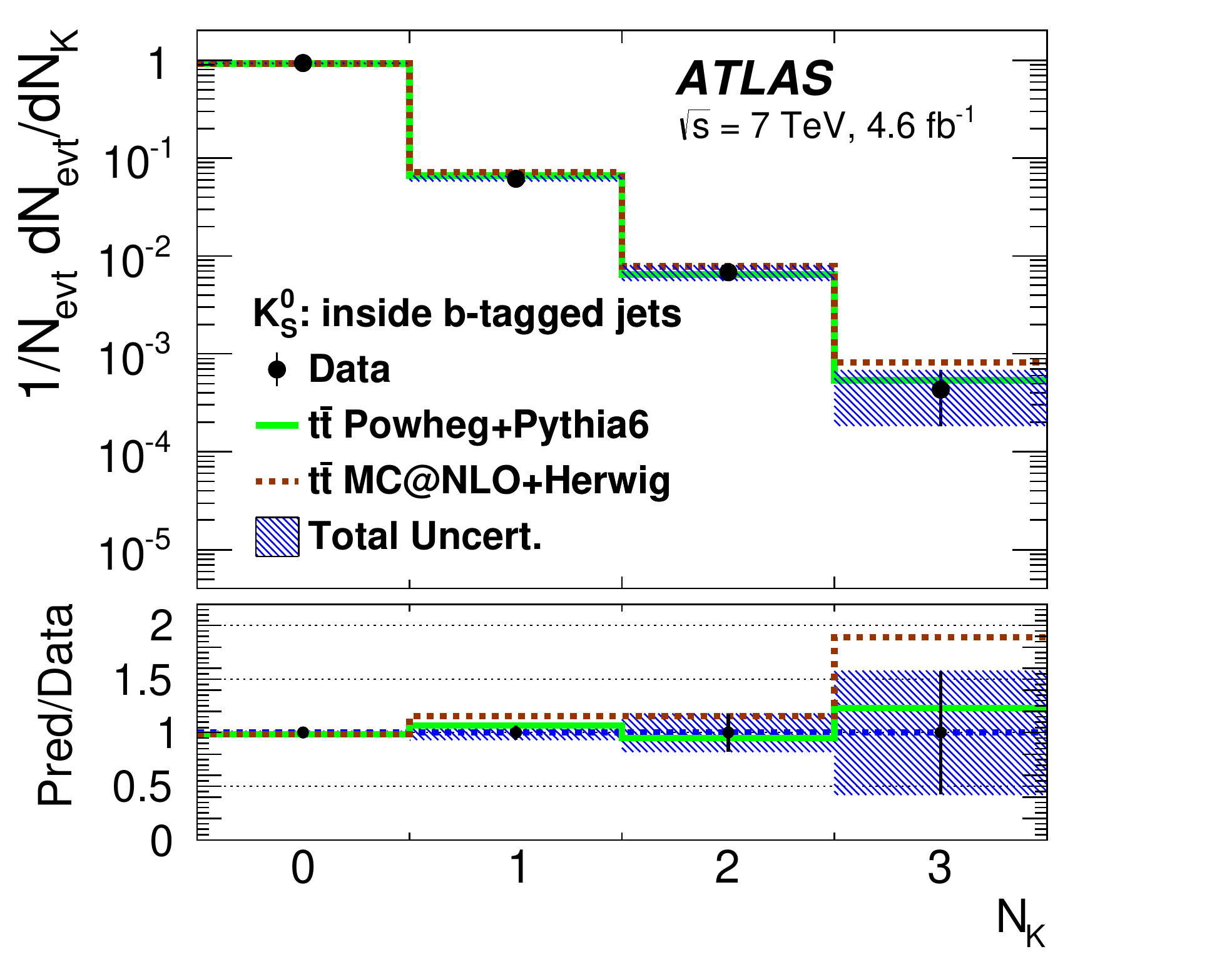}
}
\end{center}
\vspace*{-0.35cm}
\caption{
Kinematic characteristics for $\Kshort$ production inside $b$-tagged jets,
for data and detector-level MC events simulated with the \textsc{Powheg}+\textsc{Pythia6} and \textsc{MC@NLO}+\textsc{Herwig} generators. Total uncertainties are represented by the shaded area. Statistical uncertainties for MC samples are negligible in comparison with data.
}
\label{fig:b_results_pythia}
\end{figure}
 
\begin{figure}[H]
\begin{center}
\vspace*{-0.65cm}
\subfloat[]{
\label{fig:j_k0Pt}
\includegraphics[height=6.5cm]{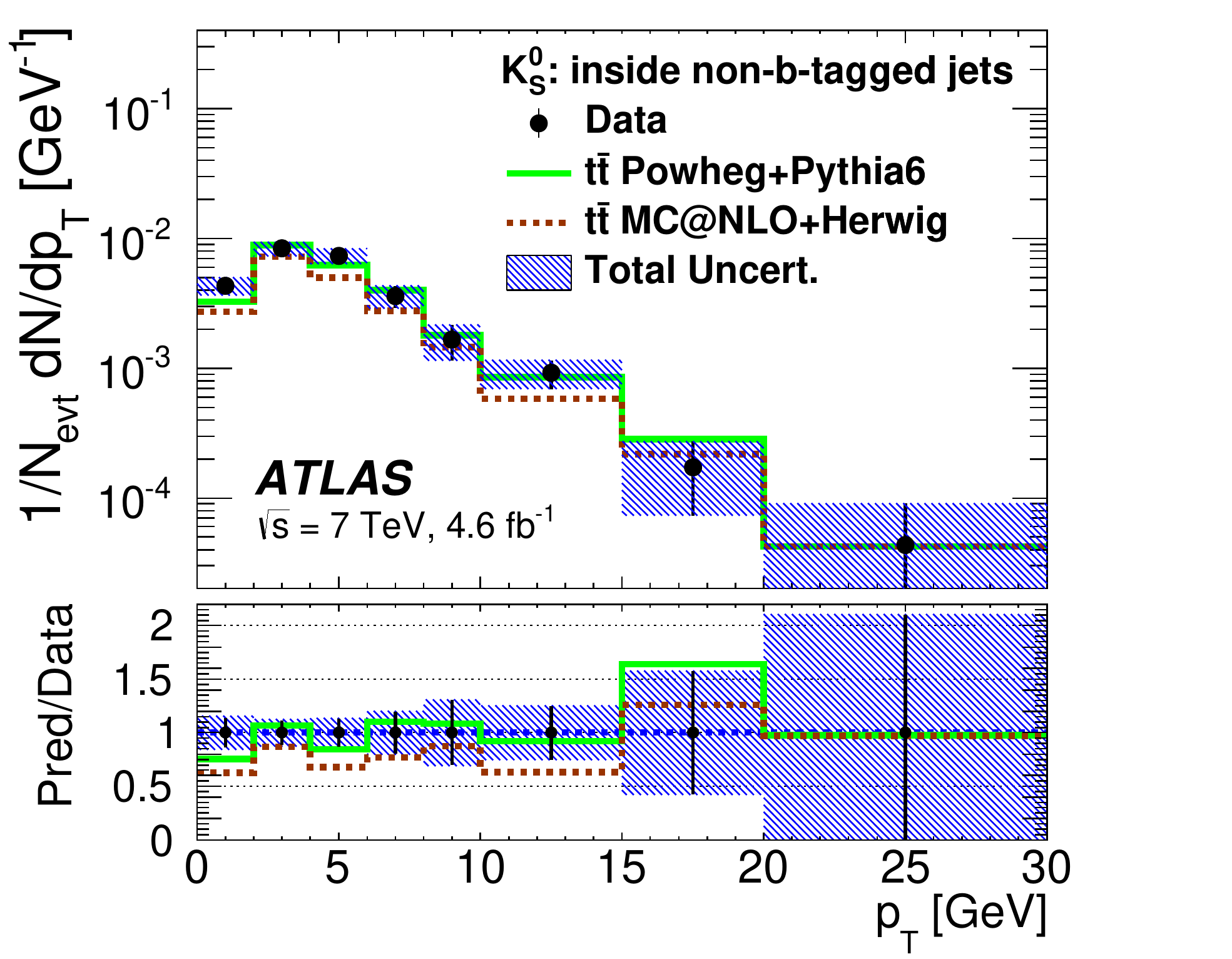}
}
\subfloat[]{
\label{fig:j_k0x}
\includegraphics[height=6.5cm]{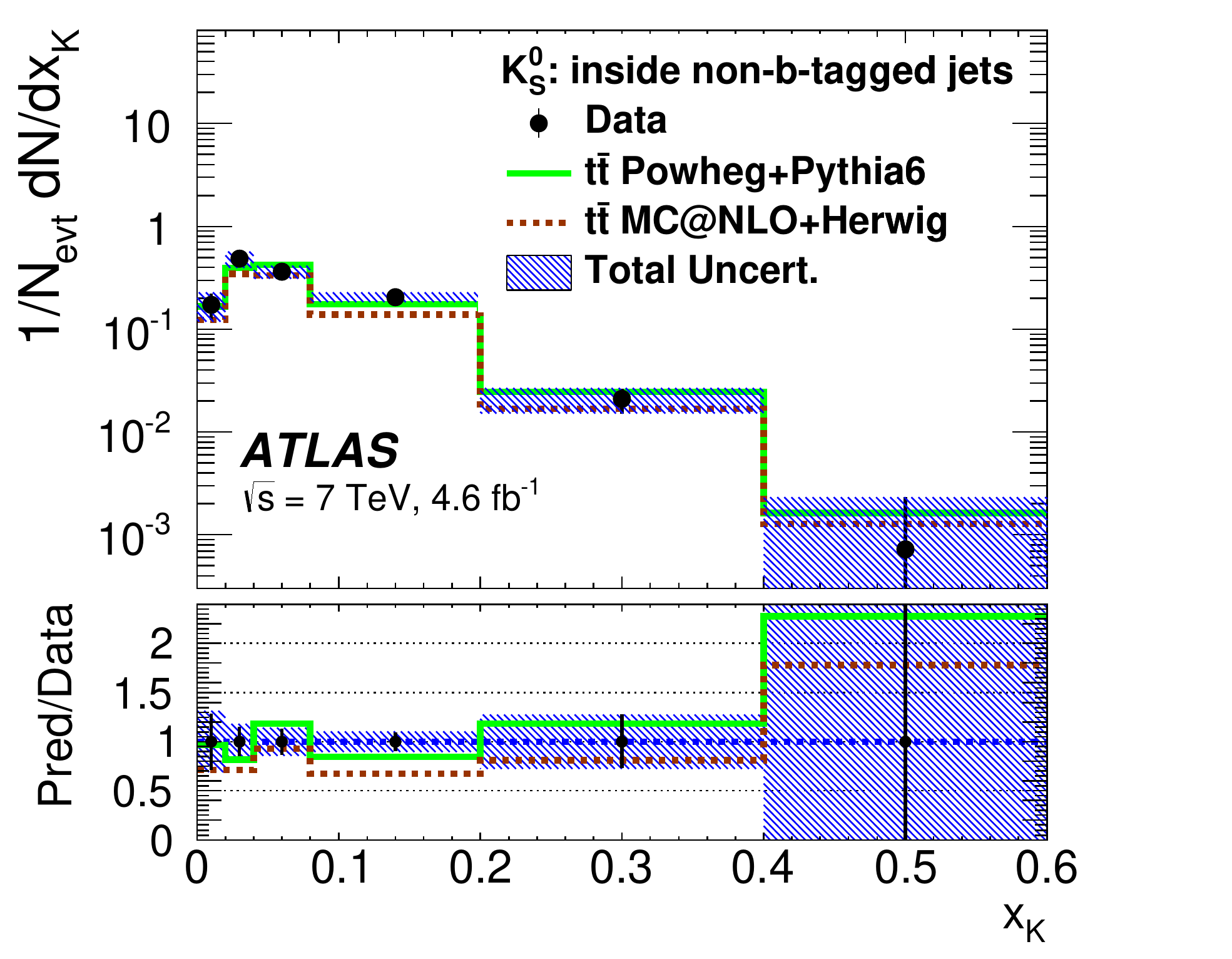}
}
\\
\subfloat[]{
\label{fig:j_k0E}
\includegraphics[height=6.5cm]{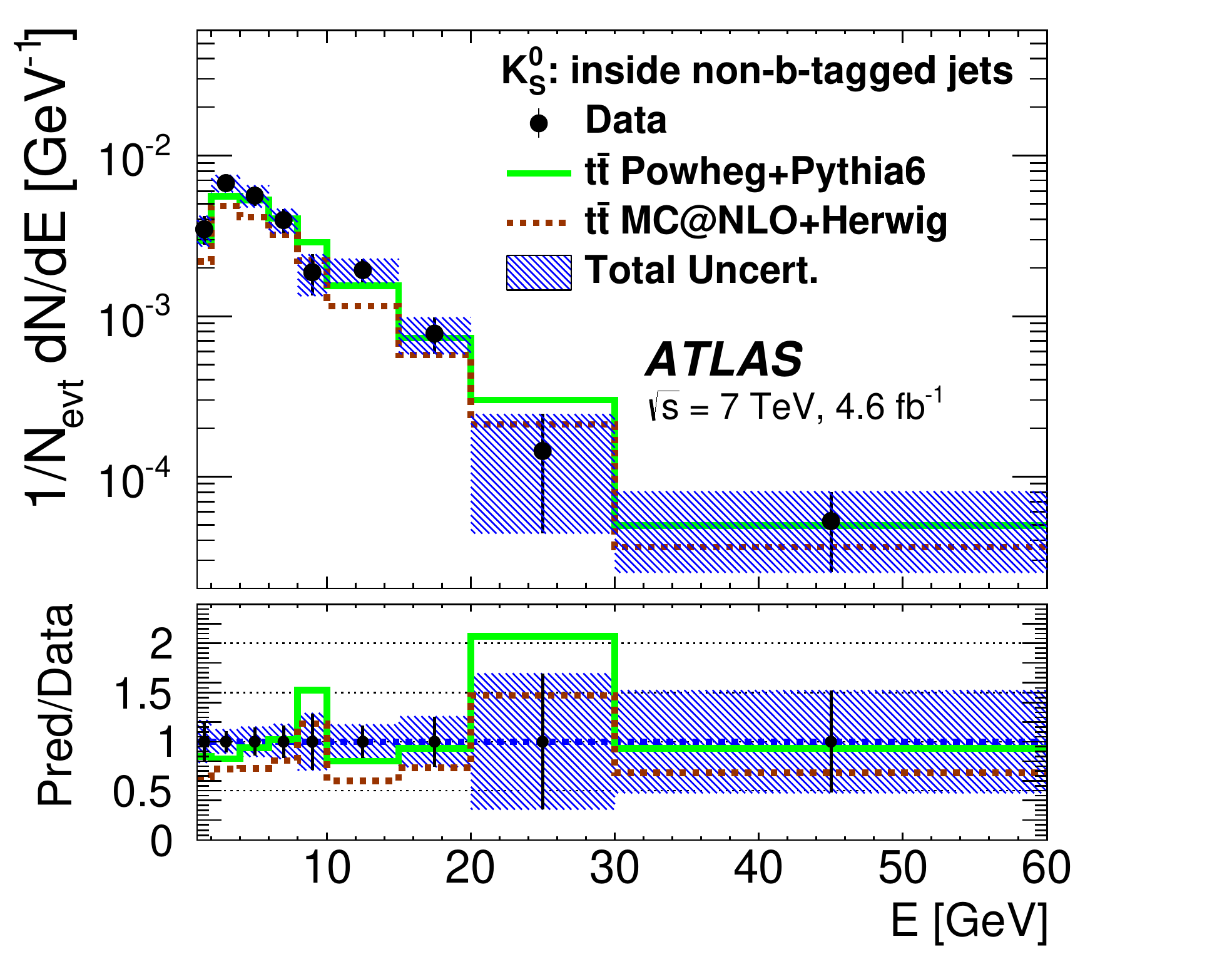}
}
\subfloat[]{
\label{fig:j_k0Eta}
\includegraphics[height=6.5cm]{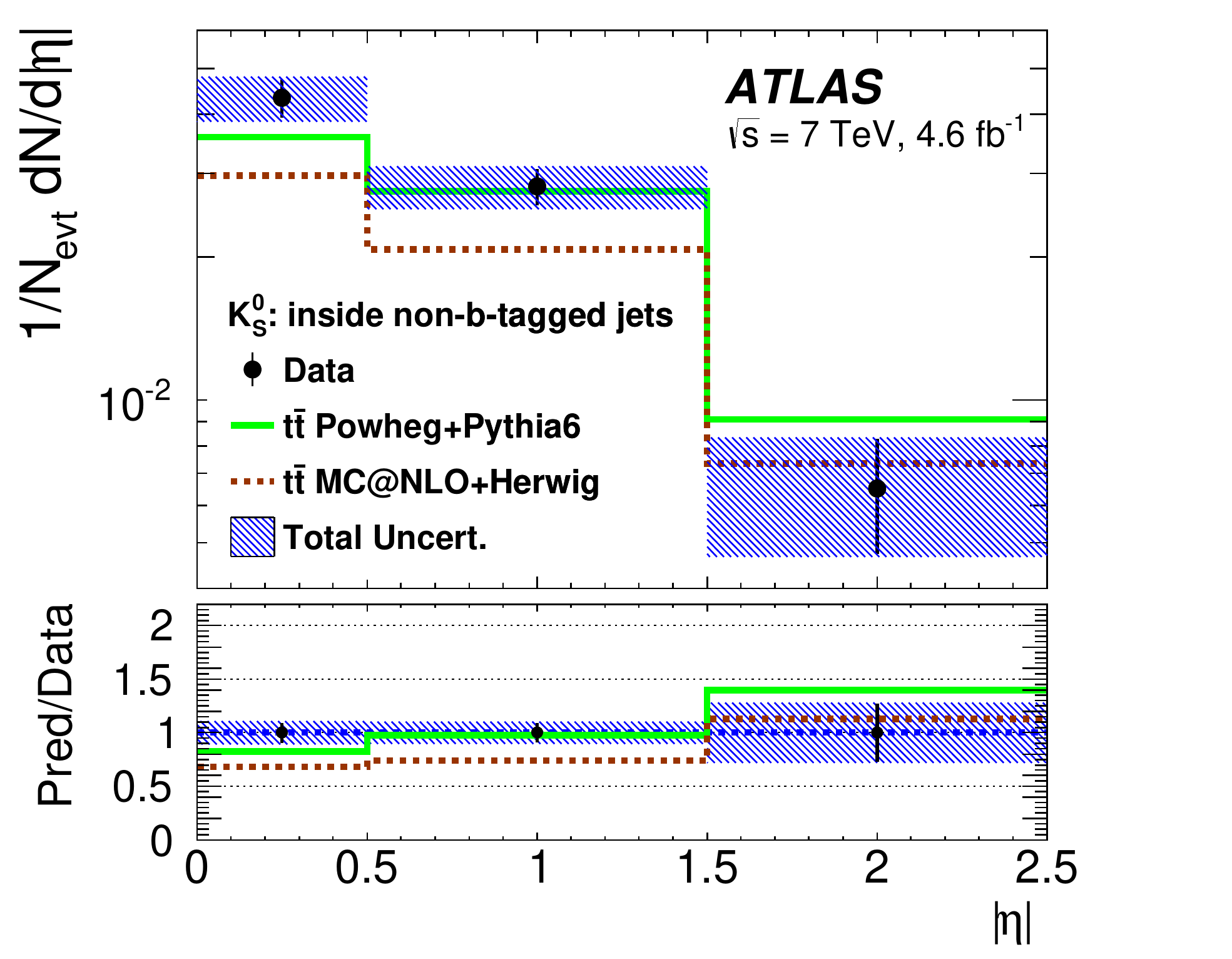}
}
\\
\subfloat[]{
\label{fig:j_k0R}
\includegraphics[height=6.5cm]{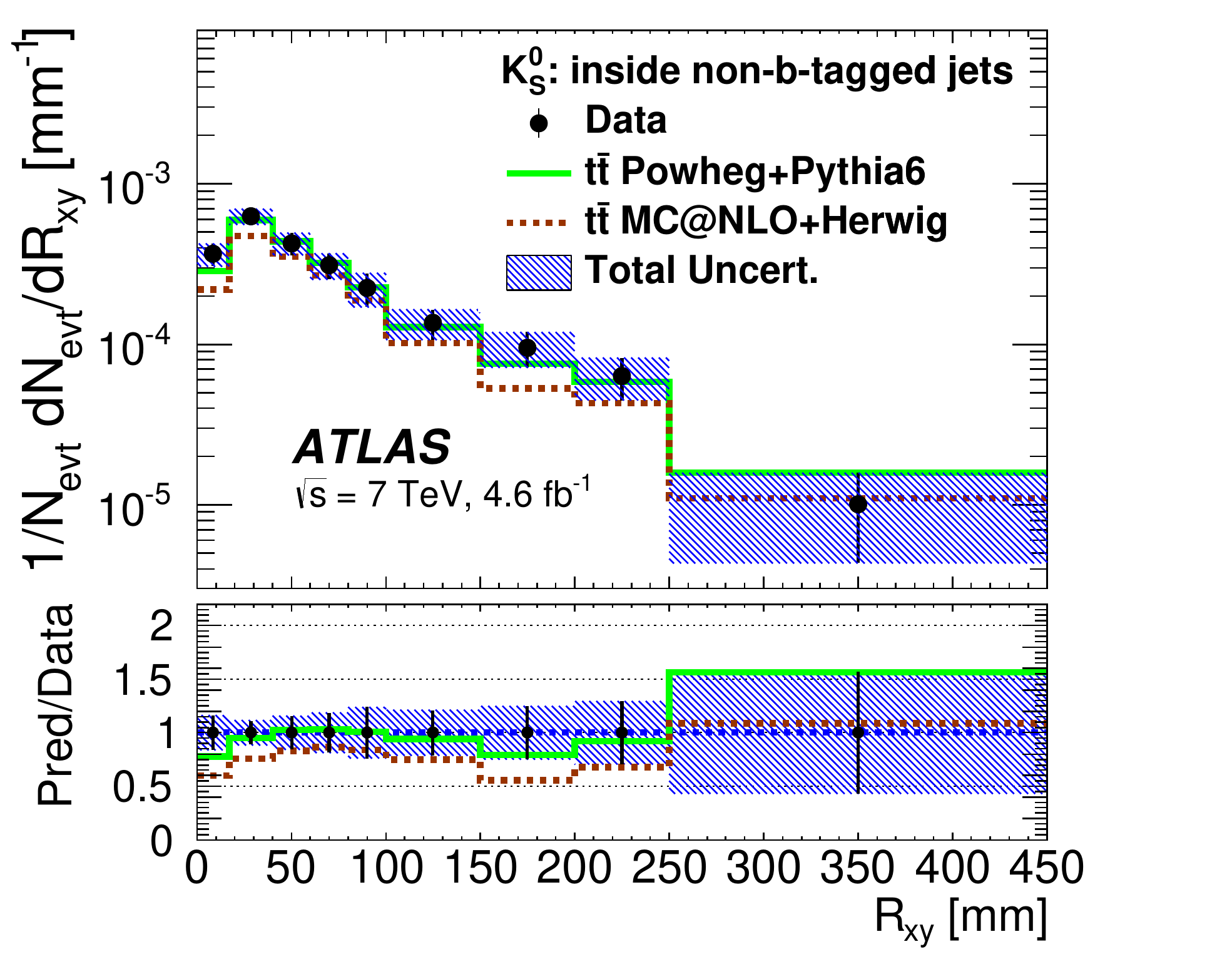}
}
\subfloat[]{
\label{fig:j_k0Multi}
\includegraphics[height=6.5cm]{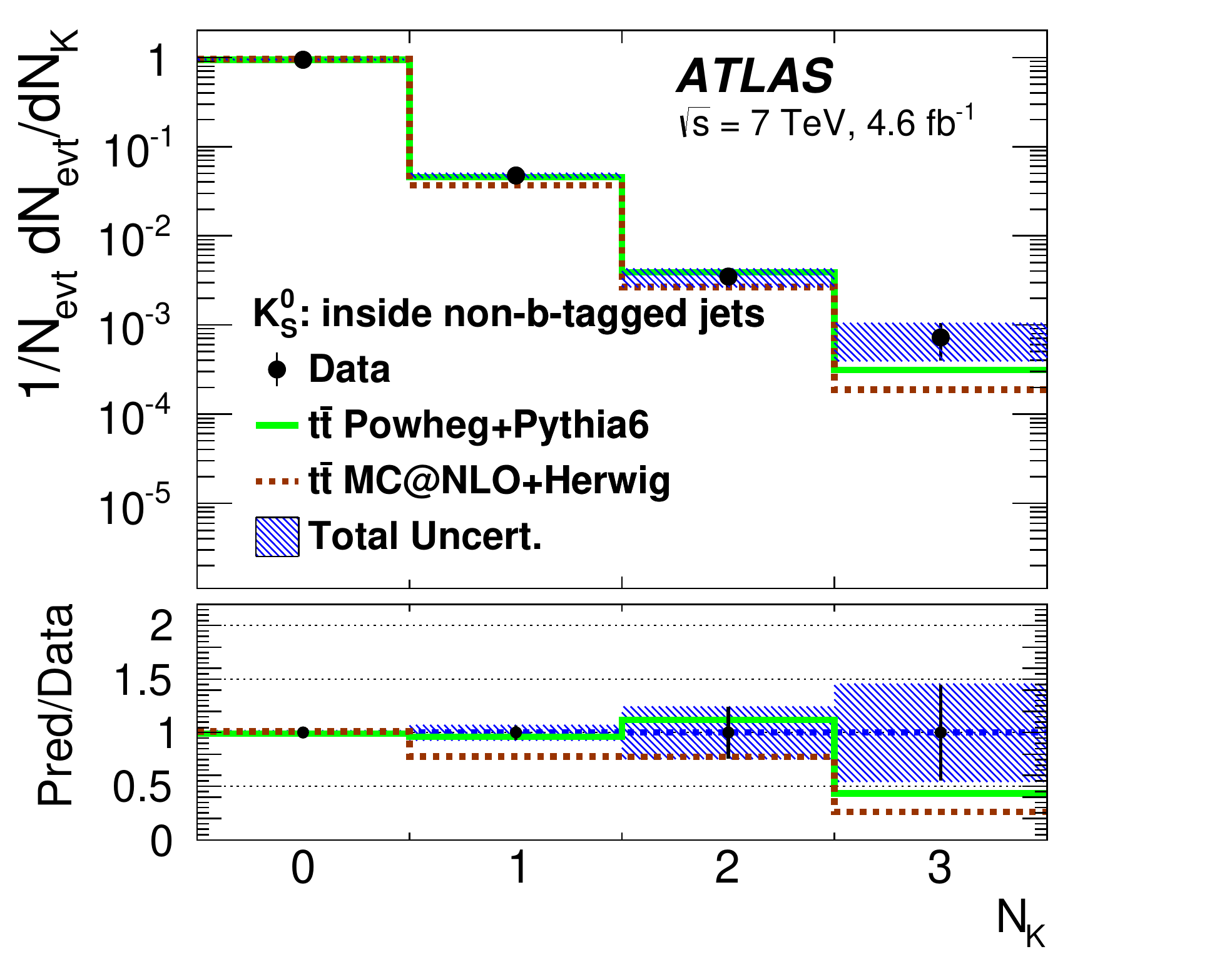}
}
\
\end{center}
\vspace*{-0.35cm}
\caption{
Kinematic characteristics for $\Kshort$ production inside non-$b$-tagged jets,
for data and detector-level MC events simulated with the \textsc{Powheg}+\textsc{Pythia6} and \textsc{MC@NLO}+\textsc{Herwig} generators. Total uncertainties are represented by the shaded area. Statistical uncertainties for MC samples are negligible in comparison with data.
}
\label{fig:j_results}
\end{figure}
 
\begin{figure}[H]
\begin{center}
\vspace*{-0.65cm}
\subfloat[]{
\label{fig:out_k0Pt}
\includegraphics[height=6.5cm]{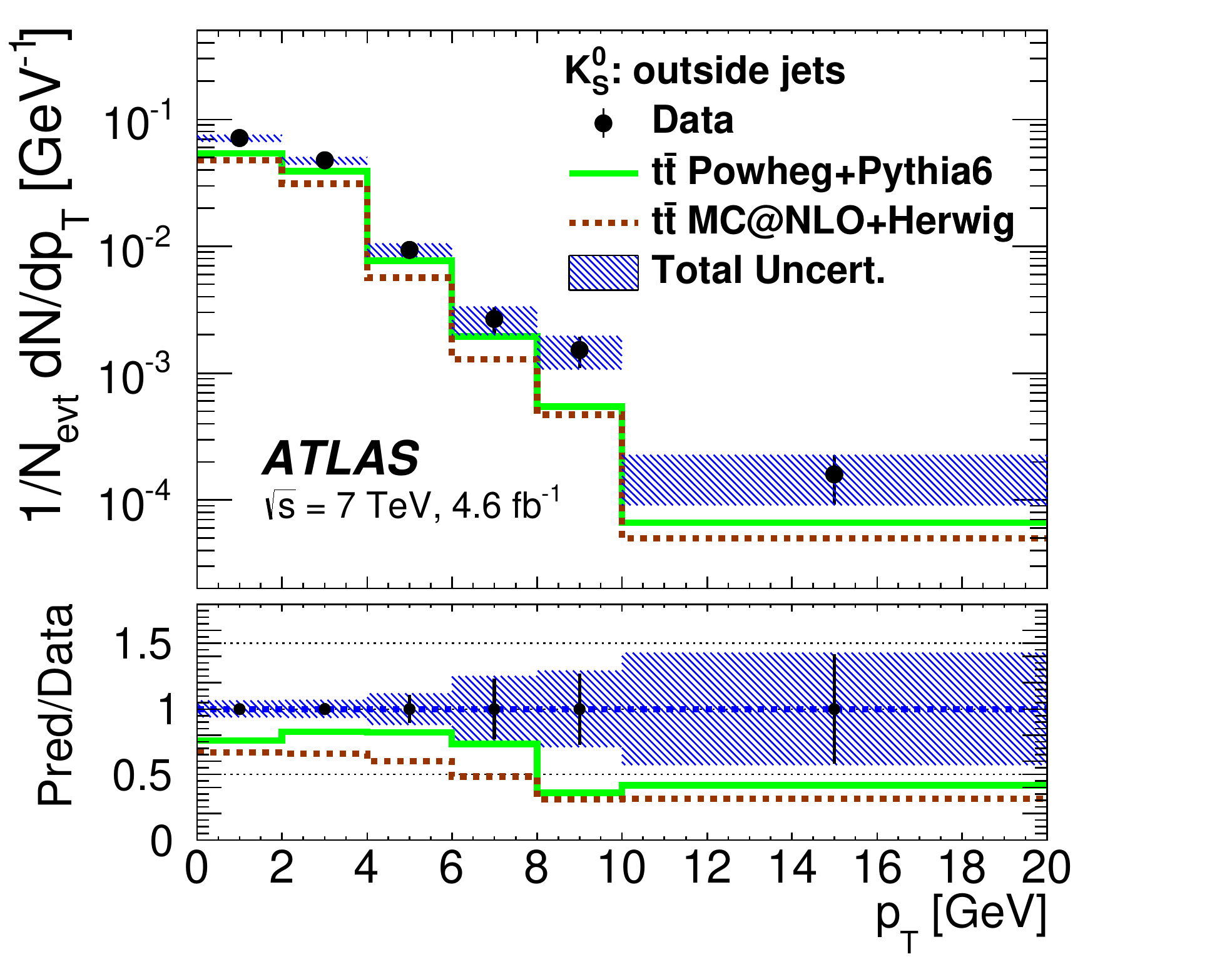}
}
\subfloat[]{
\label{fig:out_k0E}
\includegraphics[height=6.5cm]{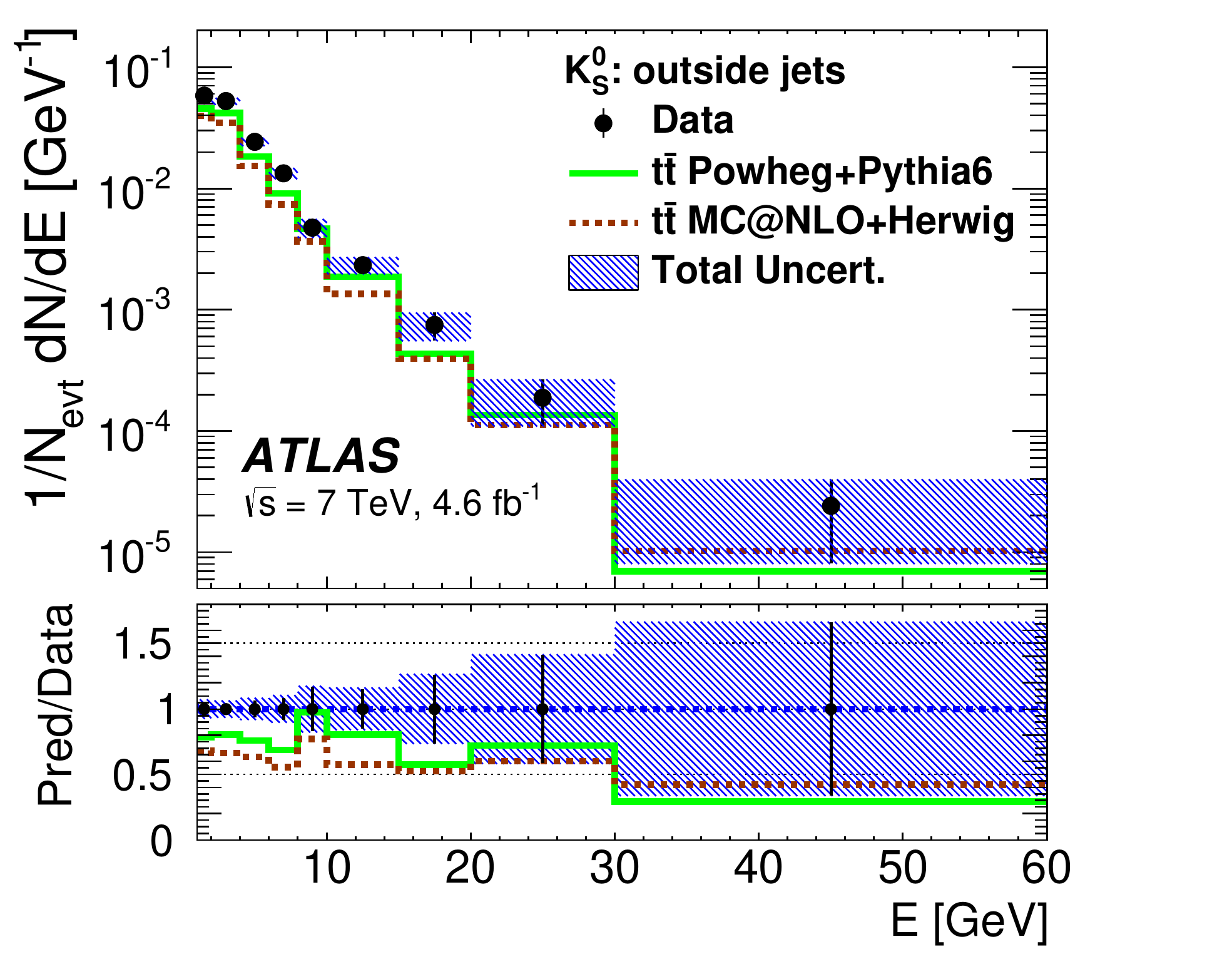}
}
\\
\subfloat[]{
\label{fig:out_k0Eta}
\includegraphics[height=6.5cm]{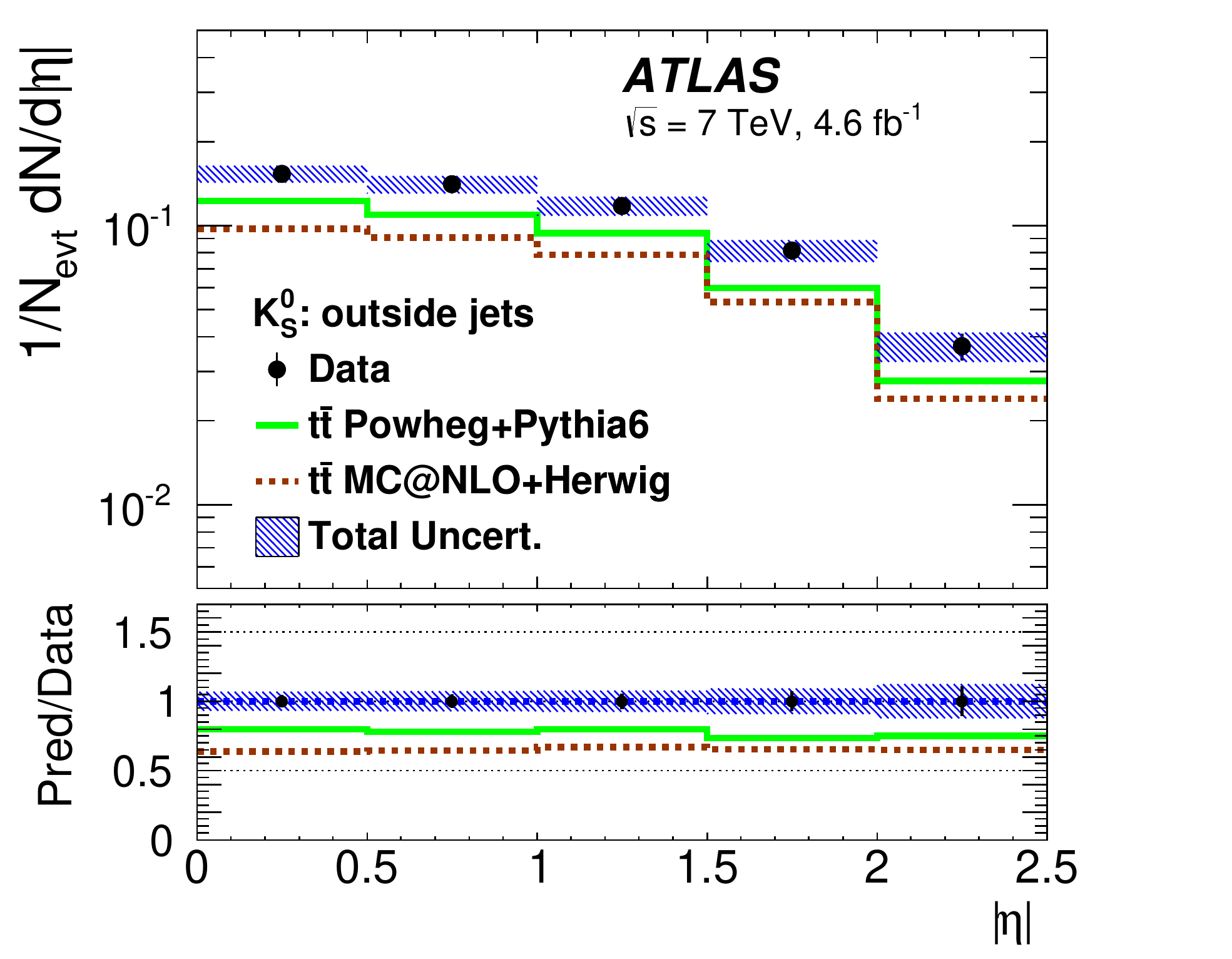}
}
\subfloat[]{
\label{fig:out_k0R}
\includegraphics[height=6.5cm]{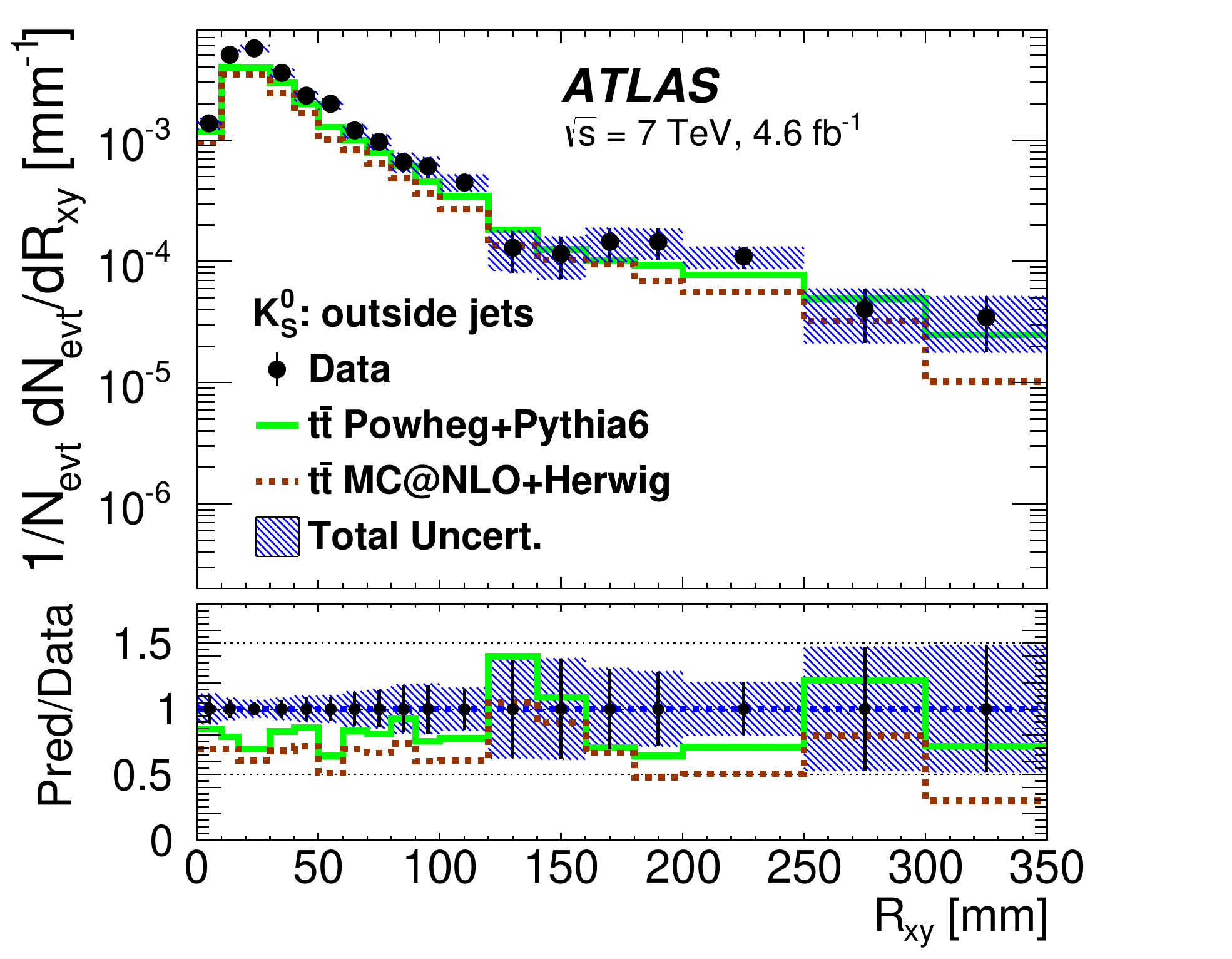}
}
\\
\end{center}
 
\subfloat[]{
\label{fig:out_k0Multi}
\includegraphics[height=6.5cm]{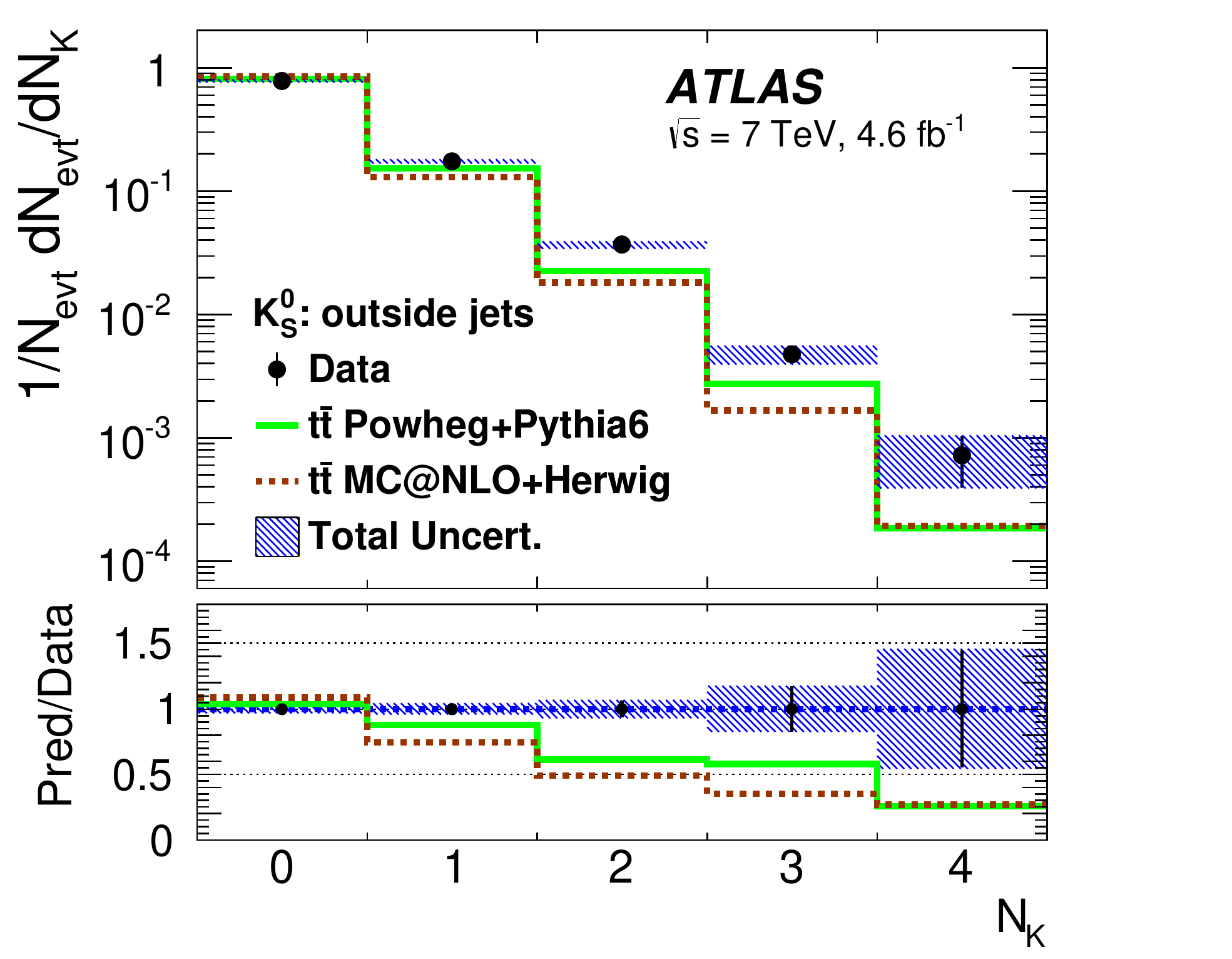}
}
\vspace*{-0.15cm}
\caption{
Kinematic characteristics for $\Kshort$ production not associated with jets,
for data and detector-level MC events simulated with the \textsc{Powheg}+\textsc{Pythia6} and \textsc{MC@NLO}+\textsc{Herwig} generators. Total uncertainties are represented by the shaded area. Statistical uncertainties for MC samples are negligible in comparison with data.
}
 
\label{fig:out_results}
\end{figure}
 
\subsection{$\Lambda$ production at detector level}
\paragraph{}
Similar distributions for $\Lambda$ production are also obtained. Due to the limited number of events, only distributions for the total sample are shown in Figure~\ref{fig:lambda_all}.
 
\begin{figure}[H]
\begin{center}
\vspace*{-0.75cm}
\subfloat[]{
\label{fig:all_lambdaPt}
\includegraphics[height=6.cm]{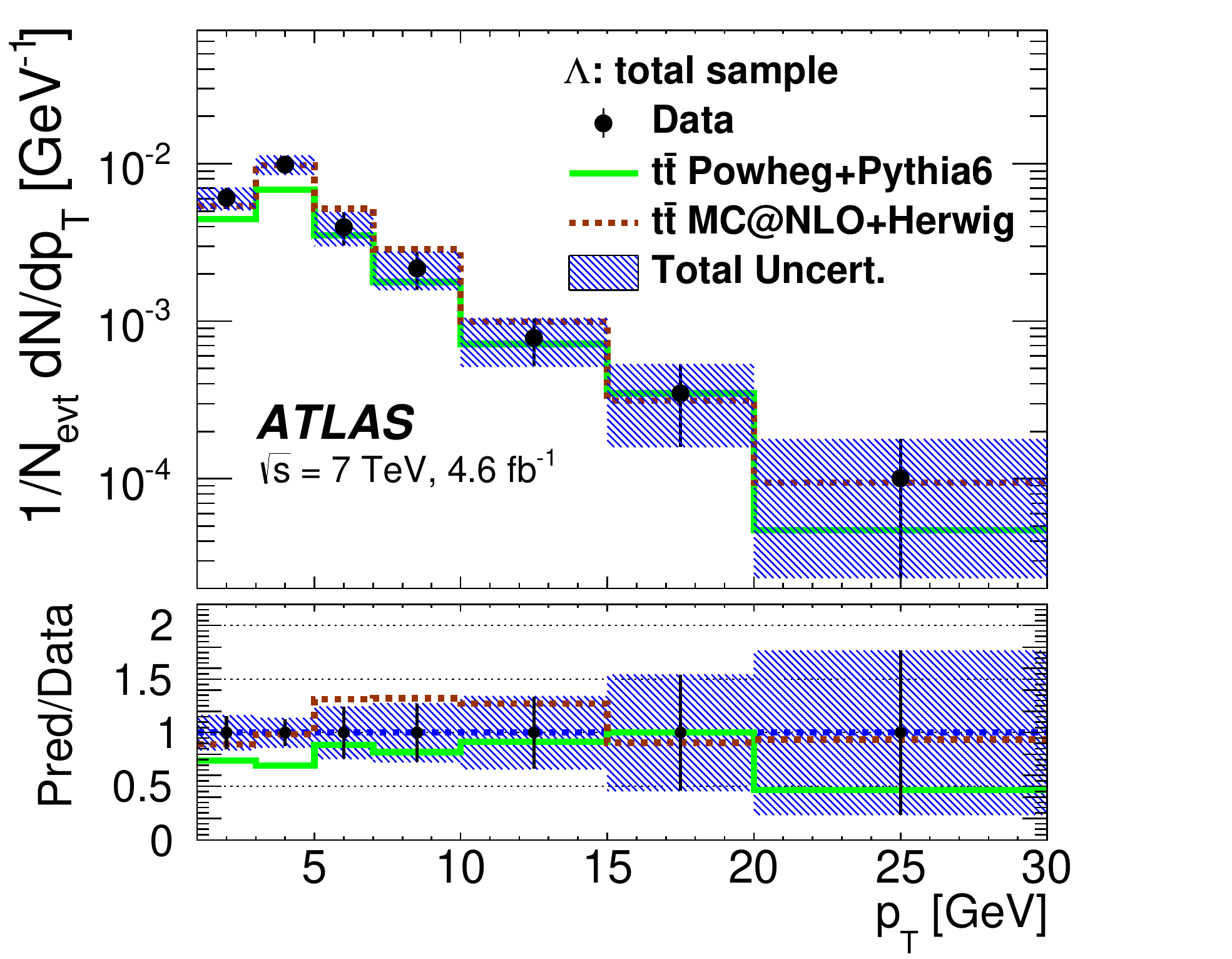}
}
\subfloat[]{
\label{fig:all_lambdaE}
\includegraphics[height=6.cm]{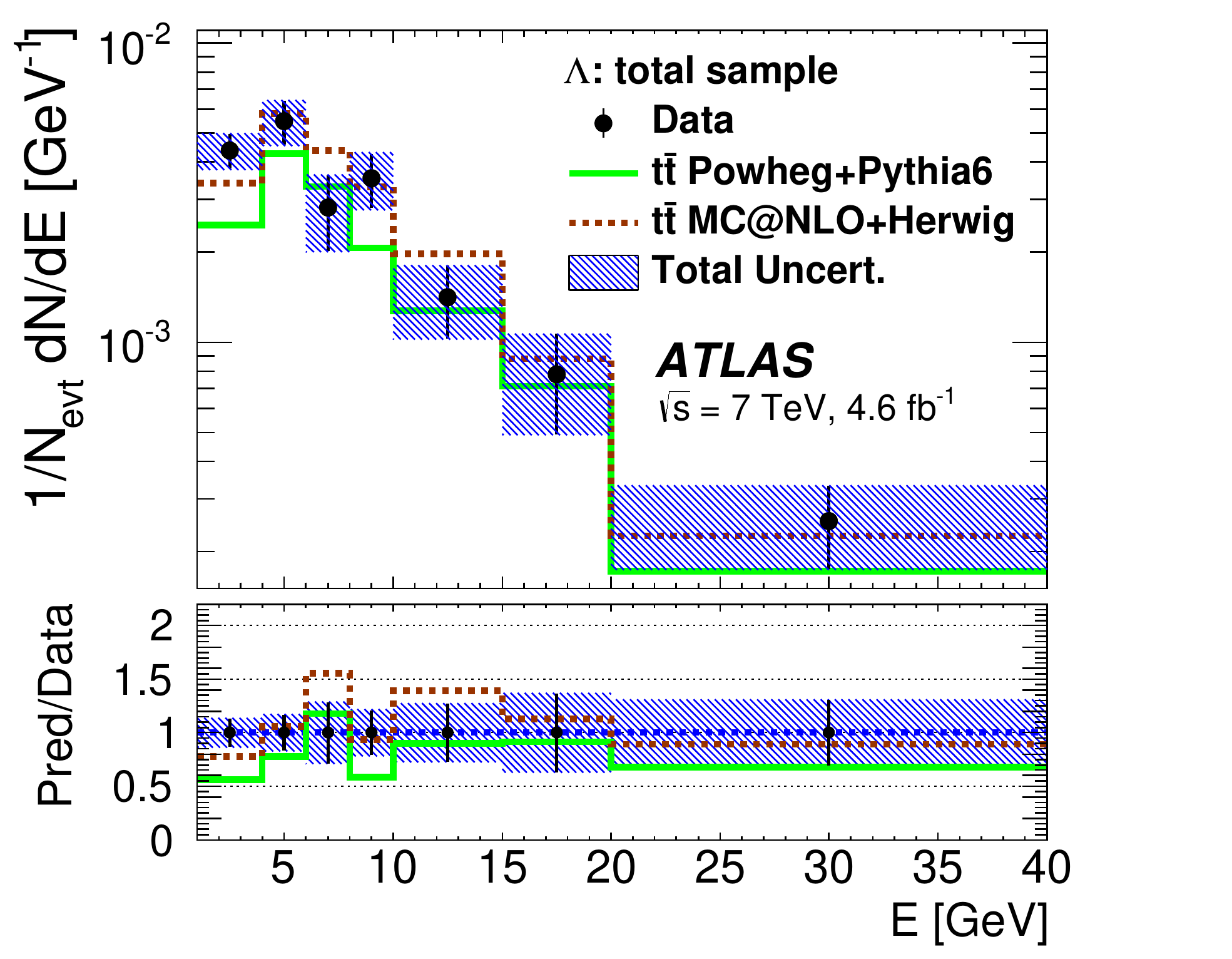}
}
\\
\vspace*{-0.45cm}
\subfloat[]{
\label{fig:all_lambdaEta}
\includegraphics[height=6.cm]{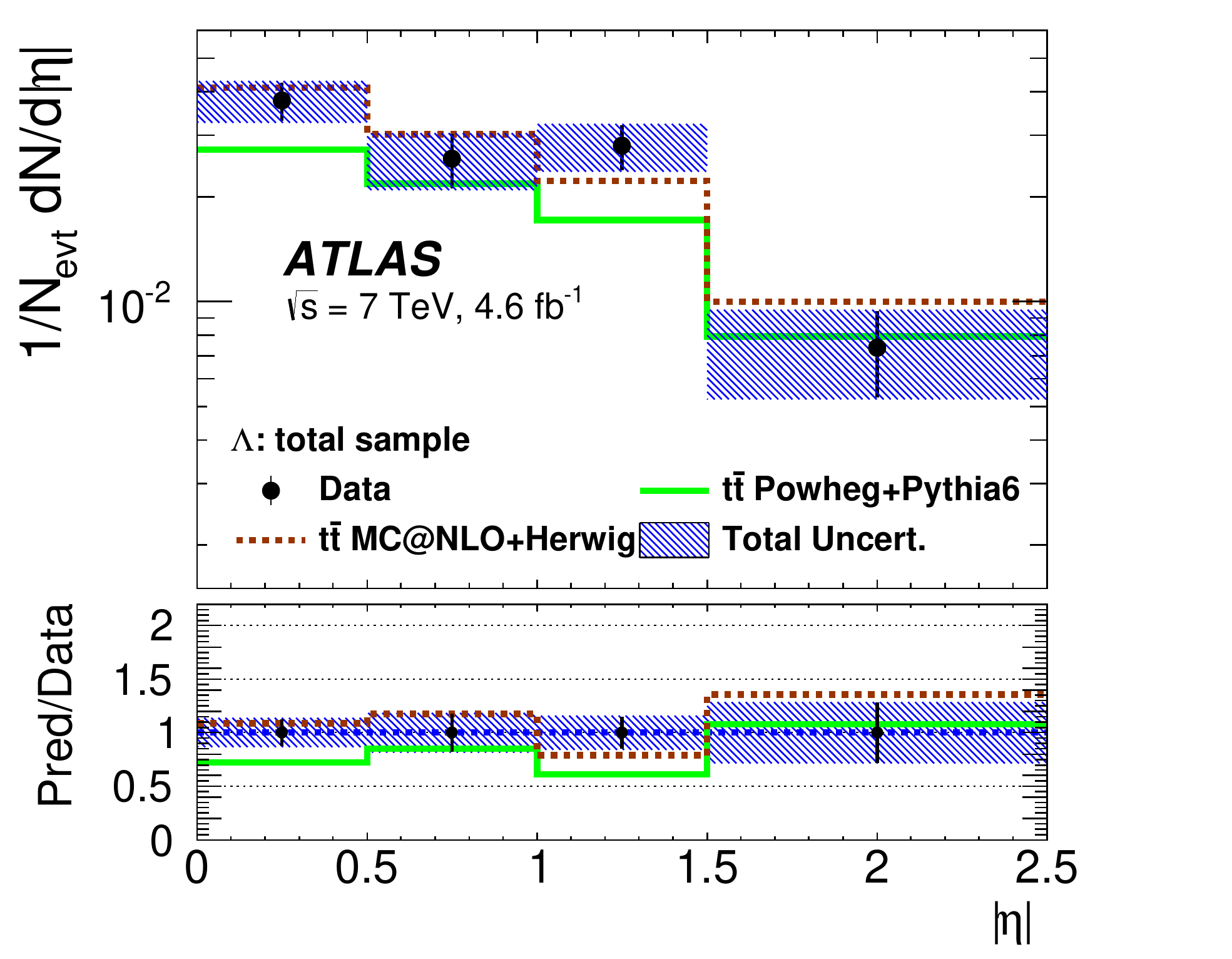}
}
\vspace*{-0.45cm}
\subfloat[]{
\label{fig:all_lambdaMulti}
\includegraphics[height=6.cm]{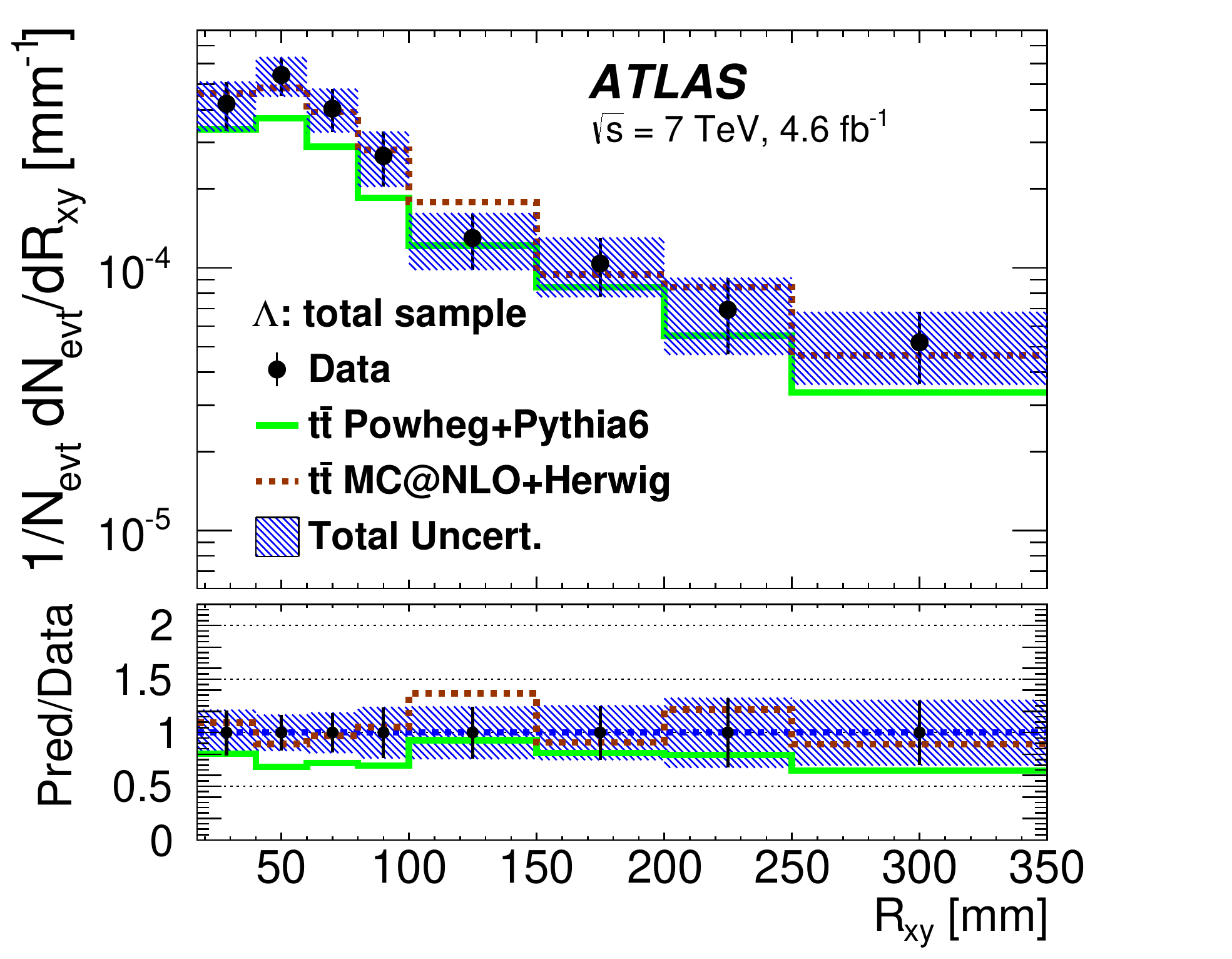}
}
\\
\end{center}
\vspace*{-0.45cm}
\hspace*{0.1cm}
\subfloat[]{
\label{fig:all_lambdaMulti}
\includegraphics[height=6.cm]{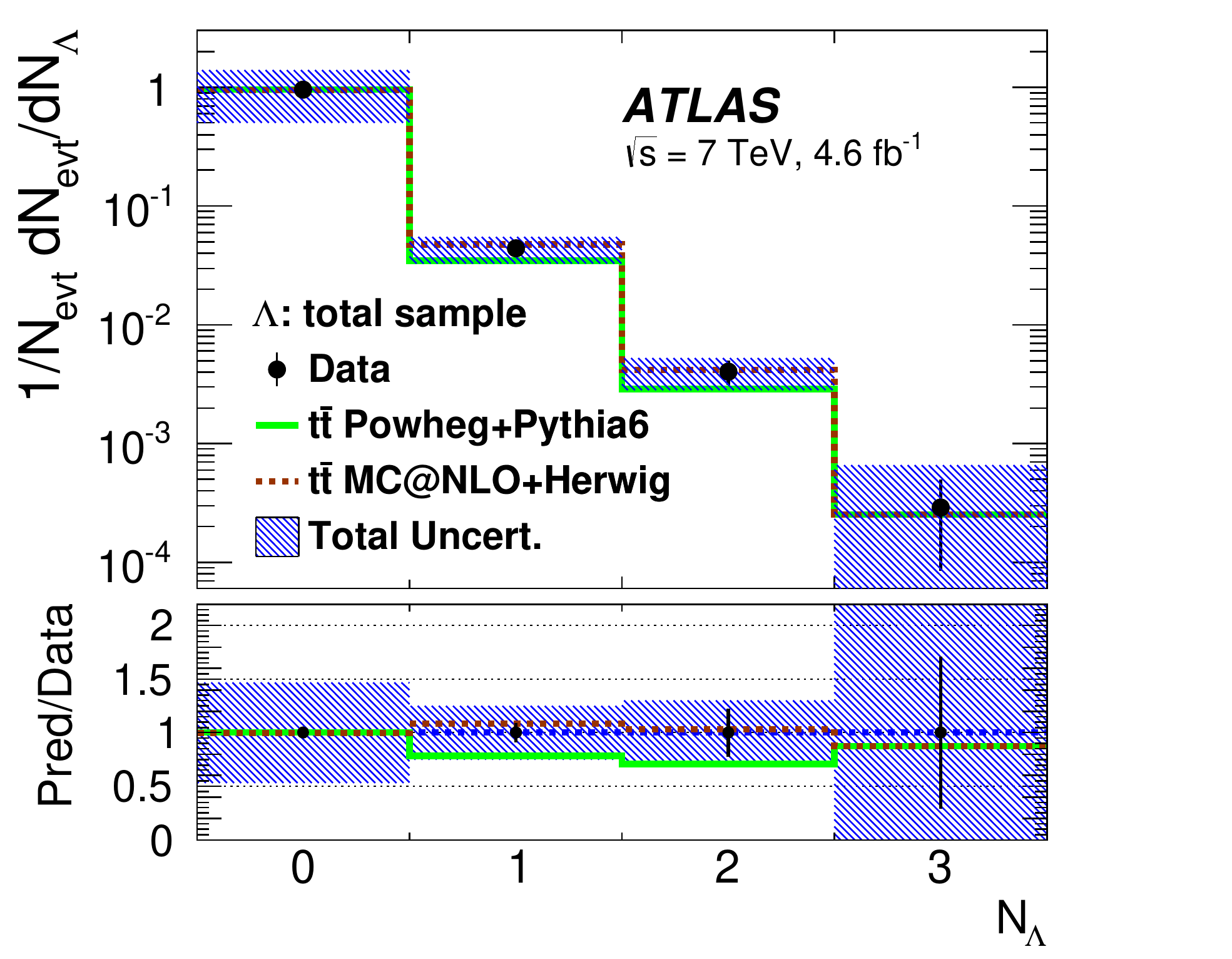}
}
\vspace*{-0.3cm}
\caption{
Kinematic characteristics for the total $\Lambda$ production,
for data and detector-level MC events simulated with the \textsc{Powheg}+\textsc{Pythia6} and \textsc{MC@NLO}+\textsc{Herwig} generators. Total uncertainties are represented by the shaded area. Statistical uncertainties for MC samples are negligible in comparison with data.
}
 
\label{fig:lambda_all}
\end{figure}
 
The gross features exhibited by the $\Lambda$ baryons are similar to those of the  $\Kshort$ mesons. The quality of the MC description of the data is also similar to that discussed in the previous subsection.

\section{Unfolding to particle level}
\label{eff}
\paragraph{}
In order to take into account detector effects, the data are unfolded to the particle level. This allows a direct comparison with theoretical calculations as well as with measurements from other experiments. For kinematic quantities such as the transverse momentum and pseudorapidity, for which migrations are negligible, this is done by computing the reconstruction efficiencies on a bin-by-bin basis (as also in Refs.~\cite{PLUTO,JADE,TPC,MARK,HRS,CELLO,TASSO,ALEPH,DELPHI,OPAL,L3,ZEUS,H1}). For the multiplicity distributions, however, a Bayesian unfolding procedure is applied because the bin-to-bin migrations are relevant.
 
\subsection{Efficiency correction}\label{eff_1}
\paragraph{}
The reconstruction efficiencies ($\epsilon$) are calculated by dividing bin-by-bin each of the distributions ($\pT$, $|\eta|$, energy and energy fraction) at detector level by the one at particle level for each of the three classes of candidates considered:
 
\begin{equation*}
\epsilon_i = \frac{\frac{1}{N^{\mathrm{det}}_{\mathrm{evt}}} \frac{\mathrm{d}N^{\mathrm{det}}_{K,\Lambda}}{\mathrm{d}x_{i}}}
{\frac{1}{N^{\mathrm{particle}}_{\mathrm{evt}}} \frac{\mathrm{d}N^{\mathrm{particle}}_{K,\Lambda}}{\mathrm{d}x_{i}}}
\end{equation*}
 
where $x_{i}$ stands for the $i$-th bin in the variable $x$ which denotes any of the kinematic
variables mentioned above. They are shown in Figure~\ref{fig:k0Eff}, for each of the classes, as well as for the total sample. For neutral strange particles embedded in $b$-tagged jets, this efficiency correction also includes the $b$-tagging efficiency. The small size of the MC sample prevents the use of a multidimensional binning for the correction procedure.
 
The particle-level distributions are obtained using leptons (from $\Wboson$ decays), jets and neutral strange particles ($\Kshort$ and $\Lambda$) in the events selected at detector level. Particle-level jets are built using all particles in MC simulation with a lifetime above $10^{-11}$~s, excluding muons and neutrinos. The kinematic criteria for jets at particle and detector level are the same, namely  $\pT>25~\GeV$ and $\abseta<2.5$. The particle-level $b$-jets are defined as those containing a $b$-hadron, with $\pT>5~\GeV$ and $\Delta R < 0.3$ from the jet axis. Particle-level $\Kshort$ and $\Lambda$ candidates, including those decaying to neutral particles, are required to be within $\abseta<2.5$ and have an energy $E>1~\GeV$, as no $\Kshort$ candidates are reconstructed below that energy at detector level. Similar to the detector level, the $\Kshort$ ($\Lambda$) candidates at particle level which fulfil these conditions are separated into three classes using the same $\Delta R$ criteria with respect to a particle-level jet.
 
MC studies show that migrations between classes when going from detector to particle level are generally smaller than 5\%. For example, $\Kshort$ candidates which are not associated with any jet at detector level, have a 1\% (3\%) probability to be classified as embedded in a $b$-jet (non-$b$-jet) at particle level. A notable exception is that of $\Kshort$ candidates inside non-$b$-tagged jets at detector level, which have a 32\%  probability to be classified as embedded in a $b$-jet at particle level. This is due to the $b$-tagging efficiency, which is included in the reconstruction efficiency as defined above.
 
The contribution of non-fiducial events, i.e. events which pass the detector-level selection but are not present at the particle level, introduces a small bias which is taken into account as a systematic uncertainty. More details are given in Section~\ref{syst}.
 
The reconstructed distributions of $\Kshort$ ($\Lambda$) are corrected with a weight given by 1/$\epsilon_i$, depending on their class. The \textsc{Powheg}+\textsc{Pythia6} MC sample was used to derive efficiencies. Since the MC simulation does not include pile-up at the particle level, the efficiency calculation effectively corrects for the pile-up effects present at the detector level. This is further discussed in Section~\ref{syst}.
 
Figure~\ref{fig:k0Eff} shows that the reconstruction efficiency inside $b$-tagged jets is lower than inside non-$b$-tagged jets, due to the fact that the average $b$-tagging efficiency is $70\%$ and the $b$-jet contamination in the non-$b$-tagged sample is around $30\%$. It was checked that the efficiency for $\Kshort$ reconstruction inside $b$-jets is independent of whether they are $b$-tagged or not at detector level. The efficiency for $\Kshort$ ($\Lambda$) outside jets peaks at lower $\pT$ values than for those inside, and falls more sharply in the distributions' tails. This can be attributed to the differences in their transverse momentum spectra.

\begin{figure}[H]
\vspace*{-0.60cm}
\begin{center}
\hspace*{-0.35cm}
\subfloat[]{
\label{fig:k0EffPt}
\includegraphics[height=6.0cm]{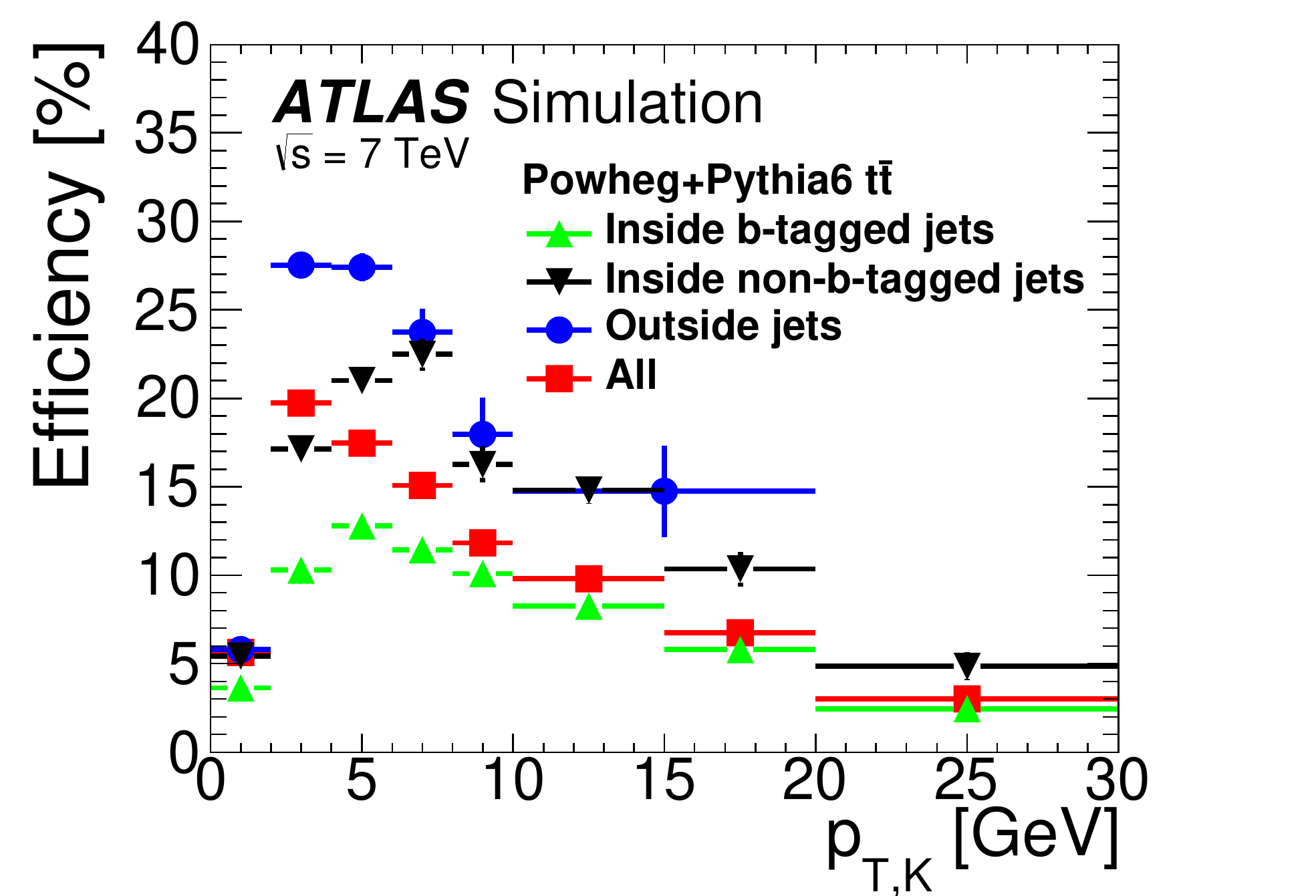}
}
\hspace*{-0.25cm}
\subfloat[]{
\label{fig:k0EffE}
\includegraphics[height=6.0cm]{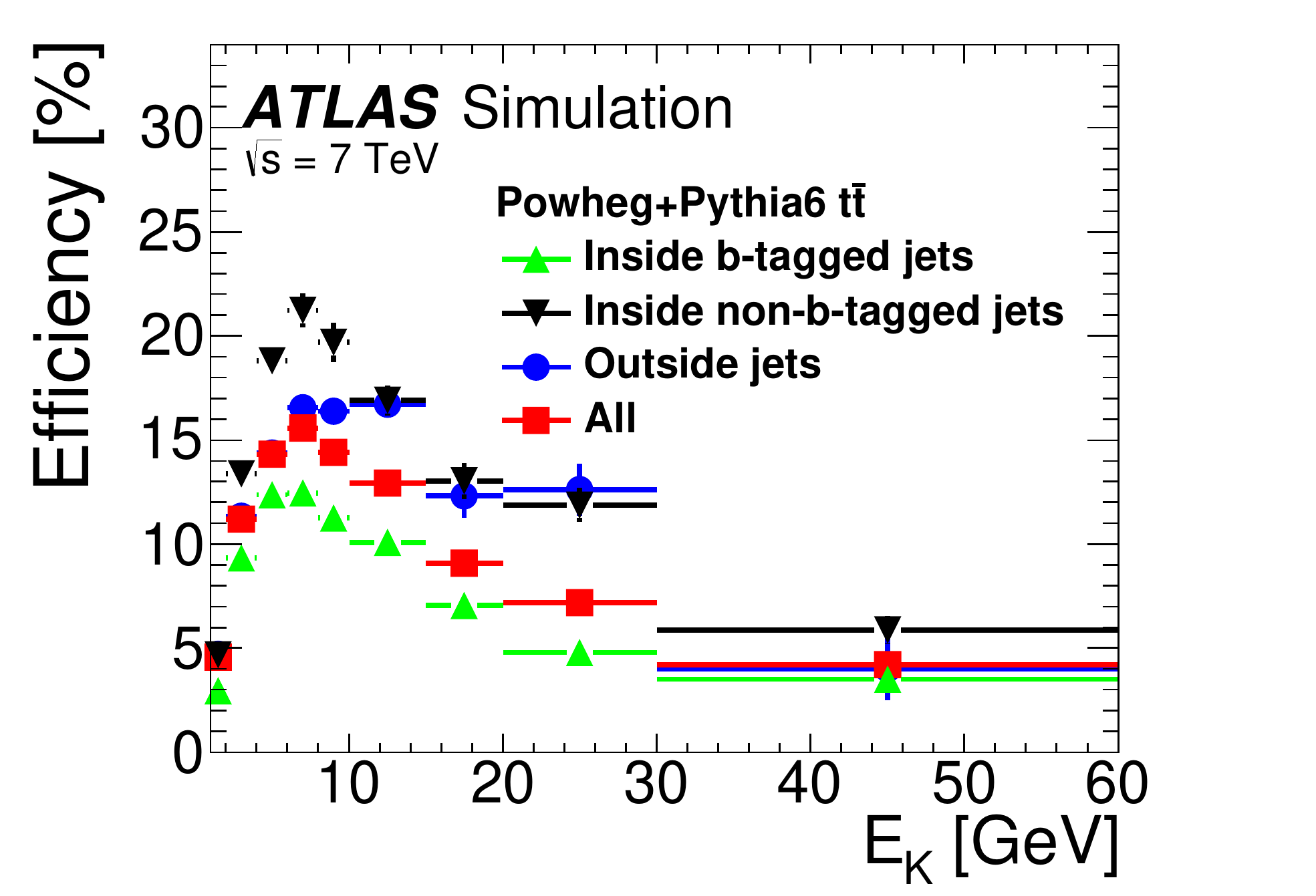}
}
\\
\vspace*{-0.45cm}
\hspace*{-0.35cm}
\subfloat[]{
\label{fig:k0EffEta}
\includegraphics[height=6.0cm]{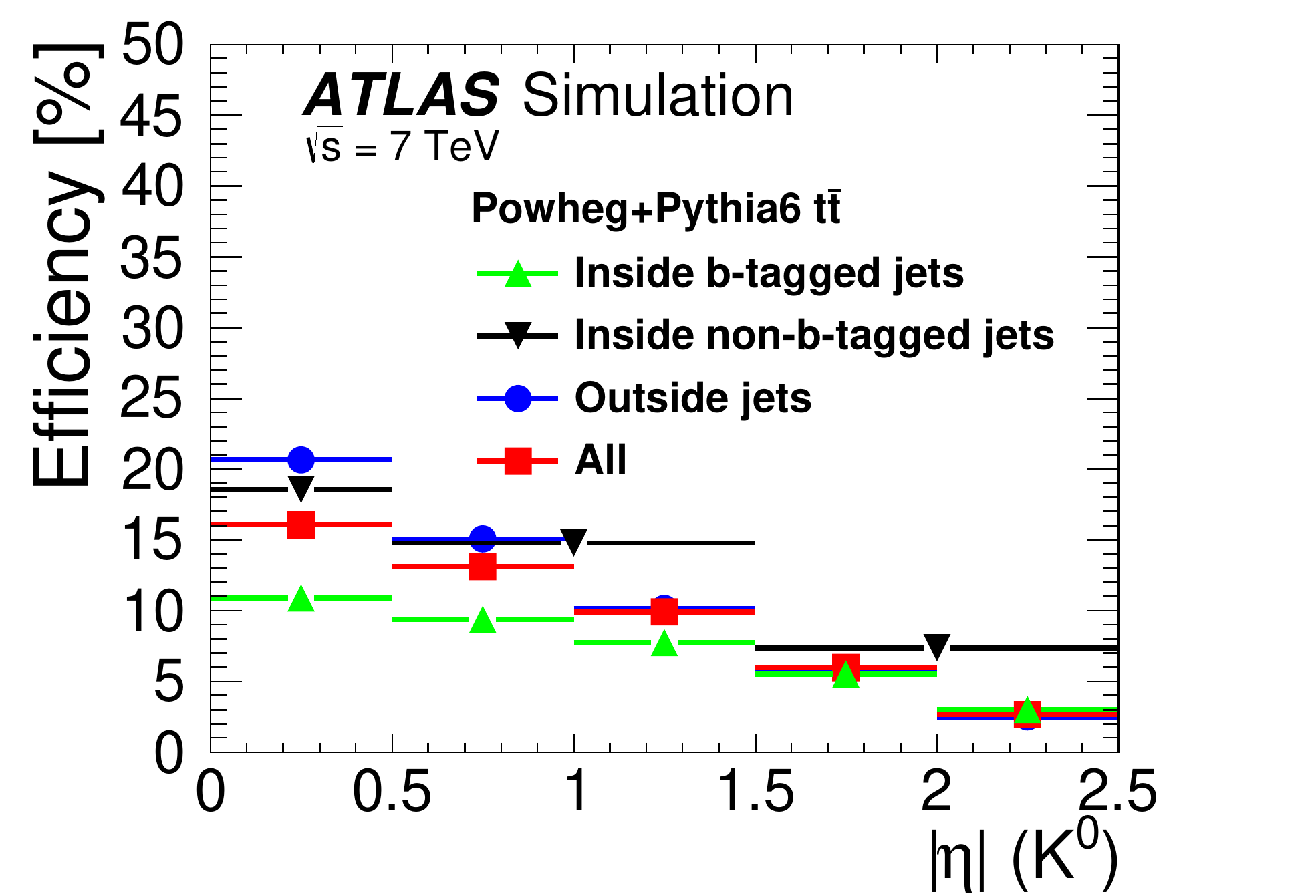}
}
\hspace*{-0.25cm}
\subfloat[]{
\label{fig:k0EffxK}
\includegraphics[height=6.0cm]{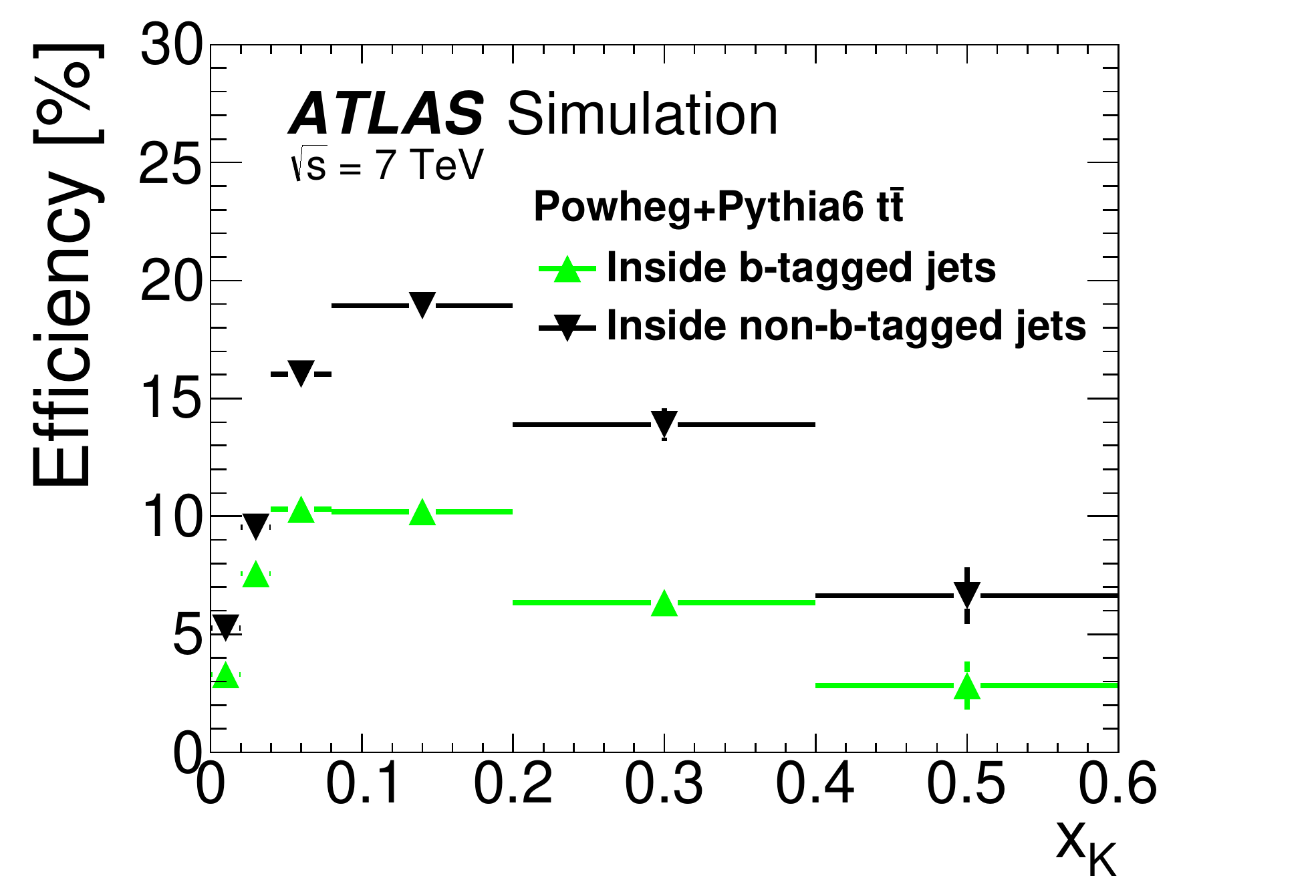}
}
\end{center}
\vspace*{-0.6cm}
\caption{
The $\Kshort$ reconstruction efficiency as a function of (a) $\pT$,
(b) energy, (c) $|\eta|$ and (d) energy fraction
for \textsc{Powheg}+\textsc{Pythia6} and four classes of $\Kshort$: inside $b$-tagged
jets (triangle), inside non-$b$-tagged jets (inverted triangle), outside any jet (circle), and the total sample (square).
}
\label{fig:k0Eff}
\end{figure}
 
The efficiency for $\Kshort$ ($\Lambda$) outside jets is lower than that reported in Ref.~\cite{minimum_bias} for a minimum-bias sample with less pile-up and restricted to lower transverse momenta.
 
In order to investigate the dependence of the efficiency corrections on the jet multiplicity, these efficiencies are derived for events with more than or at most four jets. They are found to agree within statistical errors. This is expected since each additional jet with $R=0.4$ represents only about $1.5\%$ of the total available phase space in the $\eta$--$\phi$ plane.
 
\subsection{Bayesian unfolding} \label{bayes}
\paragraph{}
The unfolding based on the efficiency calculations discussed so far relies on the assumption that the neutral strange particles, once reconstructed, are  measured to a precision which is much smaller than the bin widths. As an alternative an iterative Bayesian unfolding \cite{bayes} as implemented in the RooUnfold program \cite{RooUnfold} was tried. The method numerically calculates the inverse of the migration matrices for each of the distributions under study. The \textsc{Powheg}+\textsc{Pythia6} MC sample is used to determine these migration matrices by matching detector-level and particle-level $\Kshort$ that are in the same class and have an angular separation $\Delta R < 0.01$. The resulting matrices exhibit a very pronounced diagonal correlation. The number of iterations is chosen such that the residual bias, evaluated through a closure test as discussed in Section~\ref{syst}, is within a tolerance of 1\% for the statistically significant bins as in Ref.~\cite{inclusive}. The Bayesian unfolded distributions and the bin-by-bin corrected ones agree with each other with a precision which is much smaller than the statistical uncertainties.
 
The $\Kshort$ multiplicity distributions cannot be unfolded using a bin-by-bin efficiency correction due to the large migrations between the particle multiplicity bins at the detector and particle levels. For the multiplicity unfolding, only the visible decays, $\Kshort \rightarrow \piplus \piminus$, are considered at particle level. This reduces the size of the non-diagonal terms in the migration matrices. For the calculation of average multiplicities, a correction factor accounting for the invisible decays, $\Kshort \rightarrow \pi^0 \pi^0$, is applied a posteriori when necessary.
 
The results ($N^{\mathrm{part},i}$) of this Bayesian unfolding procedure are given by:
 
\begin{equation*}
N^{\mathrm{part},i} = \sum_{j} N_{\mathrm{detec},j} \times \epsilon_{\mathrm{detec}, j} A^{\mathrm{part}, i}_{\mathrm{detec}, j} / \epsilon^{\mathrm{part}, i}
\end{equation*}
 
where $i$ and $j$ are the particle and detector level bin indices, respectively, $N_{\mathrm{detec},j}$ is the data result at detector level, $A^{\mathrm{part}, i}_{\mathrm{detec}, j}$ is the migration matrix refined through iteration as explained above, and $\epsilon_{\mathrm{detec}, j}$ and $\epsilon^{\mathrm{part}, i}$ the matching efficiencies for detector and particle level neutral strange particles.
 
The statistical uncertainties in data and MC simulation are propagated simultaneously through the unfolding procedure by using pseudo-experiments.  A set of $10^3$ replicas is created for each measured distribution by applying a Poisson-distributed fluctuation. Each replica is then unfolded using a statistically independent fluctuated migration matrix. The statistical uncertainty of the unfolded distribution is defined as the standard deviation of the $10^3$ unfolded replicas. As a cross-check, pulls are obtained from these replicas and found to follow a normal distribution, as expected.

\section{Systematic uncertainties}
\label{syst}
Since the present analysis is concerned with the measurement of normalised distributions, many systematic uncertainty sources considered in top quark cross-section measurements \cite{dileptonic}, particularly those related to the lepton and $b$-tagging efficiency scale factors, cancel out. Similarly, the systematic uncertainty due to non-$t \bar{t}$ processes is expected to be very small and therefore not taken into account. The following systematic uncertainties are considered in this analysis:
 
\begin{itemize}
 
\item The systematic uncertainties due to tracking inefficiencies: they are taken from minimum-bias events \cite{minimum_bias} in bins of the track transverse momentum and pseudorapidity and found to be below $2\%$ and dominated by the uncertainties in the modelling of the detector material. They result in an estimated 4--5\% uncertainty for two-body decays, which is the case for $\Kshort$ and $\Lambda$ production. This relies on the assumption that the uncertainties from minimum-bias events are also valid in a dense environment as given by a jet. It was checked that there are no systematic effects in the MC description of the neutral strange particle production as a function of the angular separation to the jet-axis.

\item The systematic uncertainty related to the choice of MC generator used
in the unfolding: the systematic uncertainties due to modelling
are calculated as the relative differences between the unfolded distributions
obtained with the nominal \textsc{Powheg}+\textsc{Pythia6} MC samples and those obtained
with the alternative \textsc{MC@NLO}+\textsc{Herwig} samples. For kinematic quantities
such as the transverse momentum and pseudorapidity, they can be expressed as the deviation
of the efficiency ratios from unity.
These systematic uncertainties range up to 20--25\%, or even to 50\% for the tails
of the multiplicity distributions, and thus represent the dominant source of systematic
uncertainty. The choice of parton shower (PS) and hadronisation scheme plays
the predominant role, as tested by comparing the efficiencies calculated with
\textsc{Powheg}+\textsc{Pythia6} and \textsc{Powheg}+\textsc{Herwig} samples. The
matrix element (ME)  calculation method plays a minor role, as seen when comparing
the efficiencies calculated with \textsc{Powheg}+\textsc{Herwig} and
\textsc{MC@NLO}+\textsc{Herwig} samples.
 
\item The systematic uncertainty related to pile-up:  as a first attempt
to check how well the MC pile-up modelling describes the data, mass distributions
for $\Kshort$ and $\Lambda$ candidates are obtained for two samples of events
depending on whether the average number of interactions per bunch crossing, $\langle\mu\rangle$, is
higher or lower  than the median ($\langle\mu\rangle = 8.36$). This exercise shows that neutral
strange particles embedded in jets, with or without a $b$-tag, are not at all affected
by pile-up. This is not the case for neutral strange particles outside jets, for which
a clear linear dependence on $\langle\mu\rangle$ is observed. In order to estimate the systematic
uncertainty associated with this class, a data-driven procedure is developed, which
compares the $\langle\mu\rangle$ dependence of the $\Kshort$ multiplicity observed in data and MC
events at high longitudinal impact parameter values. The resulting systematic
uncertainty is found to be of the order of 8\%.
 
\item The systematic uncertainty related to the JES and JER:
the propagation of the JES \cite{JES} and JER \cite{JER} uncertainties is taken into account. They affect most of the distributions indirectly
through changes in the number of jets satisfying the selection criteria. The only distribution directly
affected by these uncertainties is the energy fraction. The resulting uncertainties from both the
JES and JER are found to be well below 5\%.
 
\item The $b$-tagging efficiency: in order to study the systematic uncertainty
due to the choice of $b$-tagging efficiency, the analysis was repeated using
two alternative $b$-tagging efficiencies of 60\% and 85\%. The obvious effect of this is that
the number of jets classified as $b$-tagged (or not) changes.
However, as shown in Figures~\ref{fig:b_results_pythia} and \ref{fig:j_results}, the
distributions for $\Kshort$ production inside $b$-tagged jets are very similar to those for
$\Kshort$ production inside non-$b$-tagged jets, so the uncertainties for normalised
distributions are expected to be small. The average multiplicity for $\Kshort$ production
per $b$-tagged jet (or non-$b$-tagged jet) is found to be independent of the choice of
$b$-tagging working point within the statistical uncertainty. Thus the systematic uncertainty
due to the choice of $b$-tagging efficiency is negligible.

\item The unfolding non-closure uncertainty, which is calculated in two steps. In the first step, the particle-level MC distributions are reweighted such that the reweighted detector-level distributions match the data. Then, these reweighted detector-level MC distributions are unfolded to the particle level using the same procedure as for the data, and compared with the reweighted particle-level MC distributions. The relative difference seen in this comparison is taken as the systematic uncertainty and is typically below 1\%.
 
\item Non-fiducial events uncertainty: it is calculated as the difference between two sets of \textsc{Powheg}+\textsc{Pythia6} $\Kshort$ and $\Lambda$ particle level distributions, normalised to the total number of selected events. One set comes from an event sample selected using detector-level criteria and the other set is selected using particle-level criteria. The same kinematic requirements are applied to leptons and jets at the detector and particle levels. Typically, these systematic uncertainties are below 5\%.
 
\end{itemize}
 
Table~\ref{tab:systs} summarises the approximate magnitude of the systematic uncertainties considered. The total systematic uncertainties are then calculated as the sum in quadrature of the systematic uncertainties due to the sources discussed above.
 
\begin{table}[H]
\caption{Summary table of systematic uncertainty sources: MC generator choice for the unfolding, pile-up (PU) mismodelling, tracking inefficiency, JES and JER, non-fiducial events, and non-closure test.
}
\label{tab:systs}
\begin{center}
\begin{tabular}{lrrrrrrr} \hline
Systematic uncertainty & MC choice & PU & Tracking & JES & JER & Fiducial & Non-closure\\ \hline
Relative values & $<$ 20--25\% & $\sim 8$\% & $\sim$ 4--5\% & $<5$\% & $<5$\% & $<5$\% & $<1$ \% \\ \hline
\end{tabular}
\end{center}
\end{table}

\section{Results at the particle level}
\label{unfold}
 
The $\Kshort$ and $\Lambda$ average multiplicities per event for corrected data and \textsc{Powheg}+\textsc{Pythia6} MC events at particle level are shown in Table~\ref{tab:MultiK0Unfold}. They are obtained from the unfolded transverse momentum ($\pT$) spectra.

\begin{table}[H]
\caption{$\Kshort$ and $\Lambda$ unfolded (particle-level) average multiplicities per event ($\langle n_{K,\Lambda}\rangle$),  including statistical and systematic uncertainties, for each class and for the total sample, along with the ones obtained from the \textsc{Powheg+Pythia6} MC generator at particle level.}\label{tab:MultiK0Unfold}
\begin{center}
\begin{tabular}{lrr} \hline
& Unfolded data from $\pT$ &  MC \textsc{Pythia6} particle \\ \hline
Class &  $\langle n_{K,\Lambda}\rangle \pm \mathrm{(stat)} \pm \mathrm{(syst)} $ & $\langle n_{K,\Lambda}\rangle \pm \mathrm{(stat)}$ \\ \hline
$\Kshort$ inside $b$-jets     &  $0.91  \pm 0.07 \pm 0.03$ &  $0.917 \pm 0.003$\\
$\Kshort$ inside non-$b$-jets &  $0.43  \pm 0.03 \pm 0.04$ &  $0.397 \pm 0.002$ \\
$\Kshort$ outside any jet   &   $2.91  \pm 0.10 \pm 0.57$ &  $2.248 \pm 0.004$\\
$\Kshort$ total sample     &  $4.26  \pm 0.14 \pm 0.59$ &  $3.563 \pm 0.005$\\
$\Lambda$ total sample     &  $0.65  \pm 0.07 \pm 0.05$ &  $0.499 \pm 0.014$\\
\hline
\end{tabular}
\end{center}
\end{table}
 
It was checked that the average multiplicities obtained from the iterative Bayesian unfolded multiplicity distributions are in good agreement within statistical uncertainties with the values in Table~\ref{tab:MultiK0Unfold}. Since the migration matrices were obtained considering only visible decays at particle level, the resulting multiplicities are corrected for the branching ratio of $\Kshort \rightarrow \pizero \pizero $.
 
A complete set of results at particle level can be found in Ref. \cite{hepdata}.
 
\subsection{$\Kshort$ unfolded distributions}
 
The unfolded distributions in $\pt$, $E$, $\abseta$ and $N_K$ for $\Kshort$
production inside $b$-jets, inside non-$b$-jets and  outside any jet are shown
in Figures~\ref{fig:b_results_unfold_2padsC} to \ref{fig:out_results_unfold_2padsC}. Furthermore,
for $\Kshort$ production inside jets, the distribution of the energy fraction, $x_{K}$,
is also shown. Numerical results are summarised in the Appendix. The unfolded
data are compared with the expectations from six different MC models:
\textsc{Powheg}+\textsc{Pythia6}, \textsc{MC@NLO}+\textsc{Herwig}, \textsc{Sherpa},
\textsc{Powheg}+\textsc{Pythia8}, \textsc{Powheg}+\textsc{Herwig7} and a\textsc{MC@NLO}+\textsc{Herwig7}.
 
To be more quantitative in the comparison between data and MC predictions, a
$\chi^2$ test is performed for the distributions shown in
Figures~\ref{fig:b_results_unfold_2padsC} to \ref{fig:out_results_unfold_2padsC}
prior to normalisation. The MC samples are then scaled to the same number of
$t\bar{t}$ dileptonic events as in the data. The $\chi^2$ is defined as:
\begin{equation*}
\chi^2 = V^{\mathrm{T}} \cdot Cov^{-1} \cdot V
\end{equation*}
where $V$ is the vector of differences between MC predictions and unfolded
data, and $Cov^{-1}$ denotes the inverse of the covariance matrix.
 
The covariance matrix is obtained by using pseudo-experiments. A set of $10^{3}$
replicas of the corresponding unfolded data distributions is created. In order
to include systematic effects, these replicas are smeared with Gaussian functions
whose widths are given by the systematic errors considered as uncorrelated. The
results of the $\chi^2$ test are summarised in Tables~\ref{tab:Chi2K01} to
\ref{tab:Chi2K03}, including the associated $p$-values. They are used to assess
the significance of the differences between the various generators and the data
for each observable. For the kinematic distributions, which are normalised to the
average multiplicities, the number of degrees of freedom has been taken as the
number of bins. While for the $N_K$ distributions, which are normalised to unit
area, the number of degrees of freedom has been reduced by one.
 
The unfolded distributions in Figures~\ref{fig:b_results_unfold_2padsC} to
\ref{fig:out_results_unfold_2padsC} and the $\chi^2$ and $p$-values in
Tables~\ref{tab:Chi2K01} to \ref{tab:Chi2K03} show that:
 
\begin{itemize}
 
\item On average, the \textsc{Powheg}+\textsc{Pythia6} or \textsc{Pythia8} and
\textsc{Sherpa} generators give a very similar description of the data,
while \textsc{MC@NLO}+\textsc{Herwig}, a\textsc{MC@NLO}+\textsc{Herwig7} and
\textsc{Powheg}+\textsc{Herwig7} are slightly disfavoured.
 
\item In general, the MC distributions reproduce the $\Kshort$ particle spectra
inside jets rather well. This is expected since jet fragmentation functions are
studied from S$p\bar{p}$S to Tevatron energies, so the MC simulations are tuned
fairly well.
 
\item The spectra for $\Kshort$ production outside jets are reproduced
in shape, but are underestimated by approximately $30\%$. This observation
is consistent with a CMS study of strange particle production in the
UE \cite{CMS}. These data could be used to improve the simulation of the UE,
especially to tune the $\gamma_{\mathrm{s}} = s/u$ parameter. This parameter
needs to be larger than 0.2, the value used in the \textsc{Pythia6}+\textsc{Perugia2011c}
tune \cite{perugia}. The \textsc{Pythia8}+\textsc{A14} predictions
come closer to the data than the \textsc{Pythia6}+ \textsc{Perugia2011c} ones. This is
attributed to the fact that the \textsc{A14} tune uses $\gamma_s$ value equal to 0.217,
as in the \textsc{Monash} tune \cite{Monash}, which is 10 \% larger than that in
the default \textsc{Pythia6}+\textsc{Perugia2011c} tune.
\textsc{Herwig}+\textsc{Jimmy} and  \textsc{Herwig7}+\textsc{H7UE} gives a somewhat worse
description than \textsc{Pythia6}+\textsc{Perugia2011c}, which indicates the
need to also tune the strangeness suppression here or even to use an improved
colour reconnection scheme for MPI as suggested in Ref.~\cite{HCR}. \textsc{Sherpa},
which uses strangeness suppression of $\gamma_{\mathrm{s}} = 0.4$, tends
to overestimate the $\Kshort$ yields outside jets.
 
\item The energy and transverse momentum spectra for $\Kshort$ mesons inside
$b$-jets are similar to the spectra for those inside non-$b$-jets. The spectra
for $\Kshort$ mesons produced outside jets are much softer than for those produced
in association with a jet.
 
\item The pseudorapidity distributions for $\Kshort$ mesons produced outside
jets are constant over a wider central plateau than for those produced in
association with a jet.
 
\end{itemize}

\begin{figure}[H]
\begin{center}
\vspace*{-0.65cm}
\subfloat[]{
\label{fig:b_k0Pt_unfold}
\includegraphics[height=6.5cm]{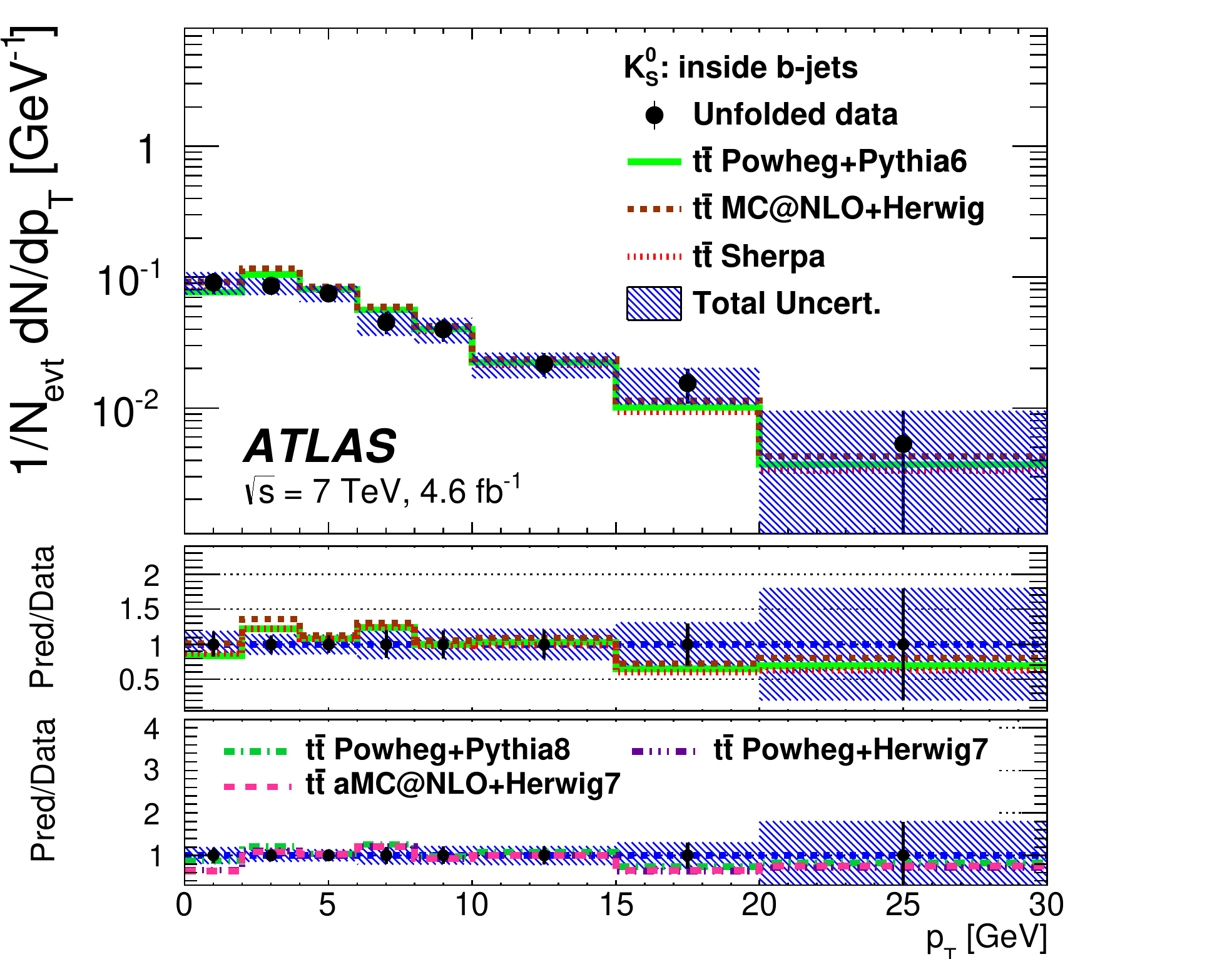}
}
\subfloat[]{
\label{fig:b_k0x_unfold}
\includegraphics[height=6.5cm]{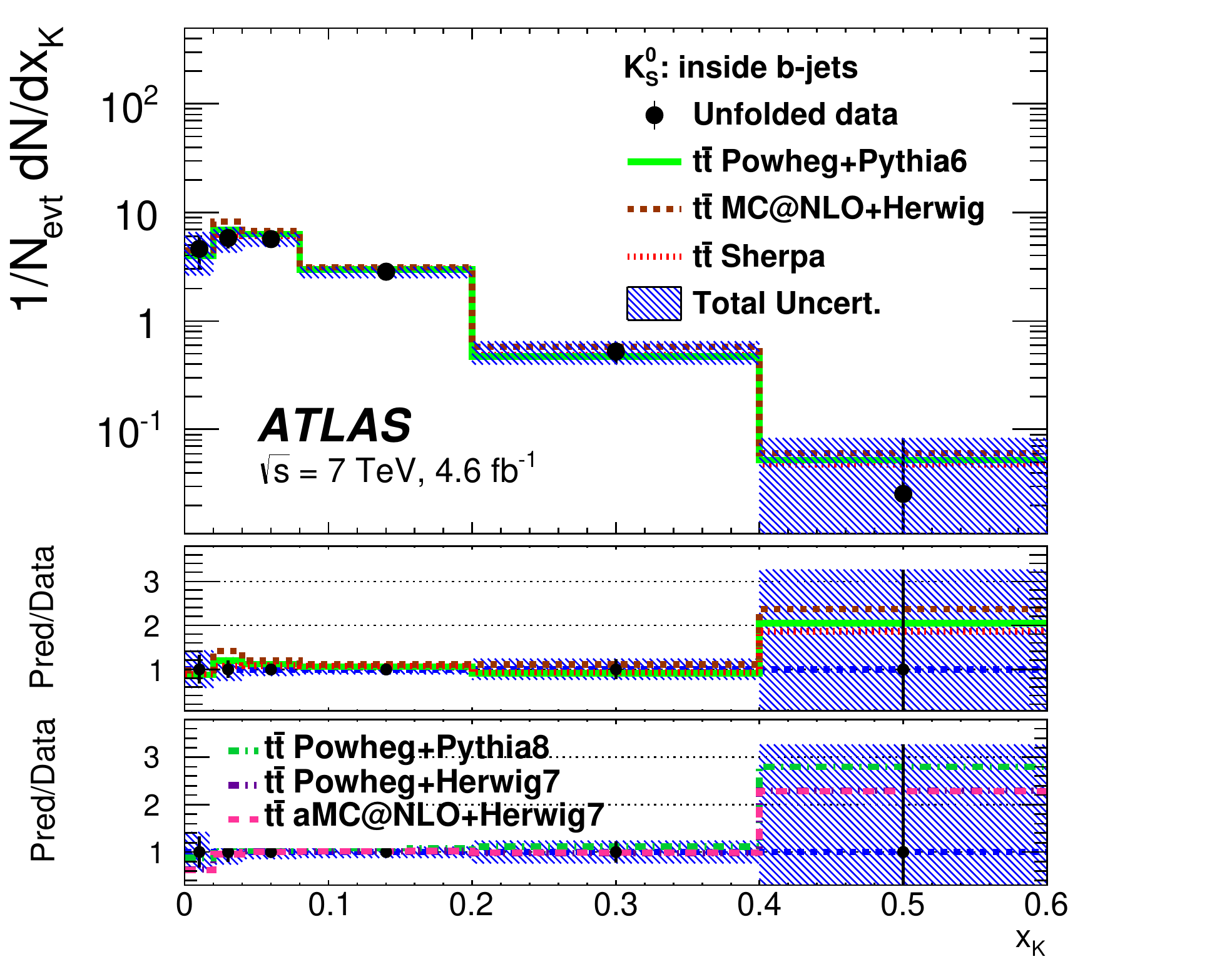}
}
\\
\subfloat[]{
\label{fig:b_k0E_unfold}
\includegraphics[height=6.5cm]{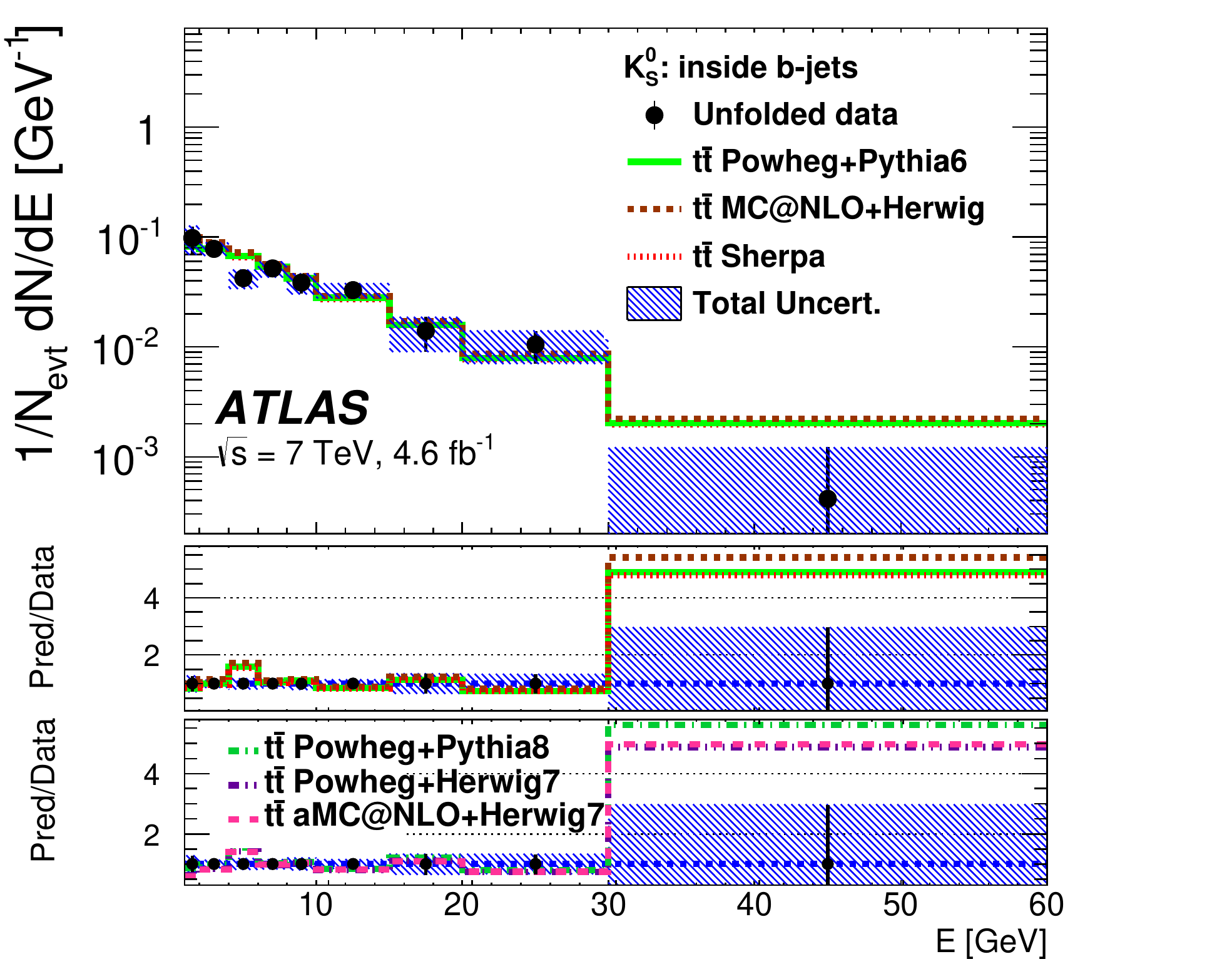}
}
\subfloat[]{
\label{fig:b_k0EtaAbs_unfold}
\includegraphics[height=6.5cm]{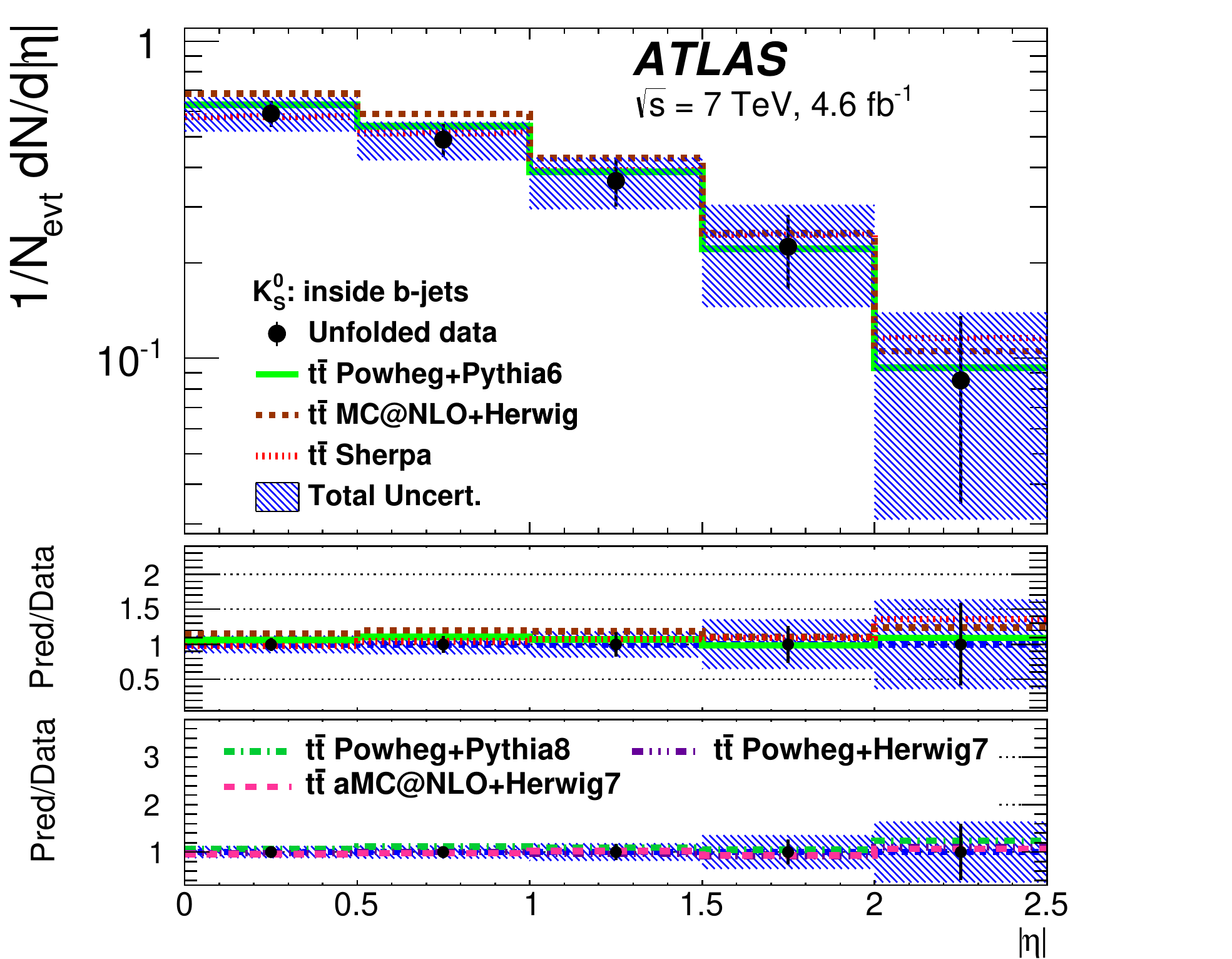}
}
\\
\end{center}
\subfloat[]{
\label{fig:b_k0Multi_unfold}
\includegraphics[height=6.5cm]{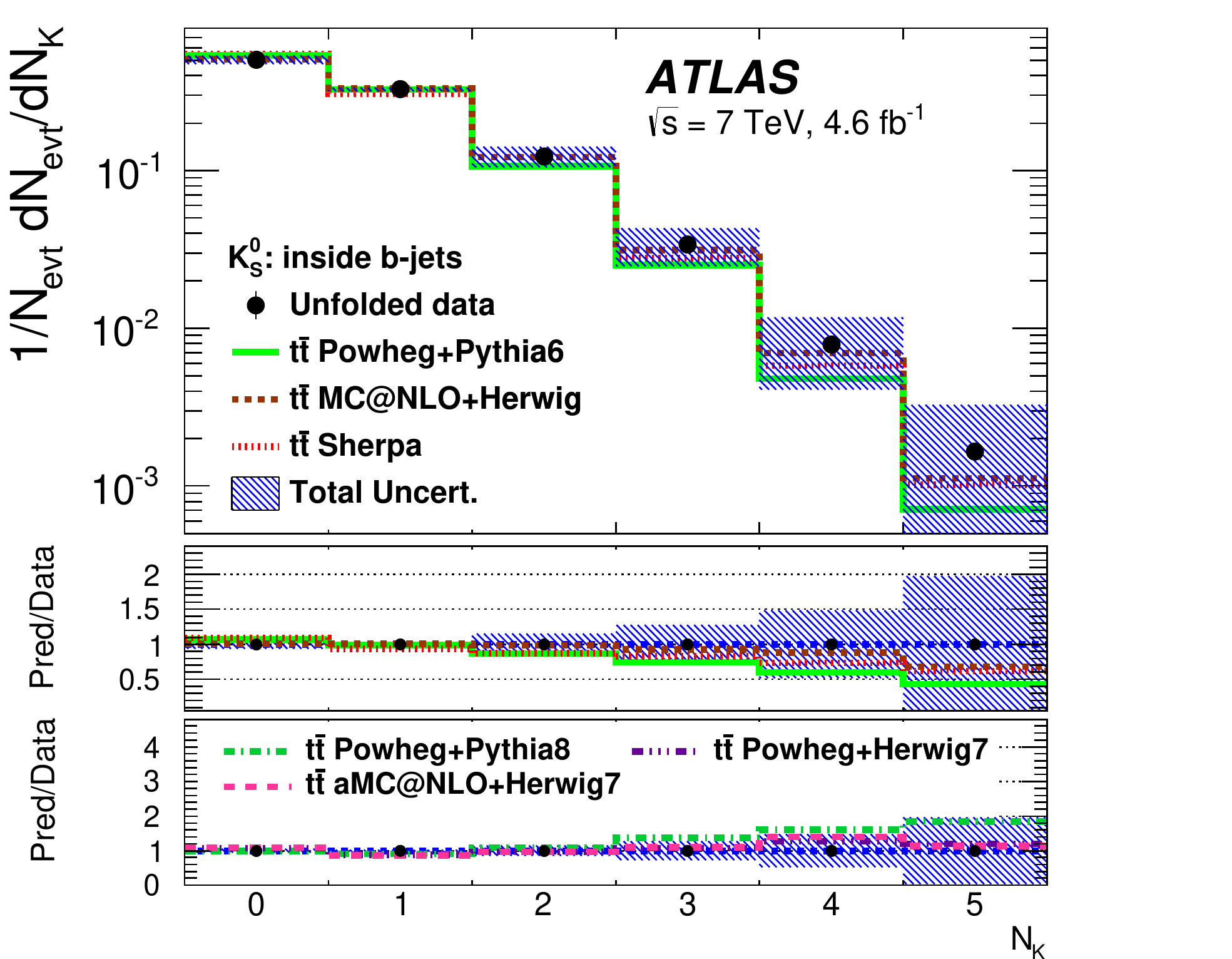}
}
\vspace*{-0.35cm}
\caption{
Kinematic characteristics for $\Kshort$ production inside $b$-jets, for corrected data and particle-level MC events simulated with the \textsc{Powheg}+\textsc{Pythia6}, \textsc{MC@NLO}+\textsc{Herwig}, \textsc{Sherpa}, \textsc{Powheg}+\textsc{Pythia8}, \textsc{Powheg}+\textsc{Herwig7} and a\textsc{MC@NLO}+\textsc{Herwig7} generators. Total uncertainties are represented by the shaded area. Statistical uncertainties for MC samples are negligible in comparison with data.}
\label{fig:b_results_unfold_2padsC}
\end{figure}

\begin{figure}[H]
\begin{center}
\vspace*{-0.65cm}
\subfloat[]{
\label{fig:j_k0Pt_unfold}
\includegraphics[height=6.5cm]{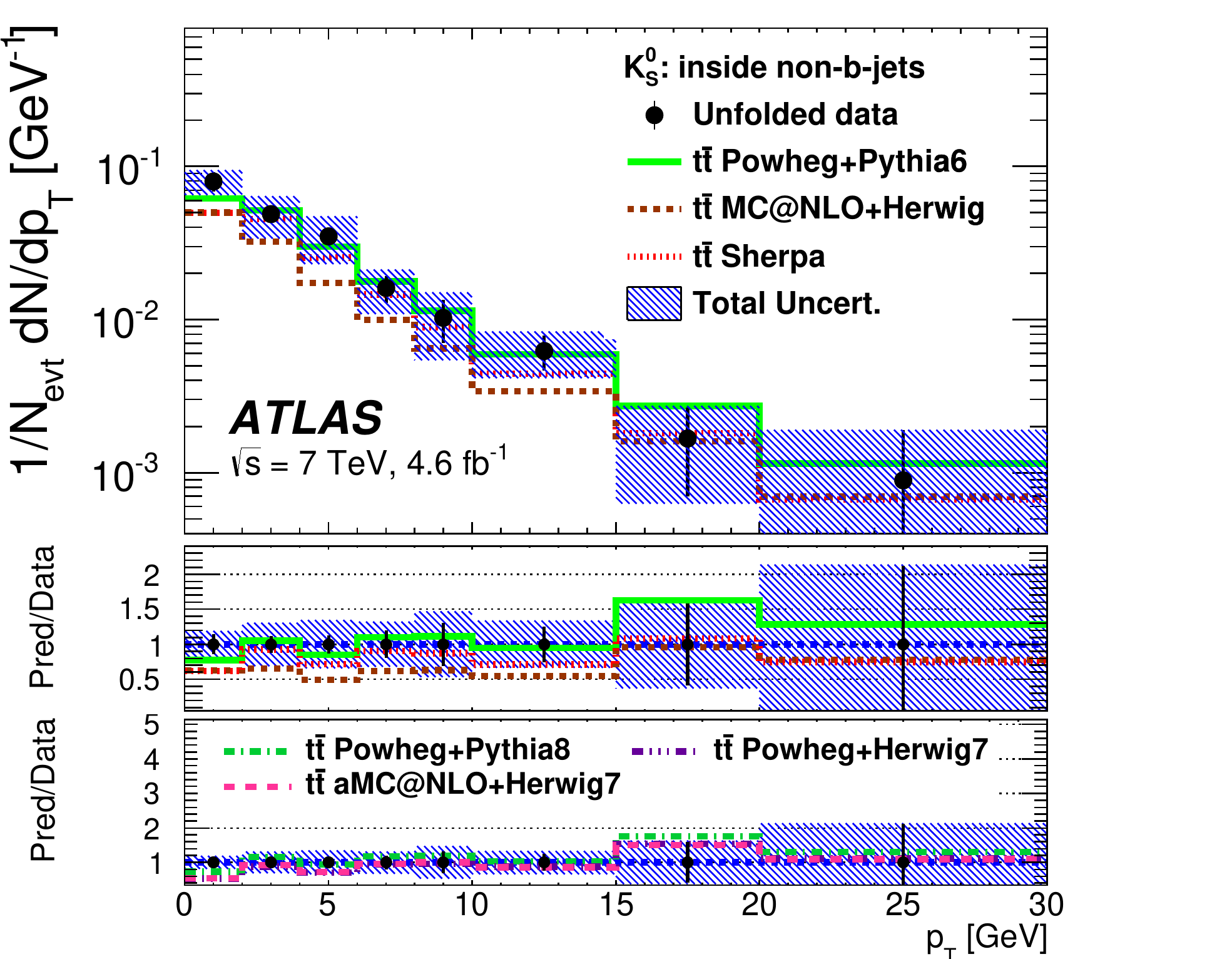}
}
\subfloat[]{
\label{fig:j_k0x_unfold}
\includegraphics[height=6.5cm]{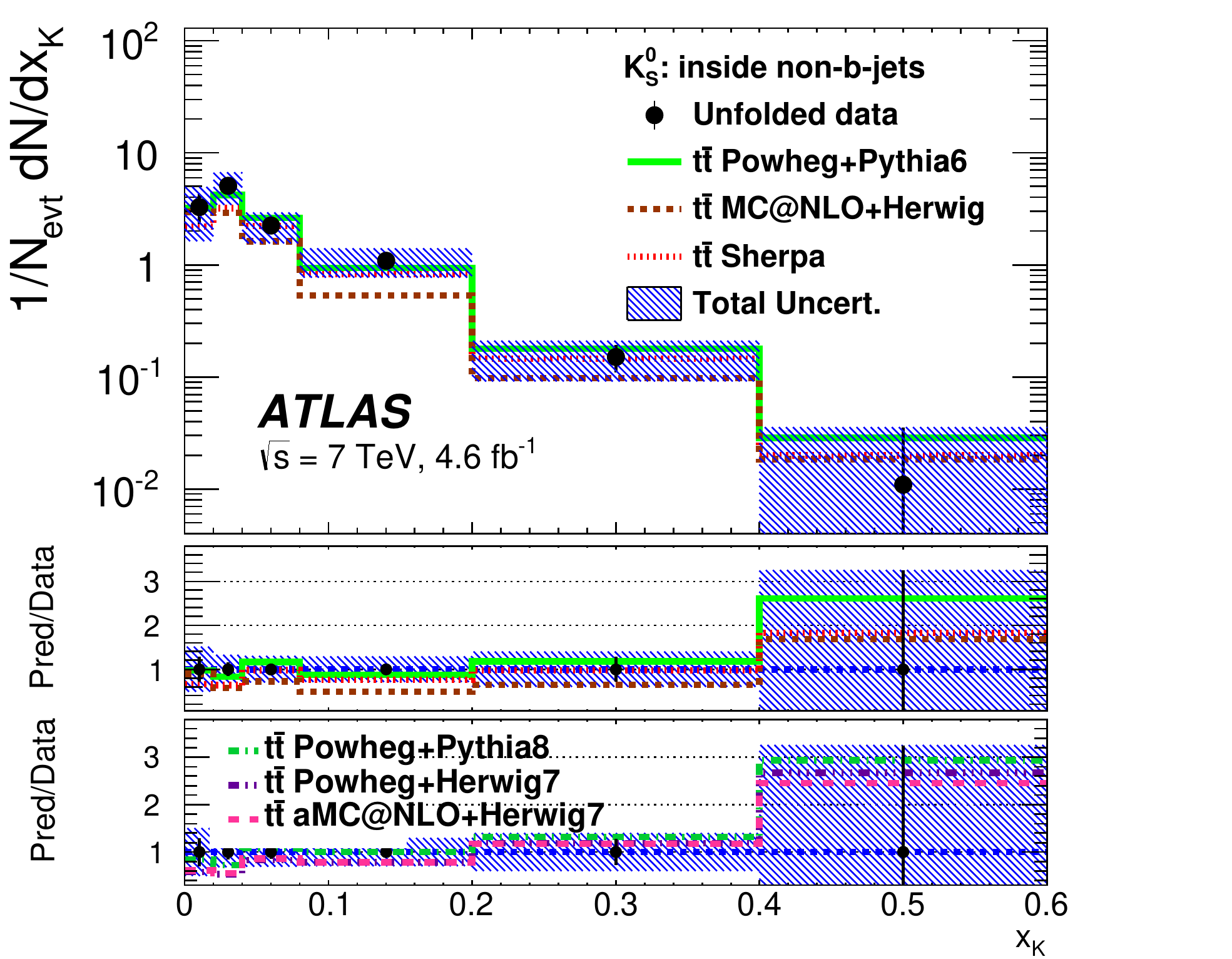}
}
\\
\subfloat[]{
\label{fig:j_k0E_unfold}
\includegraphics[height=6.5cm]{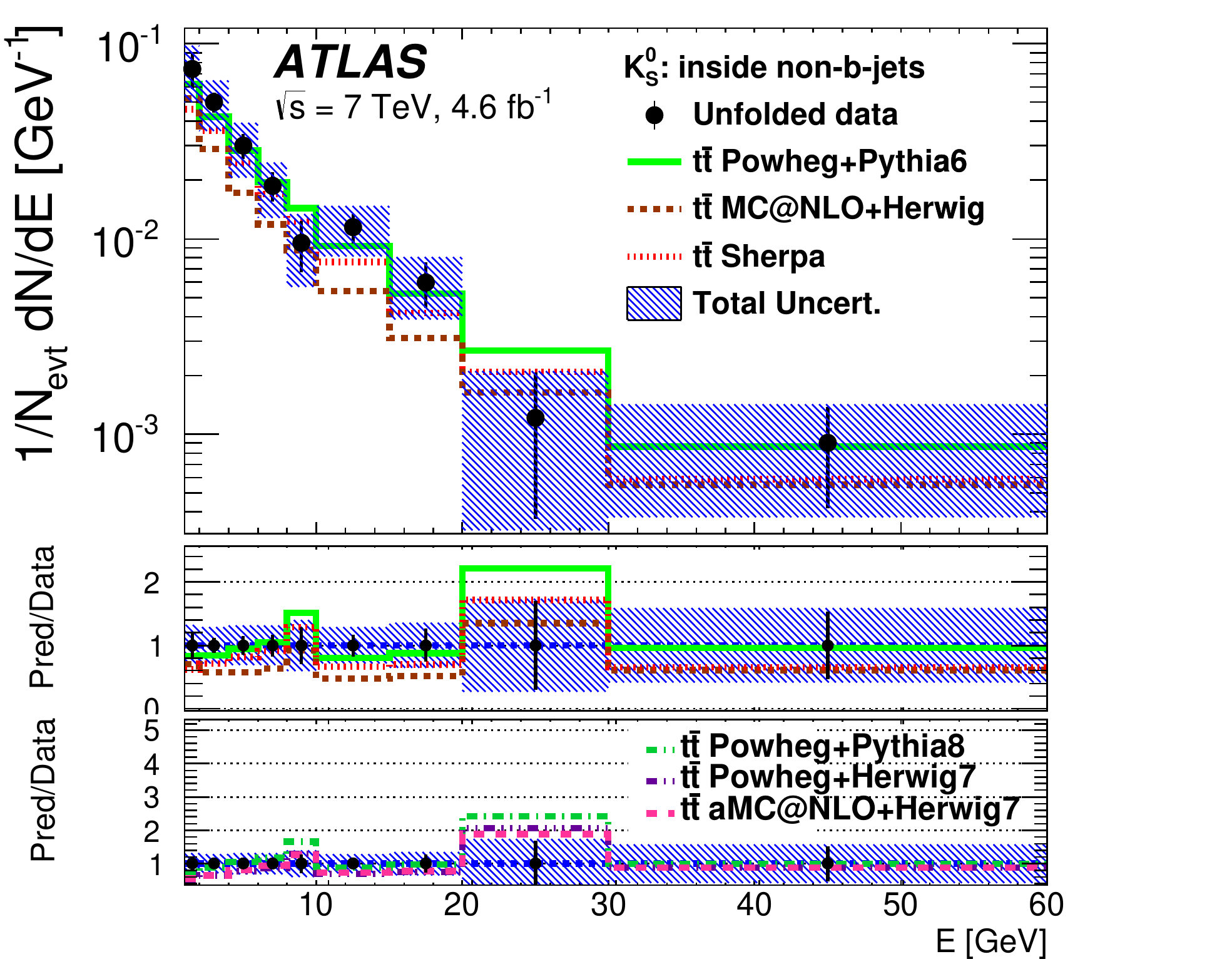}
}
\subfloat[]{
\label{fig:j_k0EtaAbs_unfold}
\includegraphics[height=6.5cm]{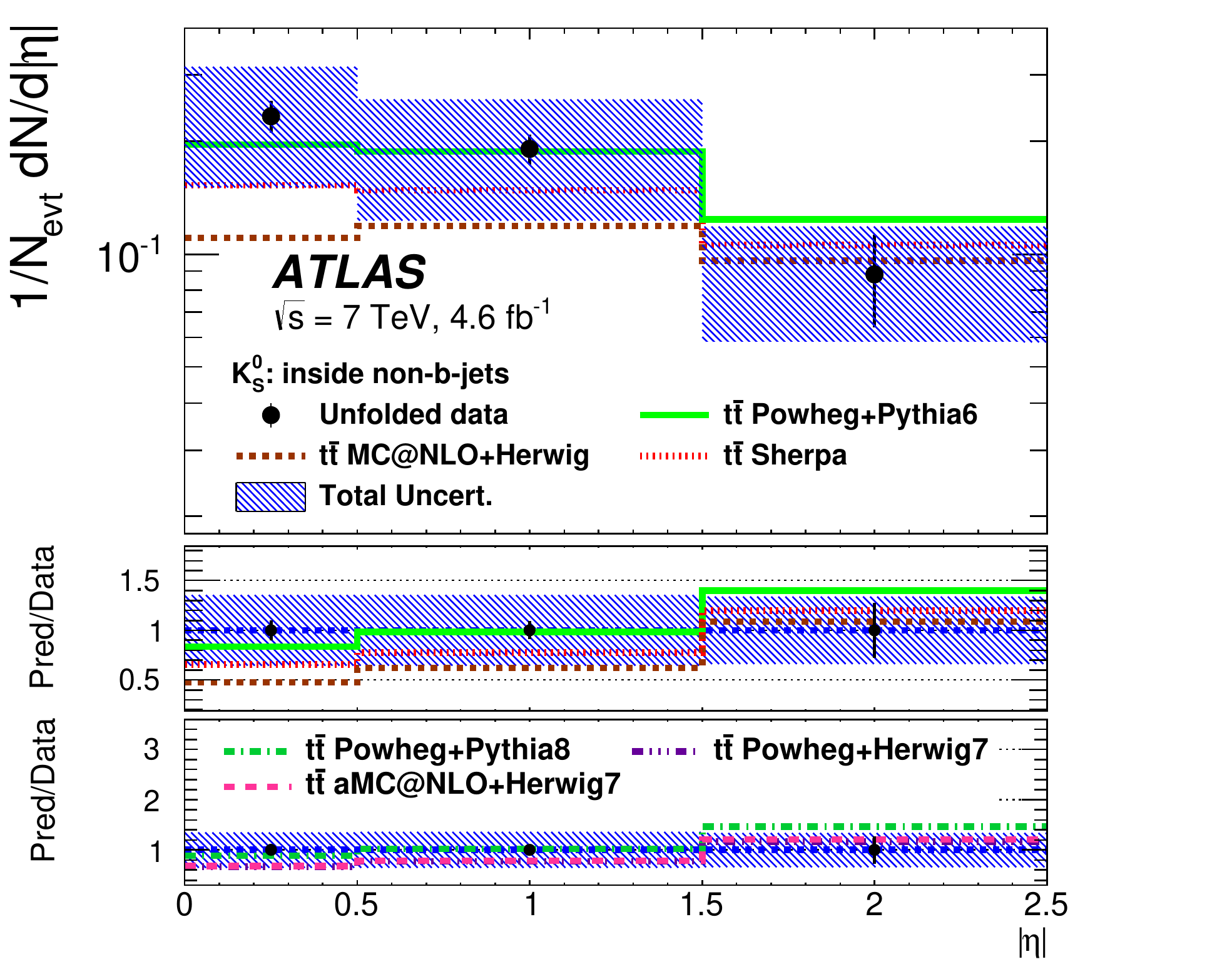}
}
\\
\end{center}
\subfloat[]{
\label{fig:j_k0Multi_unfold}
\includegraphics[height=6.5cm]{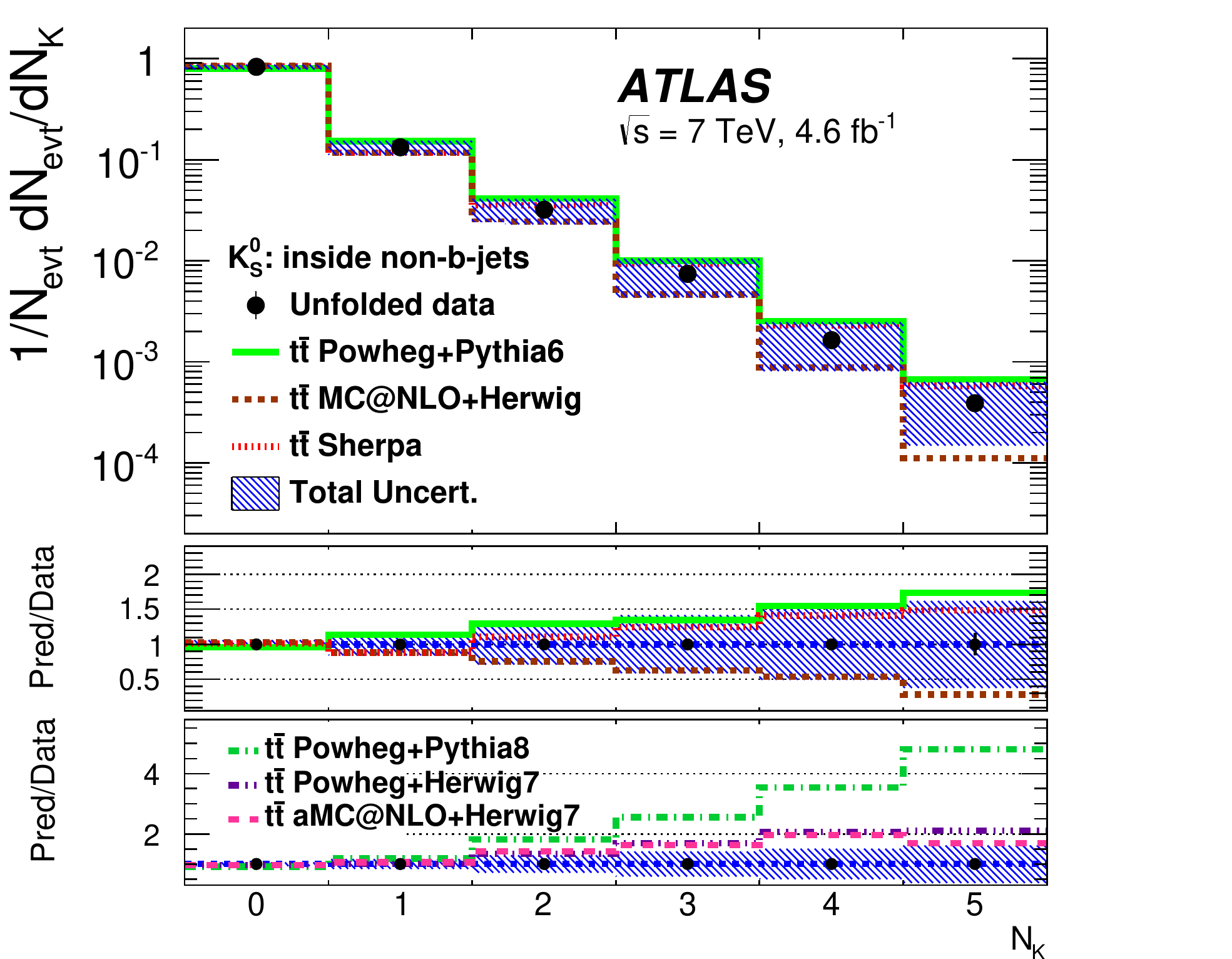}
}
\vspace*{-0.35cm}
\caption{
Kinematic characteristics for $\Kshort$ production inside non-$b$-jets, for corrected data and particle-level MC events simulated with the \textsc{Powheg}+\textsc{Pythia6}, \textsc{MC@NLO}+\textsc{Herwig}, \textsc{Sherpa}, \textsc{Powheg}+\textsc{Pythia8}, \textsc{Powheg}+\textsc{Herwig7} generators. Total uncertainties are represented by the shaded area. Statistical uncertainties for MC samples are negligible in comparison with data.}
\label{fig:j_results_unfold_2padsC}
\end{figure}

\begin{figure}[H]
\begin{center}
\vspace*{-0.65cm}
\hspace*{-0.35cm}
\subfloat[]{
\label{fig:out_k0Pt_unfold}
\includegraphics[height=8.cm]{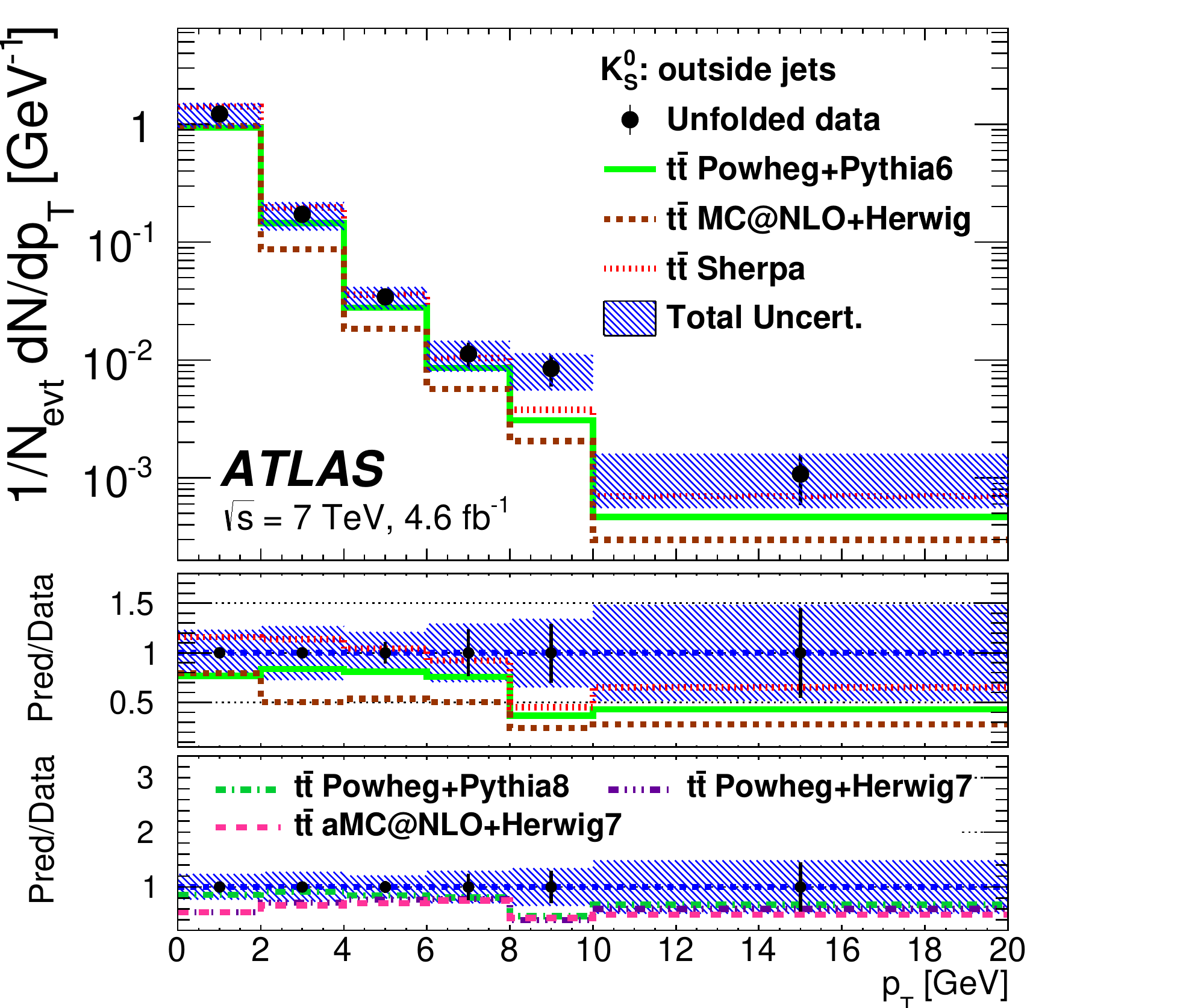}
}
\hspace*{-1.4cm}
\subfloat[]{
\label{fig:out_k0E_unfold}
\includegraphics[height=8.cm]{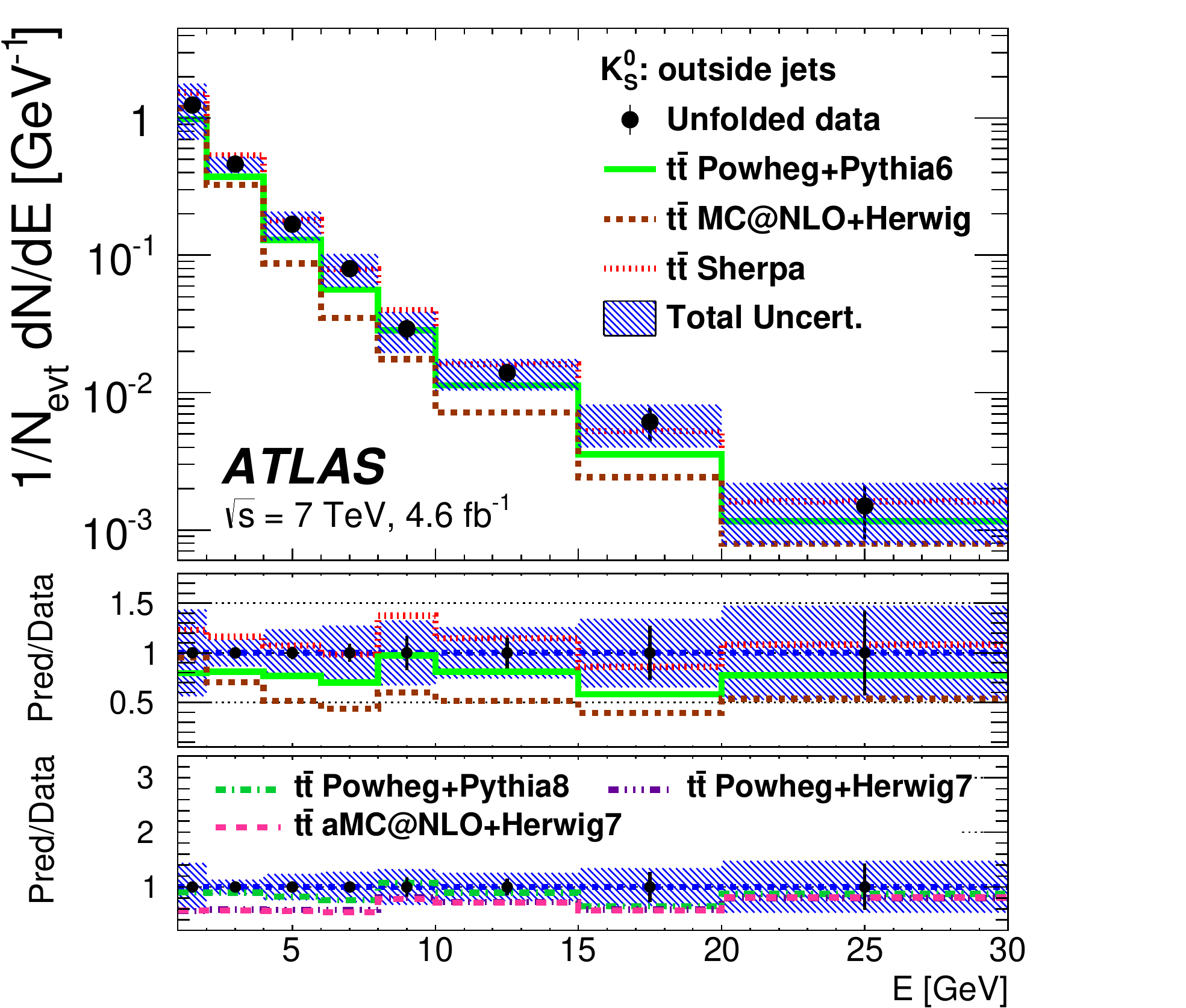}
}
\\
\hspace*{-0.35cm}
\subfloat[]{
\label{fig:out_k0EtaAbs_unfold}
\includegraphics[height=8.cm]{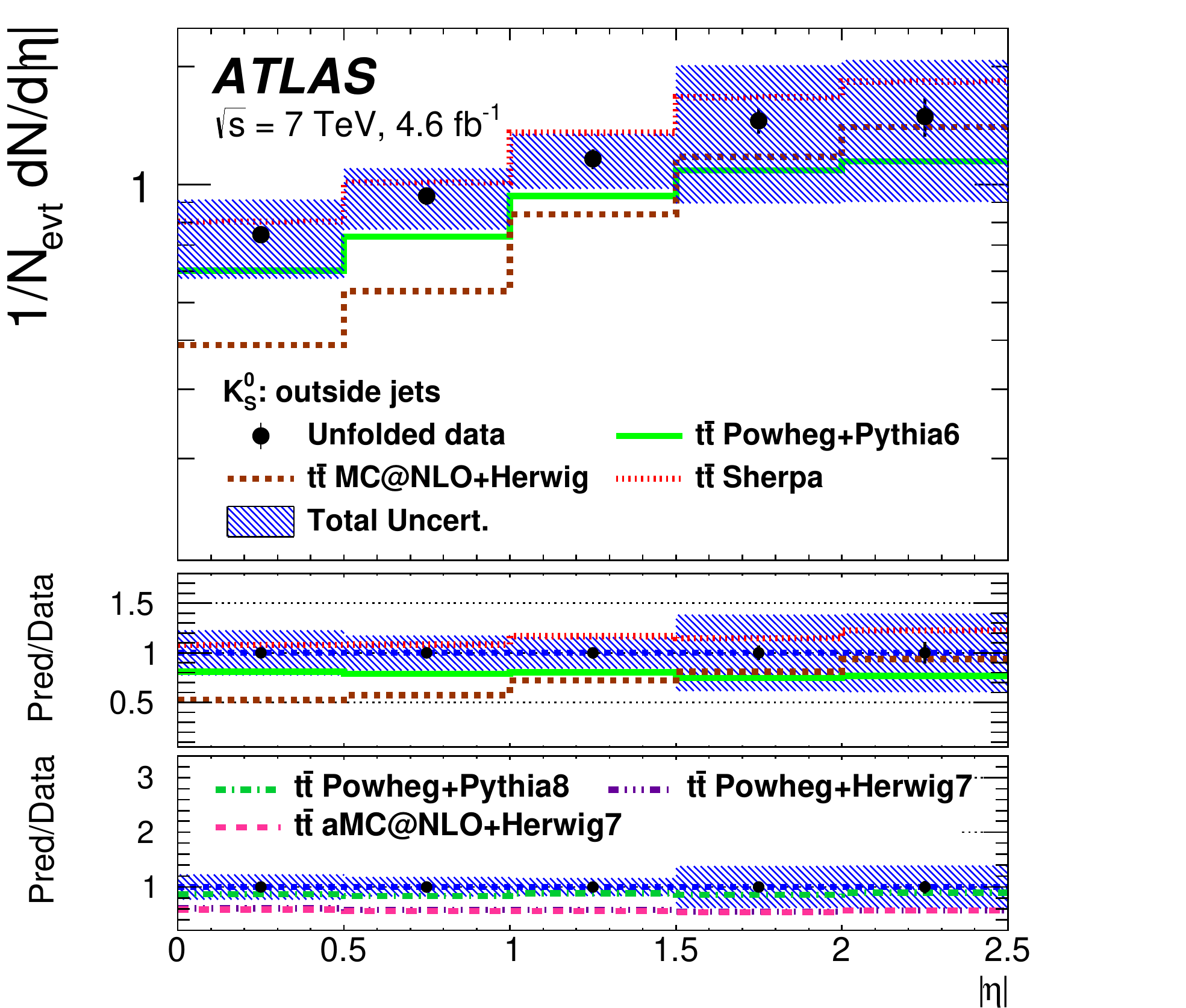}
}
\hspace*{-1.4cm}
\subfloat[]{
\label{fig:out_k0Multi_unfold}
\includegraphics[height=8.cm]{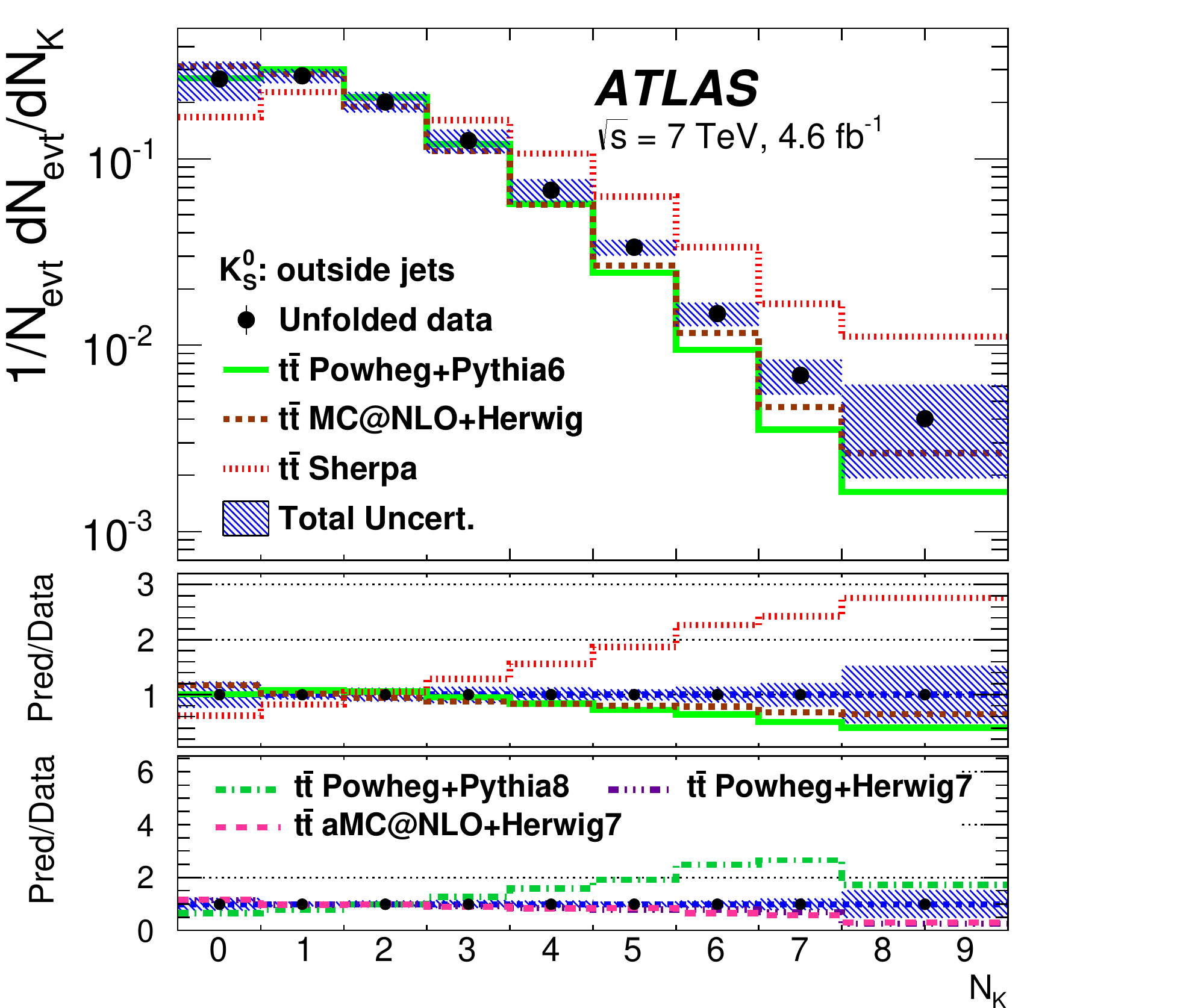}
}
\\
\end{center}
\vspace*{-0.15cm}
\caption{
Kinematic characteristics for $\Kshort$ production not associated with jets,
for corrected data and particle-level MC events simulated with the \textsc{Powheg}+\textsc{Pythia6}, \textsc{MC@NLO}+\textsc{Herwig}, \textsc{Sherpa}, \textsc{Powheg}+\textsc{Pythia8}, \textsc{Powheg}+\textsc{Herwig7} and a\textsc{MC@NLO}+\textsc{Herwig7} generators.
Total uncertainties are represented by the shaded area. Statistical uncertainties for MC samples are negligible in comparison with data.
}
 
\label{fig:out_results_unfold_2padsC}
\end{figure}

\begin{table}[H]
\caption{Values of the $\chi^2$ per degree of freedom and their corresponding $p$-values, for  $\Kshort$ production inside $b$-jets, for \textsc{Powheg}+\textsc{Pythia6}, \textsc{MC@NLO}+\textsc{Herwig}, \textsc{Sherpa}, \textsc{Powheg}+\textsc{Pythia8}, \textsc{Powheg}+\textsc{Herwig7} and a\textsc{MC@NLO}+\textsc{Herwig7} predictions. } \
\label{tab:Chi2K01}
\begin{center}
\begin{tabular}{l r l r  r r r} \hline
&  \multicolumn{6}{ c }{  $\chi^2/n.d.f.~~$ ($p$-value) $\Kshort$ inside $b$-jets} \\ \hline
& \textsc{Pw+Pythia6}  & \textsc{Mc+Herwig} &  \textsc{Sherpa}  &  \textsc{Pw+Pythia8} &  \textsc{Pw+Herwig7} & a\textsc{Mc+Herwig7} \\ \hline
$\pt$     & 0.77 (0.63) & 1.31 (0.23)  & 0.80 (0.61) & 0.59 (0.79) & 0.64 (0.74) & 0.66 (0.73) \\
$E$       & 1.81 (0.06) & 2.54 (0.007) & 1.67 (0.09) & 1.80 (0.06) & 1.54 (0.13) & 1.53 (0.13) \\
$\abseta$ & 0.56 (0.73) & 1.44 (0.21)  & 0.33 (0.90) & 0.61 (0.69) & 0.25 (0.64) & 0.23 (0.65) \\
$x_{K}$   & 0.40 (0.88) & 1.08 (0.37)  & 0.19 (0.98) & 0.25 (0.96) & 0.14 (0.99) & 0.15 (0.99) \\
$N_{K}$   & 1.25 (0.28) & 0.62 (0.68)  & 1.94 (0.08) & 2.15 (0.06) & 3.72 (0.002) & 5.18 (0.00) \\ \hline
\end{tabular}
\end{center}
\end{table}
 
\begin{table}[H]
\caption{Values of the $\chi^2$ per degree of freedom and their corresponding $p$-values, for  $\Kshort$ production inside non-$b$-jets, for \textsc{Powheg}+\textsc{Pythia6}, \textsc{MC@NLO}+\textsc{Herwig}, \textsc{Sherpa}, \textsc{Powheg}+\textsc{Pythia8}, \textsc{Powheg}+\textsc{Herwig7} and a\textsc{MC@NLO}+\textsc{Herwig7} predictions. } \
\label{tab:Chi2K02}
\begin{center}
\begin{tabular}{l r l r r r r} \hline
&   \multicolumn{6}{ c }{ $\chi^2/n.d.f.~~$ ($p$-value) $\Kshort$ inside non-$b$-jets} \\ \hline
& \textsc{Pw+Pythia6}  & \textsc{Mc+Herwig} &  \textsc{Sherpa}  &  \textsc{Pw+Pythia8} &  \textsc{Pw+Herwig7} & a\textsc{Mc+Herwig7} \\ \hline
$\pt$      &  0.42 (0.91) & 1.32 (0.23) & 0.71 (0.69) &  0.60 (0.78) & 0.92 (0.50) & 0.88 (0.53) \\
$E$        &  1.25 (0.26) & 1.58 (0.11) & 1.12 (0.34) &  1.51 (0.14) & 1.44 (0.16) & 1.16 (0.31) \\
$\abseta$  &  0.90 (0.44) & 0.64 (0.59) & 0.43 (0.73) &  1.20 (0.30) & 0.37 (0.78) & 0.43 (0.73) \\
$x_{K}$    &  0.82 (0.55) & 1.51 (0.17) & 0.78 (0.58) &  0.72 (0.64) & 0.98 (0.44) & 0.93 (0.48) \\
$N_{K}$ &  1.50 (0.19) & 1.21 (0.30) & 0.70 (0.62) &  14.29 (0.00) & 2.28 (0.04) & 2.14 (0.06) \\ \hline
\end{tabular}
\end{center}
\end{table}
 
\begin{table}[H]
\caption{Values of the $\chi^2$ per degree of freedom and their corresponding $p$-values, for  $\Kshort$ production outside jets, for \textsc{Powheg}+\textsc{Pythia6}, \textsc{MC@NLO}+\textsc{Herwig}, \textsc{Sherpa}, \textsc{Powheg}+\textsc{Pythia8}, \textsc{Powheg}+\textsc{Herwig7} and a\textsc{MC@NLO}+\textsc{Herwig7} predictions. }  \label{tab:Chi2K03}
\begin{center}
\begin{tabular}{l l l r  r l r} \hline
&  \multicolumn{6}{ c }{$\chi^2/n.d.f.~~$ ($p$-value) $\Kshort$ outside jets} \\ \hline
& \textsc{Pw+Pythia6}  & \textsc{Mc+Herwig} &  \textsc{Sherpa}  &  \textsc{Pw+Pythia8} &  \textsc{Pw+Herwig7} & a\textsc{Mc+Herwig7} \\ \hline
$\pt$     & 0.98 (0.44) & 2.64 (0.015) & 0.93 (0.47) & 0.60 (0.73) & 1.22 (0.29)  & 1.44 (0.19)  \\
$E$       & 1.05 (0.40) & 2.67 (0.004) & 1.14 (0.33) & 0.80 (0.61) & 2.25 (0.02)  & 2.47 (0.008) \\
$\abseta$ & 0.94 (0.45) & 2.30 (0.04)  & 1.36 (0.24) & 0.57 (0.71) & 3.28 (0.006) & 3.55 (0.003) \\
$N_{K}$   & 2.77 (0.005) &  1.85 (0.06) & 30.5 (0.00)& 34.59 (0.00) & 1.41 (0.18)  & 1.76 (0.08)  \\ \hline
\end{tabular}
\end{center}
\end{table}

\subsection{$\Lambda$ unfolded distributions}
 
The same distributions studied for $\Kshort$ production are now presented for the
total $\Lambda$ production. Numerical results are summarised in the Appendix. Comparisons
with MC predictions are made in Figure~\ref{fig:all_lambdaresults_unfold_2padsC}. The
$\Lambda$ production is suppressed relative to $\Kshort$ production as expected. Due
to poor statistics, $\Lambda$ production cannot be divided into classes.
 
The results of a $\chi^2$ test for the comparison between unfolded data and MC
predictions are summarised in Table~\ref{tab:Chi2Lambda}. \textsc{Powheg}+\textsc{Pythia6} and
\textsc{Sherpa} generators give a similar fair description of the data, while
\textsc{MC@NLO}+\textsc{Herwig} is somewhat disfavoured. \textsc{Powheg}+\textsc{Pythia8},
\textsc{Powheg}+\textsc{Herwig7} and a\textsc{MC@NLO}+\textsc{Herwig7} are even more disfavoured.

\begin{figure}[H]
\begin{center}
\vspace*{-0.65cm}
\hspace*{-0.35cm}
\subfloat[]{
\label{fig:all_lambdaPt_unfold}
\includegraphics[height=8.cm]{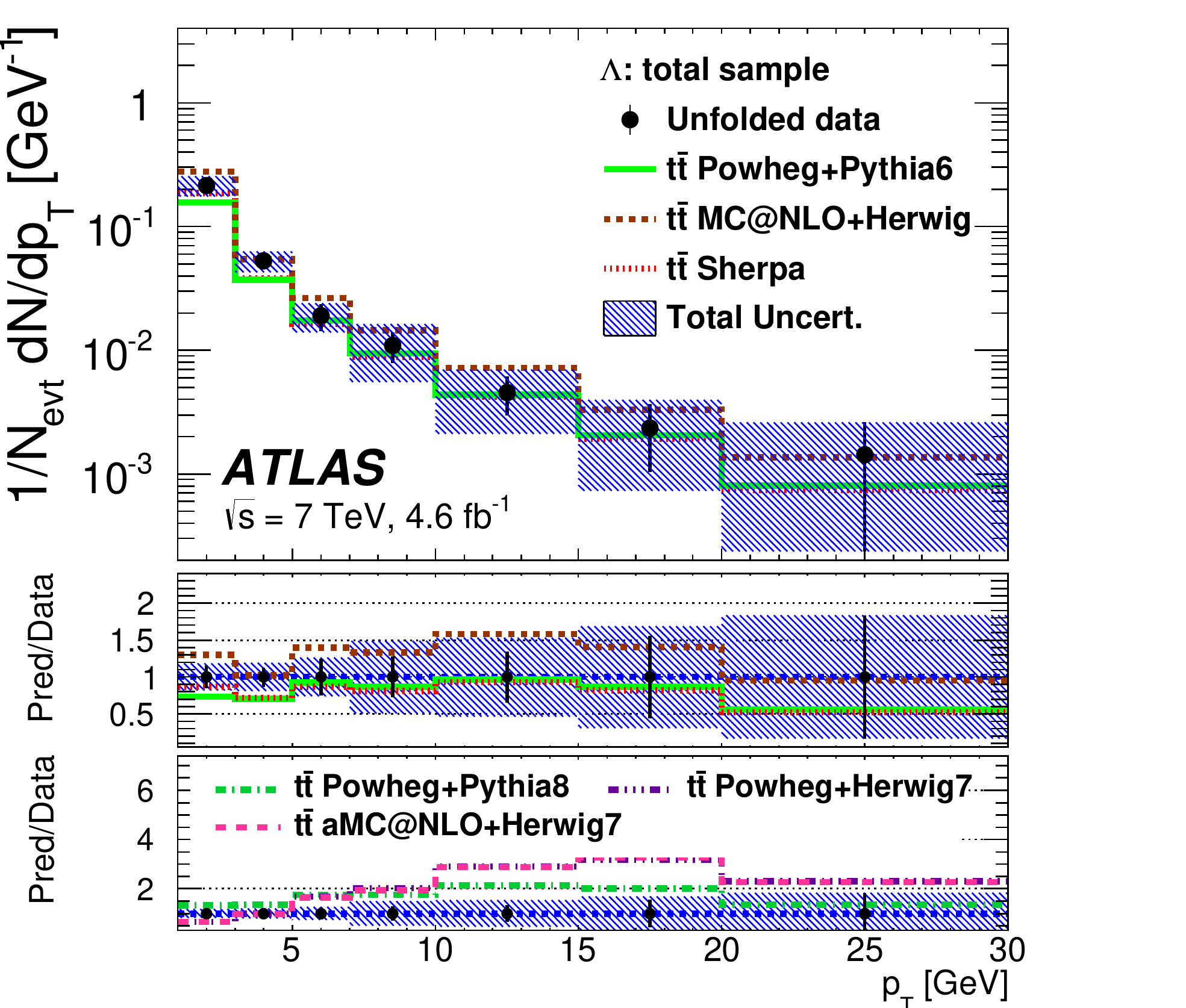}
}
\hspace*{-1.4cm}
\subfloat[]{
\label{fig:all_lambdaE_unfold}
\includegraphics[height=8.cm]{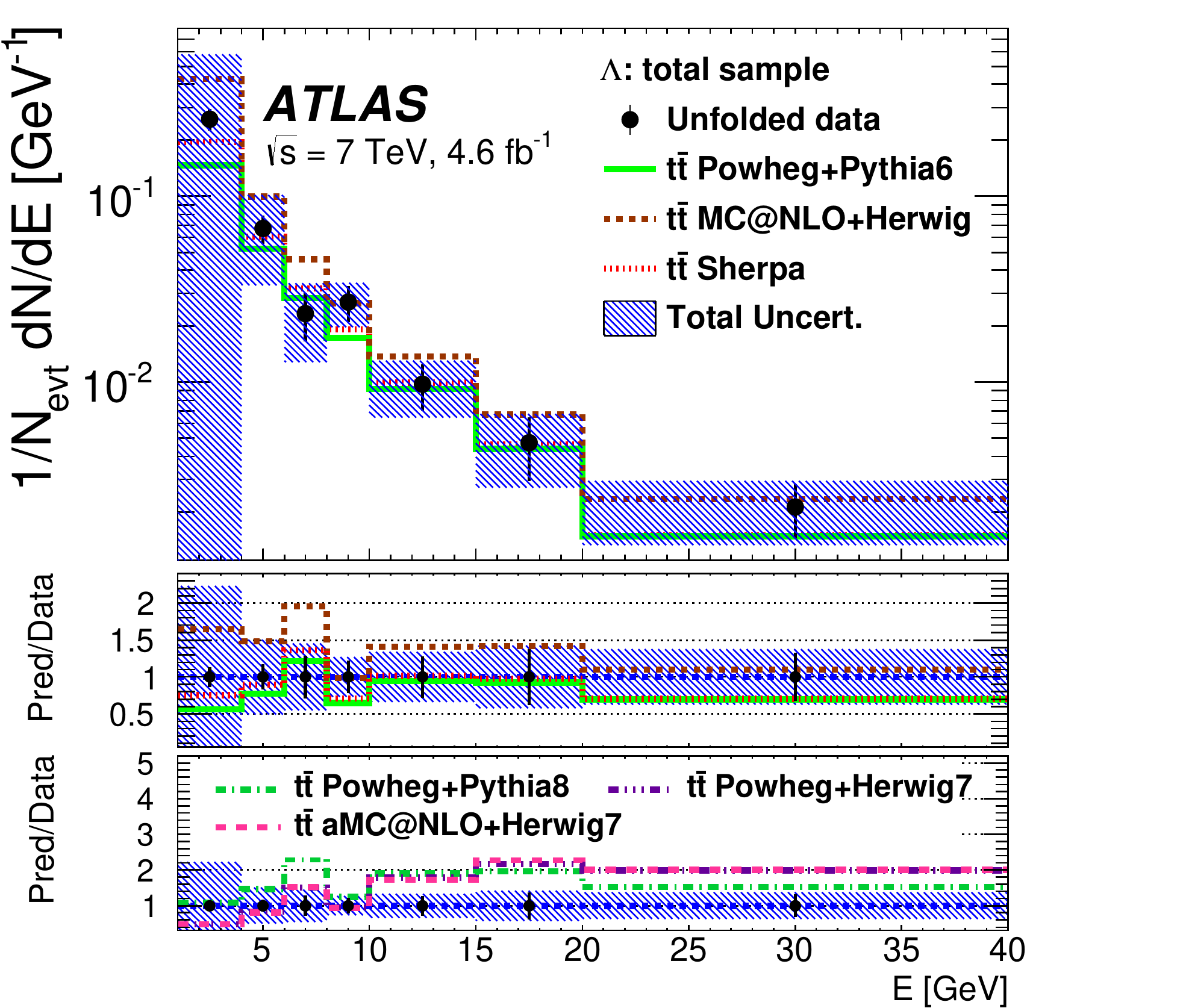}
}
\\
\end{center}
\hspace*{-0.35cm}
\subfloat[]{
\label{fig:all_lambdaEtaAbs_unfold}
\includegraphics[height=8.cm]{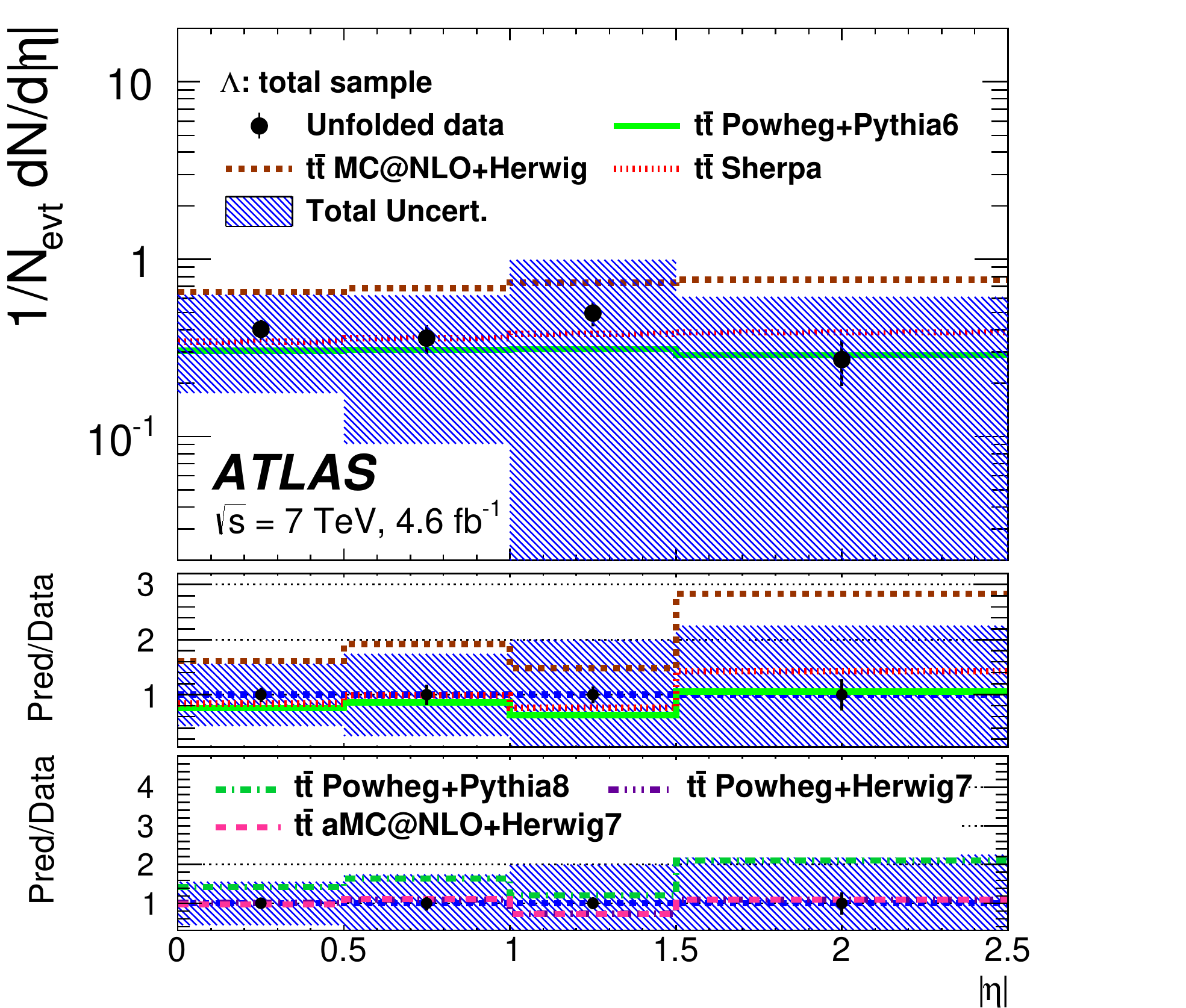}
}
\vspace*{-0.15cm}
\caption{
Kinematic characteristics for the total $\Lambda$ production,
for corrected data and particle-level MC events simulated with the \textsc{Powheg}+\textsc{Pythia6}, \textsc{MC@NLO}+\textsc{Herwig}, \textsc{Sherpa}, \textsc{Powheg}+\textsc{Pythia8}, \textsc{Powheg}+\textsc{Herwig7} and a\textsc{MC@NLO}+\textsc{Herwig7} generators. Total uncertainties are represented by the shaded area. Statistical uncertainties for MC samples are negligible in comparison with data.
}
\label{fig:all_lambdaresults_unfold_2padsC}
\end{figure}
 
\begin{table}[H]
\caption{Values of the $\chi^2$ per degree of freedom and their corresponding $p$-values, for
the total $\Lambda$ sample, for \textsc{Powheg}+\textsc{Pythia6}, \textsc{MC@NLO}+\textsc{Herwig},
\textsc{Sherpa}, \textsc{Powheg}+\textsc{Pythia8}, \textsc{Powheg}+\textsc{Herwig7} and a\textsc{MC@NLO}+\textsc{Herwig7} predictions. }  \label{tab:Chi2Lambda}
\begin{center}
\begin{tabular}{l c c c c c c} \hline
&  \multicolumn{6}{ c }{$\chi^2/n.d.f.~~$ ($p$-value) $\Lambda$ total sample} \\ \hline
& \textsc{Pw+Pythia6}  & \textsc{Mc+Herwig} &  \textsc{Sherpa}  & \textsc{Pw+Pythia8}  & \textsc{Pw+Herwig7} &  a\textsc{Mc+Herwig7} \\ \hline
$\pt$     & 0.50 (0.83) & 0.93 (0.48) & 0.52 (0.82) & 3.87 (0.0003) & 5.56 (0.00)  & 5.36 (0.00) \\
$E$       & 0.89 (0.51) & 1.90 (0.06) & 0.89 (0.51) & 3.97 (0.0002) & 3.18 (0.002) & 3.28 (0.002) \\
$\abseta$ & 0.33 (0.86) & 1.62 (0.17) & 0.25 (0.91) & 0.96 (0.43)   & 0.44 (0.78)  & 0.43 (0.79) \\ \hline
\end{tabular}
\end{center}
\end{table}
 
\subsection{Comparison with other MC generators}
 
Following Ref.~\cite{papershapes}, the sensitivity of the total neutral strange particle production to different underlying-event tunes and colour reconnection schemes was studied. A comparison with the \textsc{Acer+Pythia6} MC generator, with two different underlying-event tunes with and without colour reconnection, is presented in Figure~\ref{fig:all_results_comparison1}. The results of a $\chi^2$ test, similar to that described in the previous subsection, are summarised in Table~\ref{tab:Chi2K0Acer}. The study shows that:
\begin{itemize}
\item Colour reconnection effects are very small, and therefore difficult
to tune with present statistics.
\item \textsc{TuneAPro} is slightly disfavoured relative to the \textsc{Perugia} tune.
\end{itemize}
 
\begin{figure}[H]
\begin{center}
\vspace*{-0.3cm}
\subfloat[]{
\label{fig:all_k0Pt_comparison1}
\includegraphics[height=6.5cm]{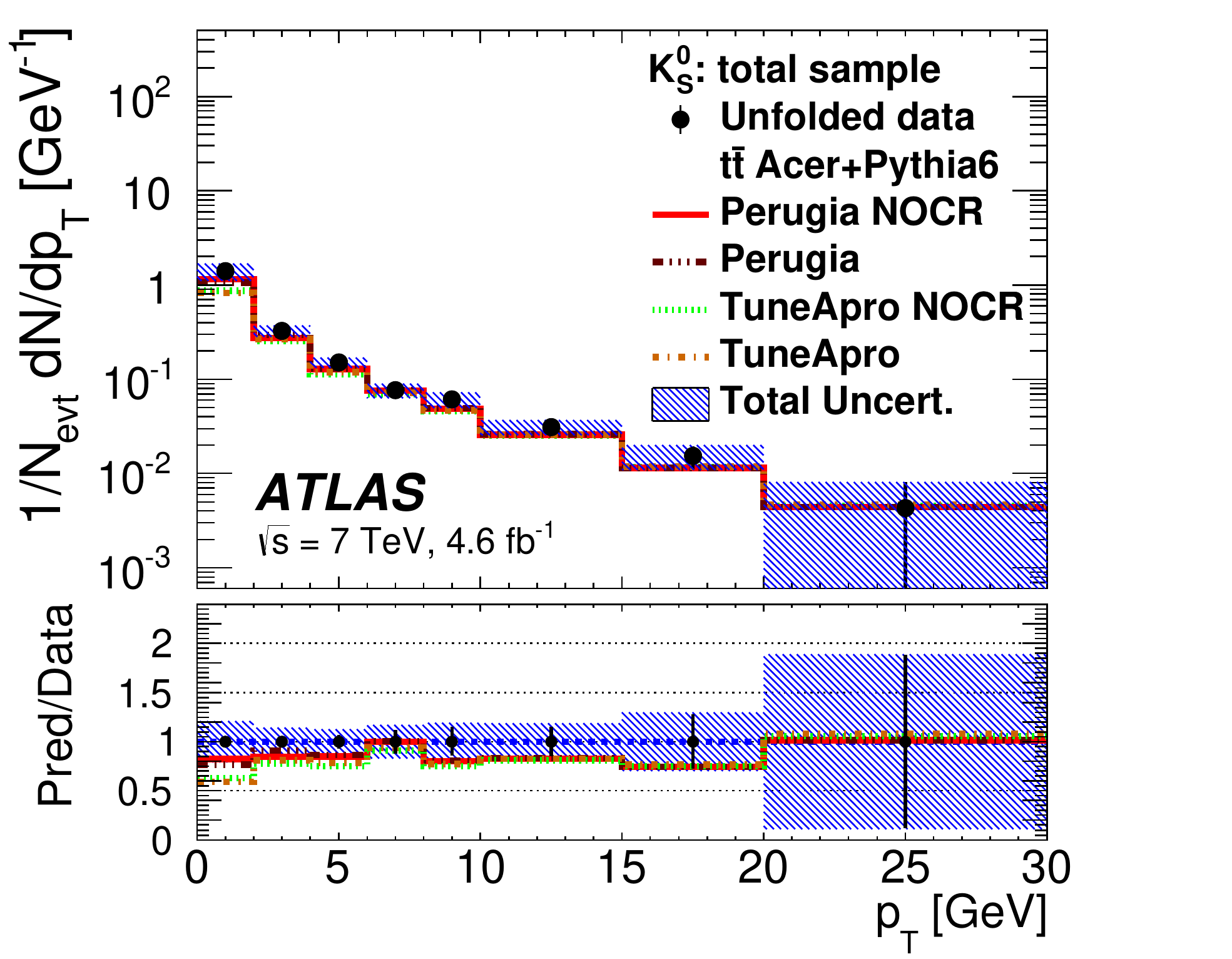}
}
\subfloat[]{
\label{fig:all_k0E_comparison1}
\includegraphics[height=6.5cm]{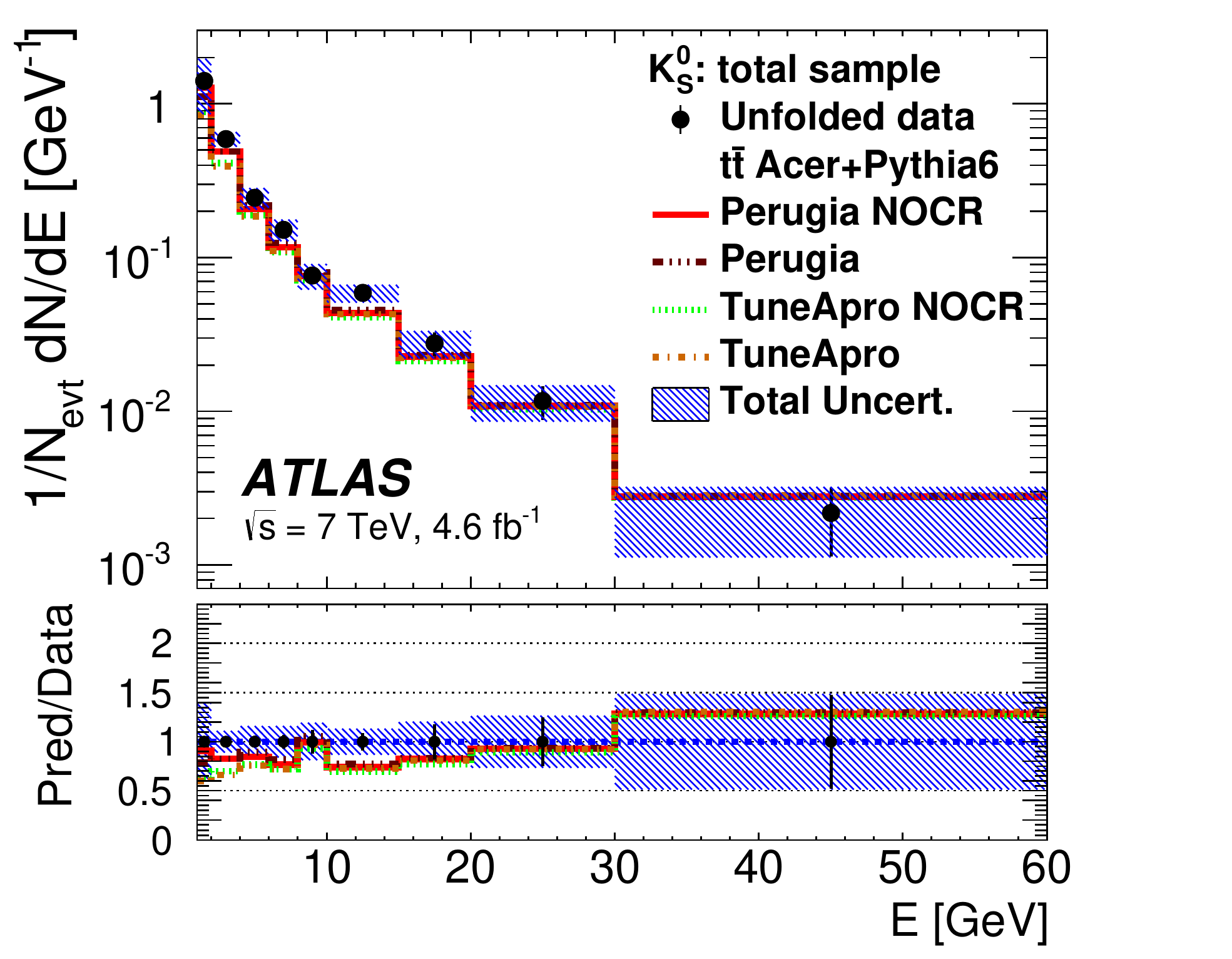}
}
\\
\subfloat[]{
\label{fig:all_k0EtaAbs_comparison1}
\includegraphics[height=6.5cm]{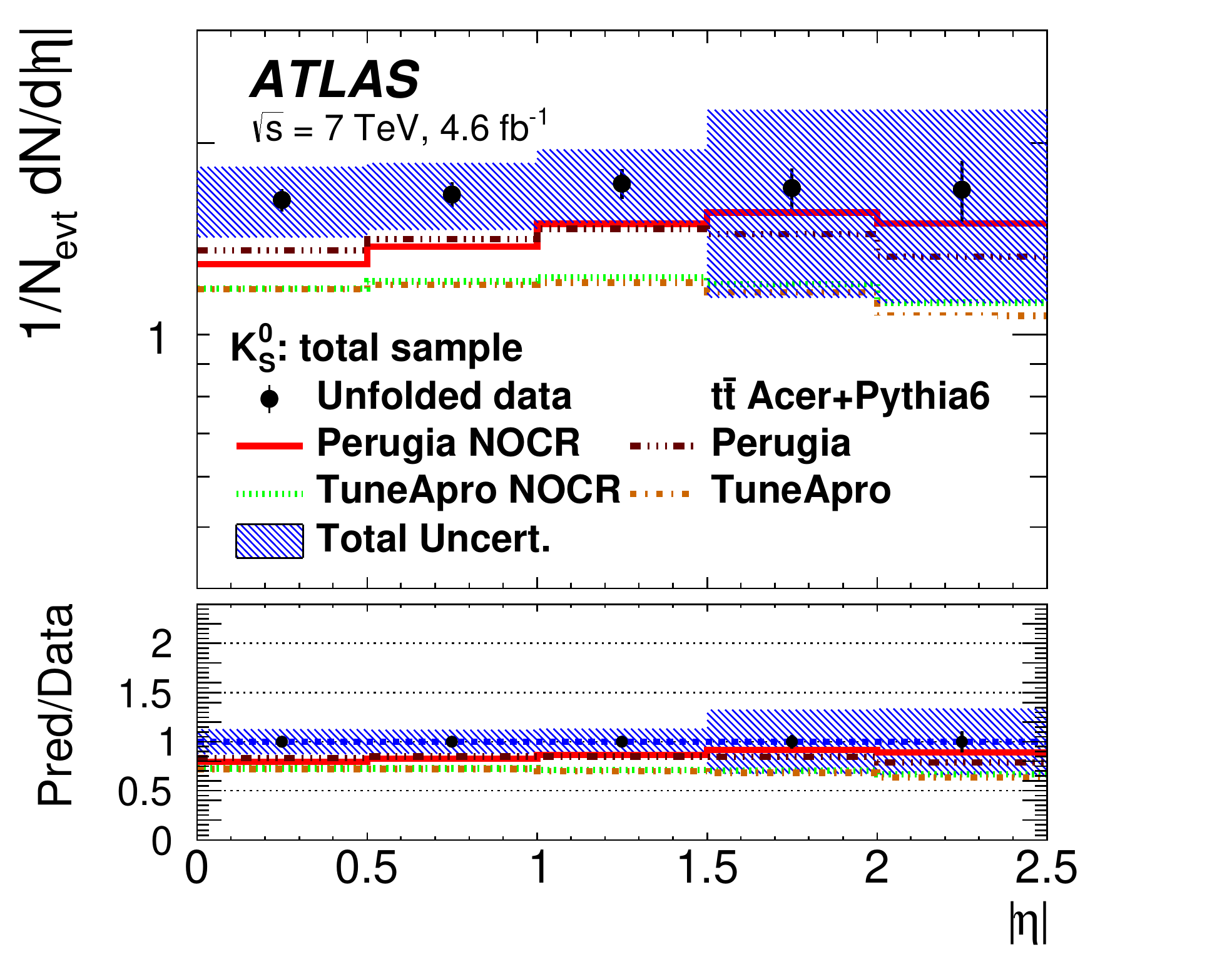}
}
\subfloat[]{
\label{fig:all_k0Multi_comparison1}
\includegraphics[height=6.5cm]{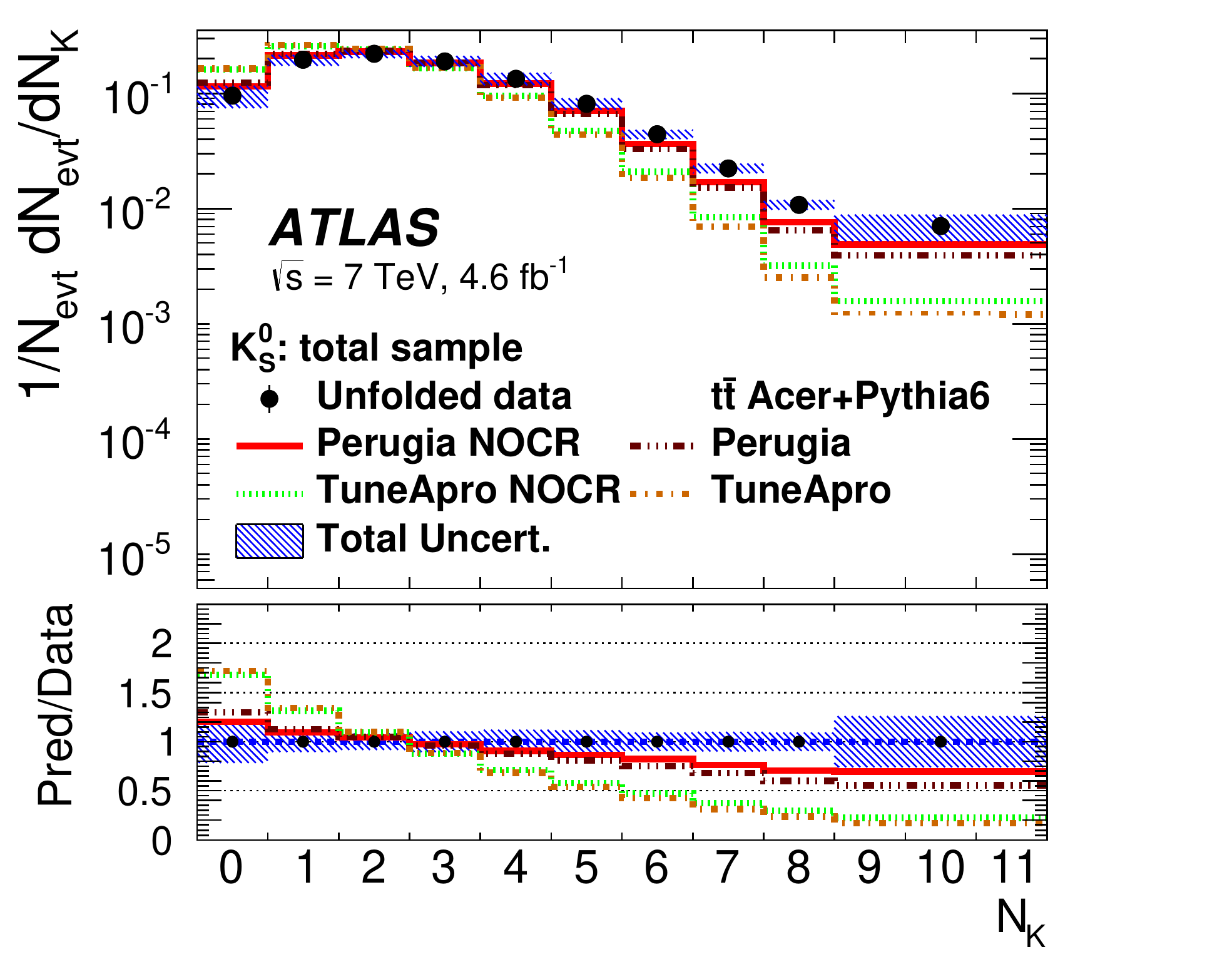}
}
\end{center}
\vspace*{-0.35cm}
\caption{
Kinematic characteristics for the total $\Kshort$ production,
for corrected data and particle-level events from the ACER+\textsc{Pythia6} generator with two different tunes: \textsc{Perugia} and
\textsc{TuneAPro}, with and without colour reconnection (CR). Total uncertainties are represented by the shaded area. Statistical uncertainties for MC samples are negligible in comparison with data.
}
 
\label{fig:all_results_comparison1}
\end{figure}
 
\begin{table}[H]
\caption{Values of the $\chi^2$ per degree of freedom and their corresponding $p$-values, for the total $\Kshort$ production, along with the \textsc{Acermc+Pythia6} predictions with the following tunes: \textsc{Perugia} and \textsc{TuneAPro} (with and without colour reconnection). }  \label{tab:Chi2K0Acer}
\begin{center}
\begin{tabular}{l r c l l} \hline
& \multicolumn{4}{ c }{$\chi^2/n.d.f.~~$ ($p$-value) $\Kshort$ total sample} \\ \hline
&  \textsc{Perugia} &  \textsc{Perugia} (no CR) & \textsc{TuneAPro} & \textsc{TuneAPro} (no CR) \\ \hline
$\pt$     & 0.42 (0.91) & 0.50 (0.85) & 1.12 (0.35)   & 1.36 (0.21) \\
$E$       & 1.54 (0.13) & 1.90 (0.05) & 3.55 (0.0002) & 3.37 (0.0004) \\
$\abseta$ & 1.15 (0.33) & 1.34 (0.24) & 4.18 (0.0008) & 3.90 (0.002) \\  \hline
\end{tabular}
\end{center}
\end{table}

\section{Summary}
\label{sum}
 
Measurements of $\Kshort$ and $\Lambda$ production in $\ttbar$ dileptonic final states are reported. They use a data sample with integrated luminosity of 4.6 $\ifb$ from proton--proton collisions at a centre-of-mass energy of 7~\TeV, collected in 2011 with the ATLAS detector at the LHC. The $\Kshort$ distributions in energy, $\pt$ and $|\eta|$ are presented for three subsamples depending on whether the $\Kshort$  is associated with a jet, with or without a $b$-tag, or is outside any selected jet. The corresponding $\Kshort$ multiplicities are also measured. The small sample size precludes such a detailed analysis for $\Lambda$ production, for which distributions are shown only for the total sample, which includes the sum of $\Lambda$ and $\bar{\Lambda}$. The results are unfolded to the particle level using the neutral strange particle reconstruction efficiencies in each distribution within the kinematic region given by $E>1~\GeV$ and $\abseta<2.5$. The measurements are compared with current MC predictions where the $\ttbar$ matrix elements are calculated at NLO accuracy with \textsc{Powheg}, \textsc{MC@NLO}, \textsc{Sherpa} and a\textsc{MC@NLO}, or at LO with \textsc{Acermc}. Several variations of the MC generators are considered:
\begin{itemize}
\item Fragmentation scheme and UE: \textsc{Pythia6+Perugia2011C}, \textsc{Pythia8+A14}, \textsc{Pythia6+TuneAPro}, \textsc{Herwig+Jimmy}, \textsc{Herwig7+H7UE} or \textsc{Sherpa}.
\item Colour reconnection effects.
\end{itemize}
 
The main conclusions to be drawn from the analysis are the following:
\begin{itemize}
\item Strange baryon production is suppressed relative to strange meson production both inside and outside jets.
\item Neutral strange particle production outside jets is much softer than inside jets, and the pseudorapidity distributions are constant over
a wider region.
\item Neutral strange particle multiplicities outside jets are larger than inside.
\item Current MC models give a fair description of the gross features exhibited
by $\Kshort$ and $\Lambda$ produced inside jets, while the observed yields for neutral strange particles outside jets lie roughly 30\% above the \textsc{Pythia6+Perugia2011C} and \textsc{Herwig+Jimmy} or \textsc{Herwig7+H7UE} MC predictions, with \textsc{Pythia8+A14} falling short of the data by 15--20 \%.
\end{itemize}
 
A better description of the yields for $\Kshort$ and $\Lambda$ outside jets in $\ttbar$ final states would require further tuning of the current MC models, particularly the strangeness suppression mechanisms, and/or more elaborate models for MPI and colour reconnection schemes. For this purpose a Rivet analysis routine and HEPData tables are provided.

\section*{Acknowledgements}
 
% The next lines are included from the .//acknowledgements/Acknowledgements.tex input file
 
We thank CERN for the very successful operation of the LHC, as well as the
support staff from our institutions without whom ATLAS could not be
operated efficiently.
 
We acknowledge the support of ANPCyT, Argentina; YerPhI, Armenia; ARC, Australia; BMWFW and FWF, Austria; ANAS, Azerbaijan; SSTC, Belarus; CNPq and FAPESP, Brazil; NSERC, NRC and CFI, Canada; CERN; CONICYT, Chile; CAS, MOST and NSFC, China; COLCIENCIAS, Colombia; MSMT CR, MPO CR and VSC CR, Czech Republic; DNRF and DNSRC, Denmark; IN2P3-CNRS, CEA-DRF/IRFU, France; SRNSFG, Georgia; BMBF, HGF, and MPG, Germany; GSRT, Greece; RGC, Hong Kong SAR, China; ISF and Benoziyo Center, Israel; INFN, Italy; MEXT and JSPS, Japan; CNRST, Morocco; NWO, Netherlands; RCN, Norway; MNiSW and NCN, Poland; FCT, Portugal; MNE/IFA, Romania; MES of Russia and NRC KI, Russian Federation; JINR; MESTD, Serbia; MSSR, Slovakia; ARRS and MIZ\v{S}, Slovenia; DST/NRF, South Africa; MINECO, Spain; SRC and Wallenberg Foundation, Sweden; SERI, SNSF and Cantons of Bern and Geneva, Switzerland; MOST, Taiwan; TAEK, Turkey; STFC, United Kingdom; DOE and NSF, United States of America. In addition, individual groups and members have received support from BCKDF, CANARIE, CRC and Compute Canada, Canada; COST, ERC, ERDF, Horizon 2020, and Marie Sk{\l}odowska-Curie Actions, European Union; Investissements d' Avenir Labex and Idex, ANR, France; DFG and AvH Foundation, Germany; Herakleitos, Thales and Aristeia programmes co-financed by EU-ESF and the Greek NSRF, Greece; BSF-NSF and GIF, Israel; CERCA Programme Generalitat de Catalunya, Spain; The Royal Society and Leverhulme Trust, United Kingdom.
 
The crucial computing support from all WLCG partners is acknowledged gratefully, in particular from CERN, the ATLAS Tier-1 facilities at TRIUMF (Canada), NDGF (Denmark, Norway, Sweden), CC-IN2P3 (France), KIT/GridKA (Germany), INFN-CNAF (Italy), NL-T1 (Netherlands), PIC (Spain), ASGC (Taiwan), RAL (UK) and BNL (USA), the Tier-2 facilities worldwide and large non-WLCG resource providers. Major contributors of computing resources are listed in Ref.~\cite{ATL-GEN-PUB-2016-002}.
 
% End of text imported from the .//acknowledgements/Acknowledgements.tex input file

\appendix
\part*{Appendix: Numerical results}
\addcontentsline{toc}{part}{Appendix}
 
Numerical values for $\pt$ and $\abseta$ $\Kshort$ unfolded distributions are presented in Tables~\ref{tab:K0PtOutUnfold} to \ref{tab:K0EtaJetUnfold}, along with statistical uncertainties and a breakdown of systematic uncertainties.
 
\vspace{-0.5cm}
\begin{table}[H]
\caption{Transverse momentum distribution unfolded to particle level for $\Kshort$ not associated with jets, and including invisible decays, along with the statistical and systematic uncertainties}\label{tab:K0PtOutUnfold}
\vspace{-0.5cm}
\begin{center}
\begin{tabular}{r l l l l l l l l l} \hline
\multicolumn{1}{l}{\pt [\GeV]} & $1 ~ ~ ~ ~ ~ ~ \mathrm{d}N$ & \multicolumn{1}{l}{Stat.} & \multicolumn{1}{l}{Model} & \multicolumn{1}{l}{Track} & \multicolumn{1}{l}{JES} & \multicolumn{1}{l}{JER} & \multicolumn{1}{l}{Pile-up} & \multicolumn{1}{l}{Fiducial} & \multicolumn{1}{l}{Unfold} \\
&   $\overline{N_{\mathrm{evt}}} ~ \overline{\mathrm{d}\pt}$ &   &    &    &   &   &  &  &  \\  \hline
\T\B (\ \ 0.0, \ \  2.0) & 1.22 & 0.04 & 0.25 & 0.06 & $^{+ 0.02 }_{- 0.02 }$ & 0.003 & 0.10 & 0.03 & 0.002 \\
\T\B (\ \ 2.0, \ \  4.0) & 0.172 & 0.008 & 0.042 & 0.009 & $^{+ 0.002 }_{- 0.002 }$ & 0.001 & 0.014 & 0.01 &  $<10^{-3}$ \\
\T\B (\ \ 4.0, \ \  6.0) & 0.034 & 0.004 & 0.005 & 0.002 & $^{+ 0.0008 }_{- 0.0006 }$ & $<10^{-3}$ & 0.003 & 0.002 &  $<10^{-3}$ \\
\T\B (\ \ 6.0, \ \  8.0) & 0.011 & 0.003 & 0.001 & 0.0006 & $^{+ 0.0008 }_{- 0.0006 }$ & $<10^{-3}$ & 0.001 & 0.001 &  $<10^{-3}$ \\
\T\B (\ \ 8.0, 10.0) & 0.0084 & 0.0025 & 0.0007 & 0.0004 & $^{+ 0.0006 }_{- 0.0004 }$ & $<10^{-3}$ & 0.0007 & 0.0006 &  $<10^{-3}$ \\
\T\B (10.0, 20.0) & 0.0011 & 0.0005 & 0.0001 & 0.0001 & $^{+ 0.0001 }_{- 0.0001 }$ & $<10^{-3}$ & 0.0001 & 0.0001 &  $<10^{-4}$ \\ \hline

\end{tabular}
\end{center}
\end{table}
 
\vspace{-0.5cm}
\begin{table}[H]
\caption{Transverse momentum distribution unfolded to the particle level for $\Kshort$ associated with $b$-jets, and including invisible decays, along with the statistical and systematic uncertainties.}\label{tab:K0PtBJetUnfold}
\vspace{-0.5cm}
\begin{center}
\begin{tabular}{r l l l l l l l l l} \hline
\multicolumn{1}{l}{\pt [\GeV]} &  $1 ~ ~ ~ ~ ~ ~ \mathrm{d}N$ & \multicolumn{1}{l}{Stat.} & \multicolumn{1}{l}{Model} & \multicolumn{1}{l}{Track} & \multicolumn{1}{l}{JES} & \multicolumn{1}{l}{JER} & \multicolumn{1}{l}{Pile-up} & \multicolumn{1}{l}{Fiducial} & \multicolumn{1}{l}{Unfold} \\ 
&   $\overline{N_{\mathrm{evt}}} ~ \overline{\mathrm{d}\pt}$ &   &    &    &   &   &  &  &  \\  \hline
\T\B (\ \ 0.0, \ \  2.0) & 0.091 & 0.017 & 0.003 & 0.005 & $^{+ 0.002 }_{- 0.002 }$ & 0.001 & $<10^{-3}$ & 0.001 & 0.001 \\
\T\B (\ \  2.0, \ \  4.0) & 0.086 & 0.011 & 0.003 & 0.004 & $^{+ 0.003 }_{- 0.003 }$ & 0.001 & $<10^{-3}$ & 0.001 & 0.001 \\
\T\B (\ \ 4.0,  \ \ 6.0) & 0.075 & 0.010 & 0.002 & 0.004 & $^{+ 0.002 }_{- 0.002 }$ & 0.001 & $<10^{-3}$ & 0.001 & 0.001 \\
\T\B (\ \ 6.0, \ \  8.0) & 0.045 & 0.009 & 0.002 & 0.002 & $^{+ 0.001 }_{- 0.001 }$ & 0.001 & $<10^{-3}$ & 0.001 & 0.001 \\
\T\B (\ \ 8.0, 10.0) & 0.040 & 0.008 & 0.002 & 0.002 & $^{+ 0.001 }_{-  <10^{-3} }$ & $<10^{-3}$ & $<10^{-3}$ & 0.001 & 0.001 \\
\T\B (10.0, 15.0) & 0.022 & 0.005 & 0.002 & 0.001 & $^{+ 0.001 }_{-  <10^{-3} }$ & $<10^{-3}$ & $<10^{-3}$ &  $<10^{-3}$ & $<10^{-3}$ \\
\T\B (15.0, 20.0) & 0.0155 & 0.005 & 0.001 & 0.0008 & $^{+  <10^{-3} }_{-  <10^{-3} }$ & $<10^{-3}$ & $<10^{-3}$ & $<10^{-3}$ & $<10^{-3}$ \\
\T\B (20.0, 30.0) & 0.0053 & 0.004 & 0.0003 & 0.0003 & $^{+ <10^{-3} }_{-  <10^{-3} }$ & $<10^{-4}$ & $<10^{-3}$ & $<10^{-4}$ & $<10^{-3}$ \\ \hline
\end{tabular}
\end{center}
\end{table}
 
\vspace{-0.5cm}
\begin{table}[H]
\caption{Transverse momentum distribution unfolded to the particle level for $\Kshort$ associated with non-$b$-jets, and including invisible decays, along with the statistical and systematic uncertainties.}\label{tab:K0PtJetUnfold}
\vspace{-0.5cm}
\begin{center}
\begin{tabular}{r l l l l l l l l l} \hline
\multicolumn{1}{l}{\pt [\GeV]} & $1 ~ ~ ~ ~ ~ ~ \mathrm{d}N$ & \multicolumn{1}{l}{Stat.} & \multicolumn{1}{l}{Model} & \multicolumn{1}{l}{Track} & \multicolumn{1}{l}{JES} & \multicolumn{1}{l}{JER} & \multicolumn{1}{l}{Pile-up} & \multicolumn{1}{l}{Fiducial} & \multicolumn{1}{l}{Unfold} \\
&   $\overline{N_{\mathrm{evt}}} ~ \overline{\mathrm{d}\pt}$ &   &    &    &   &   &  &  &  \\  \hline
\T\B (\ \ 0.0, \ \  2.0) & 0.080 & 0.012 & 0.003 & 0.003 & $^{+ 0.004 }_{- 0.005 }$ & 0.002 & $<10^{-3}$ & 0.006 & 0.002 \\
\T\B (\ \ 2.0, \ \  4.0) & 0.049 & 0.006 & 0.013 & 0.002 & $^{+ 0.002 }_{- 0.002 }$ & 0.001 & $<10^{-3}$ & 0.004 & 0.001 \\
\T\B (\ \ 4.0,  \ \ 6.0) & 0.035 & 0.005 & 0.011 & 0.002 & $^{+ 0.001 }_{- 0.001 }$ & 0.001 & $<10^{-3}$ & 0.003 & 0.001 \\
\T\B (\ \ 6.0, \ \  8.0) & 0.016 & 0.003 & 0.004 & 0.001 & $^{+ 0.001 }_{- 0.0005 }$ & 0.001 & $<10^{-3}$ & $<10^{-3}$ & $<10^{-3}$ \\
\T\B (\ \  8.0, 10.0) & 0.010 & 0.003 & 0.004 & 0.0004 & $^{+ 0.0003 }_{- 0.0002 }$ & $<10^{-3}$ & $<10^{-3}$ & $<10^{-3}$ & $<10^{-3}$ \\
\T\B (10.0, 15.0) & 0.0062 & 0.002 & 0.0013 & 0.0003 & $^{+ 0.0001 }_{- 0.0001 }$ & $<10^{-4}$ & $<10^{-3}$ &$<10^{-3}$ & $<10^{-3}$ \\
\T\B (15.0, 20.0) & 0.0017 & 0.001 & 0.0004 & 0.0001 & $^{+ <10^{-4} }_{- <10^{-4} }$ & $<10^{-4}$ & $<10^{-3}$ & $<10^{-4}$ & $<10^{-4}$ \\
\T\B (20.0, 30.0) & 0.0009 & 0.001 & 0.0002 & $<10^{-4}$ & $^{+ <10^{-4} }_{- <10^{-4} }$ & $<10^{-4}$ & $<10^{-3}$ & $<10^{-4}$ & $<10^{-4}$ \\ \hline
\end{tabular}
\end{center}
\end{table}

\begin{table}[H]
\caption{Pseudorapidity distribution unfolded to the particle level for $\Kshort$ not associated with jets, and including invisible decays, along with the statistical and systematic uncertainties.}\label{tab:K0EtaOutUnfold}
\vspace{-0.5cm}
\begin{center}
\begin{tabular}{r l l l l l l l l r} \hline
\multicolumn{1}{c}{$\abseta$} & $1 ~ ~ ~ ~ ~ ~ \mathrm{d}N$ & \multicolumn{1}{l}{Stat.} & \multicolumn{1}{l}{Model} & \multicolumn{1}{l}{Track} & \multicolumn{1}{l}{JES} & \multicolumn{1}{l}{JER} & \multicolumn{1}{l}{Pile-up} & \multicolumn{1}{l}{Fiducial} & \multicolumn{1}{l}{Unfold} \\ 
&   $\overline{N_{\mathrm{evt}}} ~ \overline{\mathrm{d}\abseta}$ &   &    &    &   &   &  &  &  \\  \hline
\T\B (0.0, 0.5) & 0.744 & 0.036 & 0.141 & 0.038 & $^{+ 0.007 }_{- 0.007 }$ & 0.004 & 0.062 & 0.046 & $<10^{-3}$ \\
\T\B (0.5, 1.0) & 0.933 & 0.048 & 0.120 & 0.048 & $^{+ 0.007 }_{- 0.009 }$ & 0.008 & 0.077 & 0.047 & $<10^{-3}$ \\
\T\B (1.0, 1.5) & 1.161 & 0.068 & 0.122 & 0.059 & $^{+ 0.010 }_{- 0.008 }$ & 0.004 & 0.097 & 0.041 & $<10^{-3}$ \\
\T\B (1.5, 2.0) & 1.454 & 0.108 & 0.529 & 0.074 & $^{+ 0.021 }_{- 0.018 }$ & 0.021 & 0.1200& 0.030 & $<10^{-3}$ \\
\T\B (2.0, 2.5) & 1.490 & 0.167 & 0.542 & 0.076 & $^{+ 0.022 }_{- 0.019 }$ & 0.022 & 0.125 & 0.031 & $<10^{-3}$ \\  \hline
\end{tabular}
\end{center}
\end{table}
 
\begin{table}[H]
\caption{Pseudorapidity distribution unfolded to the particle level for $\Kshort$ associated with $b$-jets, and including invisible decays, along with the statistical and systematic uncertainties.}\label{tab:K0EtaBJettUnfold}
\vspace{-0.5cm}
\begin{center}
\begin{tabular}{r l l l l l l l l r} \hline
\multicolumn{1}{c}{$\abseta$} & $1 ~ ~ ~ ~ ~ ~ \mathrm{d}N$ & \multicolumn{1}{l}{Stat.} & \multicolumn{1}{l}{Model} & \multicolumn{1}{l}{Track} & \multicolumn{1}{l}{JES} & \multicolumn{1}{l}{JER} & \multicolumn{1}{l}{Pile-up} & \multicolumn{1}{l}{Fiducial} & \multicolumn{1}{l}{Unfold} \\ 
&   $\overline{N_{\mathrm{evt}}} ~ \overline{\mathrm{d}\abseta}$ &   &    &    &   &   &  &  &  \\  \hline
\T\B (0.0, 0.5) & 0.592 & 0.055 & 0.015 & 0.029 & $^{+ 0.012 }_{- 0.012 }$ & 0.009 & $<10^{-3}$ & 0.032 & $<10^{-3}$ \\
\T\B (0.5, 1.0) & 0.490 & 0.059 & 0.011 & 0.024 & $^{+ 0.009 }_{- 0.007 }$ & 0.007 & $<10^{-3}$ & 0.019 & $<10^{-3}$ \\
\T\B (1.0, 1.5) & 0.363 & 0.062 & 0.015 & 0.018 & $^{+ 0.010 }_{- 0.009 }$ & 0.007 & $<10^{-3}$ & 0.002 & $<10^{-3}$ \\
\T\B (1.5, 2.0) & 0.225 & 0.059 & 0.049 & 0.011 & $^{+ 0.005 }_{- 0.008 }$ & 0.002 & $<10^{-3}$ & 0.016 & $<10^{-3}$ \\
\T\B (2.0, 2.5) & 0.085 & 0.051 & 0.018 & 0.004 & $^{+ 0.002 }_{- 0.003 }$ & $<10^{-3}$ & $<10^{-3}$ & 0.006 & $<10^{-3}$ \\  \hline
\end{tabular}
\end{center}
\end{table}
 
\begin{table}[H]
\caption{Pseudorapidity distribution unfolded to the particle level for $\Kshort$ associated with non-$b$-jets, and including invisible decays, along with the statistical and systematic uncertainties.}\label{tab:K0EtaJetUnfold}
\vspace{-0.5cm}
\begin{center}
\begin{tabular}{r l l l l l l l l r} \hline
\multicolumn{1}{c}{$\abseta$} & $ 1 ~ ~ ~ ~ ~ ~ \mathrm{d}N$ & \multicolumn{1}{l}{Stat.} & \multicolumn{1}{l}{Model} & \multicolumn{1}{l}{Track} & \multicolumn{1}{l}{JES} & \multicolumn{1}{l}{JER} & \multicolumn{1}{l}{Pile-up} & \multicolumn{1}{l}{Fiducial} & \multicolumn{1}{l}{Unfold} \\ 
&   $\overline{N_{\mathrm{evt}}} ~ \overline{\mathrm{d}\abseta}$ &   &    &    &   &   &  &  &  \\  \hline
\T\B (0.0, 0.5) & 0.233 & 0.023 & 0.077 & 0.010 & $^{+ 0.007 }_{- 0.009 }$ & 0.002 & $<10^{-3}$ & 0.013 & $<10^{-3}$ \\
\T\B (0.5, 1.5) & 0.191 & 0.017 & 0.063 & 0.008 & $^{+ 0.006 }_{- 0.006 }$ & 0.001 & $<10^{-3}$ & 0.014 & $<10^{-3}$ \\
\T\B (1.5, 2.5) & 0.088 & 0.024 & 0.016 & 0.004 & $^{+ 0.003 }_{- 0.004 }$ & 0.003 & $<10^{-3}$ & 0.005 & $<10^{-3}$ \\  \hline
\end{tabular}
\end{center}
\end{table}
 
Numerical values for the $\pt$ and $|\eta|$ $\Lambda$ unfolded distributions are presented in Tables~\ref{tab:lambdaPtJetUnfold} and \ref{tab:lambdaEtaJetUnfold}, along with statistical uncertainties and a breakdown of systematic uncertainties.

\begin{table}[H]
\caption{Transverse momentum distribution unfolded to the particle level for the $\Lambda$ total sample, and including invisible decays, along with the statistical and systematic uncertainties.}\label{tab:lambdaPtJetUnfold}
\vspace{-0.5cm}
\begin{center}
\begin{tabular}{r l l l l l l l l r} \hline
\multicolumn{1}{l}{\pt [\GeV]} & $1 ~ ~ ~ ~ ~ ~ \mathrm{d}N$ & \multicolumn{1}{l}{Stat.} & \multicolumn{1}{l}{Model} & \multicolumn{1}{l}{Track} & \multicolumn{1}{l}{JES} & \multicolumn{1}{l}{JER} & \multicolumn{1}{l}{Pile-up} & \multicolumn{1}{l}{Fiducial} & \multicolumn{1}{l}{Unfold} \\ 
&   $\overline{N_{\mathrm{evt}}} ~ \overline{\mathrm{d}\pt}$ &   &    &    &   &   &  &  &  \\  \hline
\T\B (\ \ 1.0,  \ \ 3.0) & 0.215 & 0.034 & 0.114 & 0.011 & $^{+ 0.003 }_{- 0.004 }$ & 0.002 & 0.017 & 0.004 & $<10^{-3}$ \\
\T\B (\ \ 3.0, \ \  5.0) & 0.053 & 0.007 & 0.004 & 0.003 & $^{+ 0.001 }_{- 0.001 }$ & 0.001 & 0.004 & 0.001 & $<10^{-3}$ \\
\T\B (\ \ 5.0, \ \  7.0) & 0.019 & 0.005 & 0.002 & 0.001 & $^{+ <10^{-3} }_{- <10^{-3} }$ & 0.001 & 0.002 & 0.001 & $<10^{-3}$ \\
\T\B (\ \ 7.0, 10.0) & 0.011 & 0.003 & 0.001 & 0.001 & $^{+ <10^{-3} }_{- <10^{-3} }$ & $<10^{-3}$ & 0.001 & $<10^{-3}$ & $<10^{-3}$ \\
\T\B (10.0, 15.0) & 0.0045 & 0.002 & 0.002 & $<10^{-3}$ & $^{+ <10^{-3} }_{- <10^{-3} }$ & $<10^{-3}$ & $<10^{-3}$ & $<10^{-3}$ & $<10^{-3}$ \\
\T\B (15.0, 20.0) & 0.0024 & 0.001 & 0.001 & $<10^{-3}$ & $^{+ <10^{-4} }_{- <10^{-4} }$ & $<10^{-3}$ & $<10^{-3}$ & $<10^{-4}$ & $<10^{-3}$ \\
\T\B (20.0, 30.0) & 0.0014 & 0.001 & 0.001 & $<10^{-3}$ & $^{+ <10^{-4} }_{- <10^{-} }$ & $<10^{-4}$ & $<10^{-3}$ & $<10^{-4}$ & $<10^{-4}$ \\ \hline
 
\end{tabular}
\end{center}
\end{table}
 
\begin{table}[H]
\caption{Pseudorapidity distribution unfolded to the particle level for the $\Lambda$ total sample, and including invisible decays, along with the statistical and systematic uncertainties.}\label{tab:lambdaEtaJetUnfold}
\vspace{-0.5cm}
\begin{center}
\begin{tabular}{r l l l l l l l l r} \hline
\multicolumn{1}{c}{$\abseta$} & $1 ~ ~ ~ ~ ~ ~ \mathrm{d}N$ & \multicolumn{1}{l}{Stat.} & \multicolumn{1}{l}{Model} & \multicolumn{1}{l}{Track} & \multicolumn{1}{l}{JES} & \multicolumn{1}{l}{JER} & \multicolumn{1}{l}{Pile-up} & \multicolumn{1}{l}{Fiducial} & \multicolumn{1}{l}{Unfold} \\ 
&   $\overline{N_{\mathrm{evt}}} ~ \overline{\mathrm{d}\abseta}$ &   &    &    &   &   &  &  &  \\  \hline
\T\B (0.0, 0.5) & 0.403 & 0.053 & 0.220 & 0.020 & $^{+ 0.009 }_{- 0.006 }$ & 0.006 & 0.032 & 0.006 & $<10^{-3}$  \\
\T\B (0.5, 1.0) & 0.358 & 0.066 & 0.258 & 0.018 & $^{+ 0.006 }_{- 0.008 }$ & 0.003 & 0.029 & 0.004 & $<10^{-3}$  \\
\T\B (1.0, 1.5) & 0.494 & 0.077 & 0.488 & 0.024 & $^{+ 0.011 }_{- 0.011 }$ & 0.008 & 0.040 & 0.011 & $<10^{-3}$  \\
\T\B (1.5, 2.5) & 0.271 & 0.077 & 0.331 & 0.013 & $^{+ 0.003 }_{- 0.006 }$ & 0.007 & 0.022 & 0.003 & $<10^{-3}$  \\ \hline
\end{tabular}
\end{center}
\end{table}

\printbibliography

\clearpage % ATLAS Collaboration author list
% Reference date of TOPQ-2016-05 is 2019-02-21
% Author list last updated on date 23-DEC-19
% Data extracted on 23-Dec-2019 for paper reference TOPQ-2016-05
% at 3:58pm
 
\begin{flushleft}
{\Large The ATLAS Collaboration}

\bigskip

G.~Aad$^\textrm{\scriptsize 101}$,    
B.~Abbott$^\textrm{\scriptsize 128}$,    
D.C.~Abbott$^\textrm{\scriptsize 102}$,    
O.~Abdinov$^\textrm{\scriptsize 13,*}$,    
A.~Abed~Abud$^\textrm{\scriptsize 70a,70b}$,    
K.~Abeling$^\textrm{\scriptsize 53}$,    
D.K.~Abhayasinghe$^\textrm{\scriptsize 93}$,    
S.H.~Abidi$^\textrm{\scriptsize 167}$,    
O.S.~AbouZeid$^\textrm{\scriptsize 40}$,    
N.L.~Abraham$^\textrm{\scriptsize 156}$
% ...
\bigskip
\\

% ...
$^{13}$Institute of Physics, Azerbaijan Academy of Sciences, Baku; Azerbaijan.\\
% ...
$^{40}$Niels Bohr Institute, University of Copenhagen, Copenhagen; Denmark.\\
% ...
$^{53}$II. Physikalisches Institut, Georg-August-Universit\"{a}t G\"ottingen, G\"ottingen; Germany.\\
% ...
$^{70}$$^{(a)}$INFN Sezione di Pavia;$^{(b)}$Dipartimento di Fisica, Universit\`a di Pavia, Pavia; Italy.\\
% ...
$^{93}$Department of Physics, Royal Holloway University of London, Egham; United Kingdom.\\
% ...
$^{101}$CPPM, Aix-Marseille Universit\'e, CNRS/IN2P3, Marseille; France.\\
$^{102}$Department of Physics, University of Massachusetts, Amherst MA; United States of America.\\
% ...
$^{128}$Homer L. Dodge Department of Physics and Astronomy, University of Oklahoma, Norman OK; United States of America.\\
% ...
$^{156}$Department of Physics and Astronomy, University of Sussex, Brighton; United Kingdom.\\
% ...
$^{167}$Department of Physics, University of Toronto, Toronto ON; Canada.\\
% ...
$^{a}$ Also at Borough of Manhattan Community College, City University of New York, New York NY; United States of America.\\
$^{b}$ Also at Centre for High Performance Computing, CSIR Campus, Rosebank, Cape Town; South Africa.\\
% ...
$^{*}$ Deceased

\end{flushleft}

% Created with Glance <Atlas.Glance@cern.ch>

\end{document}